\documentclass{aa}  
\usepackage{graphicx}
\pdfoutput=1
\usepackage{txfonts}
\usepackage[font=small, labelfont=bf]{caption}
\usepackage{subcaption}
\usepackage{booktabs,siunitx}
\usepackage{babel,blindtext}
\usepackage{natbib}
\bibpunct{(}{)}{;}{a}{}{,} 
\usepackage{enumitem}
\usepackage{subcaption}
\usepackage{graphicx} 
\usepackage[colorlinks=true, linkcolor = blue, urlcolor=blue, citecolor = blue]{hyperref}
\usepackage{longtable}
\begin{document} 

   \title{Towards a census of high-redshift dusty galaxies with $\mathit{Herschel}$}

   \subtitle{A selection of $"500\:\mu\rm m$-risers$"$} 

   \author{D. Donevski\inst{1}
   \and V. Buat\inst{1}
   \and F. Boone\inst{2}
   \and C. Pappalardo\inst{3,4}
   \and M. Bethermin\inst{1,7}
   \and C. Schreiber\inst{5}
   \and F. Mazyed\inst{1}
   \and J. Alvarez-Marquez\inst{6}
   \and S. Duivenvoorden\inst{8}
          }

   \institute{Aix Marseille Univ, CNRS, LAM, Laboratoire d’Astrophysique de Marseille, Marseille, France,\\
              \email{darko.donevski@lam.fr}
               \and
               Universite de Toulouse; UPS-OMP; IRAP; Toulouse, France
            	\and
            Centro de Astronomia e Astrofísica da Universidade de Lisboa, Observatório Astronómico de Lisboa, Tapada da Ajuda,
            1349-018 Lisboa, Portugal
            \and 
            Instituto de Astrofísica e Ciencias do Espaço, Universidade de Lisboa, OAL, Tapada da Ajuda, 1349-018 Lisboa, Portugal
            	\and
            	Leiden Observatory, Leiden University, 2300 RA Leiden, The Netherlands
            	\and
            Departamento de astrofisica, Centro de Astrobiologia (CAB, CSIC-INTA), Carretera de Ajalvir, 28850 Torrejón de Ardoz, Madrid, Spain
            \and
            European Southern Observatory, Karl-Schwarzschild-Str. 2, 85748 Garching, Germany
            \and
            Astronomy Centre, Department of Physics and Astronomy, University of Sussex, Brighton BN1 9QH, UK
             }

   \date{Received ; accepted }

 
  \abstract
   {
   Over the last decade a large number of dusty star forming galaxies has been discovered up to redshift $z=2-3$ and recent studies have attempted to push the highly-confused $\mathit{Herschel}$ SPIRE surveys beyond that distance. 
   To search for $z\geq4$ galaxies they often consider the sources with fluxes rising from 250 $\mu$m to $500\:\mu$m (so-called "500 $\mu$m-risers"). $\mathit{Herschel}$ surveys offer a unique opportunity to efficiently select a large number of these rare objects, and thus gain insight into the prodigious star-forming activity that takes place in the very distant Universe.} 
   {
   {We aim to implement a novel method to obtain a statistical sample of  "500 $\mu$m-risers" and fully evaluate our selection inspecting different models of galaxy evolution.}}
   {We consider one of the largest and deepest ${\it Herschel}$ surveys, the Herschel Virgo Cluster Survey. We develop a novel selection algorithm which links the source extraction and spectral energy distribution fitting. To fully quantify selection biases 
   we make end-to-end simulations including clustering and lensing.}
   {We select 133 "500 $\mu$m-risers" over 55 deg$^{2}$, imposing the criteria: $S_{500}>S_{350}>S_{250}$, $S_{250}>13.2$ mJy and $S_{500}>$ 30 mJy. 
   Differential number counts are in a fairly good agreement with models, displaying better match than other existing samples. The estimated fraction of strongly lensed sources is $24^{+6}_{-5}\%$ based on models.} 
{ We present the faintest sample of "500 $\mu$m-risers" down to $S_{250}=13.2$ mJy. We show that noise and strong lensing have an important impact on measured counts and redshift distribution of selected sources. We estimate the flux-corrected star formation rate density at $4<z<5$ with the "500 $\mu$m-risers" and found it close to the total value measured in far-infrared. It indicates that colour selection is not a limiting effect to search for the most massive, dusty $z>4$ sources.}
	

\maketitle

\section{Introduction}
The abundance of dusty galaxies at high-redshifts ($z>4$) constrains our theories about early galaxy formation, since it is generally stated they are the progenitors of massive ellipticals seen in overdense regions of the local Universe (e.g. \citealt{eales17}, \citealt{toft15}, \citealt{simpson14}, \citealt{casey14}). 

The widely accepted picture is that dusty star-forming galaxies (DSFGs) occupy most massive halos in early epochs and lie on the most extreme tail of the galaxy stellar mass function (e.g. {\citealt{ikarashi17}, \citealt{fudamoto17}, \citealt{oteo16a}, \citealt{mancuso16}, \citealt{michalowski14}).
These DSFGs are usually selected in the far-infrared (FIR) regime where the star formation rates can be directly measured. 

Large FIR surveys, such as those conducted with the ${\it Herschel}$ Space Observatory (\citealt{pilbratt10}), provide an opportunity to build a thorough census of prodigious starbursts over cosmological redshifts using wide and blind concept of searches. The $\mathit{Herschel}$ SPIRE photometer (\citealt{griffin10}) was often used for mapping large areas at wavelengths of 250 $\mu$m, 350 $\mu$m and 500 $\mu$m. The redshift peak of most $\mathit{Herschel}$ detected sources matches with the redshift where galaxies have formed most of their stars ($z\sim2$, \citealt{pearson13}, \citealt{lapi11}, \citealt{amblard10}). Considering that rest-frame dust Spectral Energy Distribution (SED) of a galaxy typically peaks between 70-100 $\mu$m, colours of sources in {\it Herschel} SPIRE bands were used to select candidate high-redshift dusty objects. To search for $z\gtrsim4$ candidates, there is a particular interest to exploit the sources having red SPIRE colours, with rising flux densites from 250 to 500 microns (so-called "500 $\mu$m-risers"). Such galaxies should lie at $z\geqslant4$ unless they have dust temperatures that are notably lower than is seen in local FIR-bright equivalents (\citealt{asboth16}, \citealt{yuan15}, \citealt{dowell14}, \citealt{roseboom12}).

If the selection of "500 $\mu$m-risers" is free of contaminants such are blended systems and powerful non-thermal sources (e.g. quasars), it is expected to offer us insight into very distant and dusty, star-forming galactic events. Recently, a rapidly flourishing literature on dusty high-$z$ galaxies has grown, including handful of  serendipitously discovered "500 $\mu$m-risers" (e.g. \citealt{daddi09}, \citealt{cox11}, \citealt{capak11}, \citealt{combes12}, \citealt{vieira13}}, \citealt{miettinen16}, \citealt{negrello17}).  
However, these findings had serious shortcomings - they were limited to few objects with red SPIRE colours.\footnote{Throughout the text we adopt following terminology regarding colours of IR-detected sources: red colours and red sources refer to "500 $\mu$m-risers", while "350 $\mu$m peakers" refer to galaxies having SED peak at 350 $\mu$m. }

Being primarily focused on individual (usually strong lensing) candidates or millimeter selected samples regardless of the galaxy colour, they poorly constrained the statistics of "500 $\mu$m-risers".  
To derive a larger number of potentially unlensed "500 $\mu$m-risers" and analyse them in a more standardized manner, several works used map-search technique (\citealt{asboth16}, \citealt{dowell14}, \citealt{ivison17}). They have led to the discovery of most distant dusty starburst galaxies known to date: SPT0311-58 at $z=6.902$ (\citealt{strandet17}), HFLS3 at $z=6.34$ (\citealt{riechers13}) and G09-83808 at $z=6.02$ (\citealt{zavala17}, \citealt{fudamoto17}). Furthermore, studies that used lowest-resolution {\it Herschel} SPIRE-maps to select "500 $\mu$m-risers" (\citealt{dowell14} and \citealt{asboth16}) claimed significant overprediction of observed number of "500 $\mu$m-risers" with those predicted by existing models (e.g. \citealt{hayward13}, \citealt{b12}). 

Even if $\mathit{Herschel}$ offers a direct insight to the IR emission of high-$z$ objects, there are critical limitations like sensitivity of detectors and low spatial resolution. These are responsible for biases such as source confusion. The sensitivity of SPIRE instrument allows to directly detect only the brightest, thus rarest objects at the tip of luminosity function (\citealt{cowley15}, \citealt{karim13}, \citealt{oliver10}) and we therefore need large surveys to increase the statistics. Study of \cite{dowell14} analyzed maps of three different extragalactic fields observed as part of the HerMES  ($\mathit{Herschel}$ Multi-tiered Extragalactic Survey, \citealt{oliver10}) program, while \cite{ivison17} and \cite{asboth16} probed much wider but shallower area of $\mathit{H}$-ATLAS and HeLMS (HerMES Large Mode Survey) field respectively (see \hyperref[sec:2.1]{Section 2.1} for details about field's properties).

In this work we aim to introduce a slightly different approach to build a statistically significant sample of red, potentially $z>4$ sources. The new selection scheme we propose is motivated by the size and the depth of the field we chose to investigate. $\mathit{Herschel}$ Virgo Cluster Survey (HeViCS, \citealt{davies10}) is deeper than the one used in the analysis of \cite{asboth16} and \cite{ivison17}, and larger than the area analysed by \cite{dowell14}.

The paper is organized as follows: in \hyperref[sec:2]{Section 2} we describe the methods we used for the data analysis and our new selection criteria for "500 $\mu$m-risers". In \hyperref[sec:3]{Section 3} and \hyperref[sec:4]{Section 4} we present expected redshift/luminosity trend of selected galaxies and differential number counts. In \hyperref[sec:5]{Section 5} we compare our results to different models. We perform simulations to review all selection biases and highlight the necessity of a further refinement of selection criteria in searching for $z\gtrsim4$ sources. The nature of "500 $\mu$m-risers" and main conclusions are outlined in \hyperref[sec:6]{Section 6} and \hyperref[sec:6]{Section 7} respectively. We assume a \cite{planck16} cosmology.  

\section{Data Analysis}
   \label{sec:2}
%
   \subsection{HeViCS field}
   \label{sec:2.1}

    HeViCS is a fully-sampled survey that covered a region centered at the Virgo cluster (\citealt{davies10}, \citealt{davies12}). It is one of the largest uniform Herschel surveys, and its main advantage is the sensitivity and the uniformity of data. In this survey, ${\it Herschel}$  observed four overlapping fields (fields V1-V4) in fast parallel-mode. The total entire survey region is 84 square degrees, where 55 square degrees are observed at unvarying depth with eight orthogonal cross scans (see \citealt{auld13} and \citealt{ciro15} for more details). 
    
     HeViCS observations reached a depth close to the confusion limit at the shortest SPIRE wavelength (250 $\mu$m). Because of the number of repeated scans, instrumental noise is significantly reduced in HeViCS maps, giving the 1$\sigma$ levels of 4.9, 4.9 and 5.5 mJy  at 250 $\mu$m, 350 $\mu$m and 500 $\mu$m respectively (\citealt{auld13}). In the overlapped, deeper regions recorded by 16 scans these values are even smaller, namely 3.5, 3.3 and 4.0 mJy. Due to the presence of bright sources (see \hyperref[sec:2.3]{Section 2.3}), global noise estimation is not a straightforward task. We excluded bright sources from the map by masking them, and after their removal the global noise is derived from the variance of the map. It reaches the 1$\sigma$ values of 6.58, 7.07 and 7.68 mJy (250 $\mu$m, 350 $\mu$m and 500 $\mu$m respectively) for a major area covered by 8 cross-scans. The extensive contributor to the overall noise measured in HeViCS maps is confusion noise, usually defined as the the variance in the sky map due to the fluctuations of unresolved sources inside the SPIRE beam. We calculate confusion using the relation $\sigma_{\rm{conf}} =
         \sqrt{\sigma_{\rm{tot}}^2-{\sigma_{\rm{inst}}^2}}$, where $\sigma_{\rm{tot}}$ is the total noise measured in the map, and $\sigma_{\rm{inst}}$ is the instrumental noise. Values determined for the confusion noise are 4.4, 5.2 and 5.5 mJy at 250 $\mu$m, 350 $\mu$m and 500 $\mu$m band respectively, and almost identical to the ones presented in \cite{auld13}. These values are also close to the confusion noise measured in HerMES maps, within twice the uncertainty of 3$\sigma$-clipping estimates from \cite{nguyen10} (3.8 mJy, 4.7 mJy and 5.2 mJy at 250 $\mu$m, 350 $\mu$m and 500 $\mu$m band respectively).
         In this work we use only SPIRE data. In parallel, each HeViCS tile has been observed by the $\mathit{Herschel}$ PACS instrument, but the depth of PACS data ($5\sigma_{\rm tot}$\:=70 mJy at $100\:\mu$m, \citealt{pappalardo16}) is not sufficient to directly detect our "FIR-rising" sources. 
PACS data at 100 $\mu$m and 160 $\mu$m will be added together with a deep optical-NIR maps from the Next Generation Virgo Cluster Survey (NGVS, \citealt{ngvs12}) in a following paper analysing ancillary data.  

\subsection{An overview of other $\mathit{Herschel}$ fields}
\label{sec:2.2}
There are several large ${\it Herschel}$  fields (e.g. with an observed area $\theta>10$ deg$^2$) used for detection of "500\:$\mu$m-risers".  Summary of their properties is shown in Table \hyperref[tab:1]{$1$}. All the values are taken from the literature (\citealt{oliver12}, \citealt{wang15}).
\begin{table*}[ht]
	\caption{Properties of different large fields mapped by ${\it Herschel}$}
	\label{tab:1}   
	\centering   
	\vspace{-0.14cm}   
	\begin{tabular}{cc c c c c c c c}     %
		\hline  
		\toprule    
		Field & Mode & N(rep) & Time & Area & Noise level in $m$Jy\\
		&  & & [hr] & [deg$^{2}$]& $5\sigma$ at 250 $\mu$m \\
		(1) & (2) & (3) & (4) & (5) & (6)\\
		\hline 
		\\
		\textbf{HeViCS}  &  Parallel & 8 & 286 & 55 &  30.5\\
		\\
		\hline 
		\\
		\textbf{H-ATLAS} &  Parallel & 2 & 600 & 550 &  56.0\\
		\\
		\textbf{HerMES}\\
		\\
		FLS & Parallel & 2 & 17.1 & 6.7 & 25.8\\
		Bootes NDWFS & Parallel & 2 & 28 & 10.5 &  25.8\\
		ELAIS-N2 &  Parallel & 2 & 28 & 12.28 &  25.8\\
		Lockman-Swire & Parallel & 4 & 71.2 & 16 &  13.6\\
		XMM-LSS SWIRE & Parallel & 4 & 71.2 & 18.87 &  25.8\\
		HeLMS & Sp.Fast & 2 & 103.4 & 274 &  64.0\\ 
		
		\hline
		\bottomrule                  
	\end{tabular}\\
	\vspace{0.33cm}

	\vspace{0.4cm} 
	
	\caption*{\textbf{Notes:} Columns (1) Name of the field observed by ${\it Herschel}$; (2) ${\it Herschel}$ observing mode ; (3) The total number of repeats of the observing mode in the set; (4) Total time of observations; (5) Field area of good pixels; (6) Total noise from the literature. Noise is calculated using the relation $\sigma_{\rm{tot}} =
		\sqrt{\sigma_{\rm{conf}}^2+{\sigma_{\rm{inst}}^2}}$; The later six fields are areas with different design levels nested as a part of HerMES: 
		Total area covered in HerMES is 380 deg$^{2}$. Shallow HeLMS field covers the area of 274 deg$^{2}$, and deeper Level 1- Level 6 fields cover the total area of about 80 deg$^{2}$. FLS and Lockman-SWIRE have been used to probe "500 $\mu$m-risers" selection by \cite{dowell14}, while \cite{asboth16} applied the same selection method in the HeLMS field.}
\end{table*}
The $H$-ATLAS survey (\citealt{eales10}) is used to select red candidates by \cite{ivison17}. $H$-ATLAS is designed to uniformly cover 600 deg$^{2}$ of sky, but with its two scans survey did not reach the level of confusion noise at 250 $\mu$m. 
Studies of \cite{asboth16} and \cite{dowell14} acquired the data from different fields which are part of HerMES survey (\citealt{oliver12}). The HerMES survey observed 380 $\rm deg^{2}$ of the sky. The survey has a hierarchical structure containing 7 levels, ranging from very deep observations of clusters to wider fields with varying size and depth. The largest (and the shallowest) observed area is HeLMS, with its 274 square degrees. 
HeViCS maps consist of two or four time more scans comparing to other fields listed in Table \hyperref[tab:1]{$1$}. It leads to a reduction of instrumental noise by a factor of $\sqrt{2}$. The global 250 $\mu$m noise in HeViCS is smaller than in $H$-ATLAS and HeLMS. However, it is still higher than in other HerMES fields, showing that confusion is a very compelling supplier to the noise budget for point sources in the HeViCS field. 
	We clarify that statement repeating our analysis on regions overlapped between the tiles, which have greater coverage. We found no significant noise reduction, implying that the maps are dominated by confusion noise. 

\subsection{Map filtering}
\label{sec:2.3}
Prior to performing source extraction on SPIRE maps, we reduce the background contamination. 250 $\mu$m map of V2 field in HeViCS is strongly affected by galactic cirrus emission. This contamination peaks at around 150-200 $\mu$m (\citealt{bracco11}, \citealt{valiante16}), implying it is the brightest in the shortest SPIRE bands. 
The main effect of cirrus emission is to increase the confusion in the maps, but the small-scale structure within it can also lead to spurious detections in the catalogues. 

We follow the method applied by \cite{ciro15}. Using the \texttt{SEXtractor} (\citealt{sex96}) we re-grid the 250 $\mu$m map into meshes larger than the pixel size. The \texttt{SEXtractor} makes an initial pass through the pixel data, and compute an estimator for the local background in each mesh of a grid that covers the whole adopted frame. We apply repetitive iterations to estimate the mean and standard deviation of the pixel value distribution in boxes, removing outlying pixels at each iteration. Important step in this procedure is the choice of the box size, since we do not want the background estimation be affected by the presence of objects and random noise. The box size should generally be larger than the typical size of sources in the image, but small enough to encapsulate any background variations. 
	We therefore fix the mesh size to 8 pixels, adopting the result from \cite{ciro15}. 
 The local background is clipped iteratively to reach $\pm3\sigma$ convergence around its median. After the background subtraction, the number of sources appears uniform for regions with different cirrus emission. We use the background subtracted map as an input for source extraction process described in next subsection.

\subsection{Extraction of sources}
\label{sec:2.4}

Blind SPIRE source catalogues have been produced for HeViCS (\citealt{ciro15}). Nonetheless, when density of sources is very high, which is the case in highly crowded HeViCS field, blind source extraction cannot separate blended point sources in an efficient way. Additionally, it remains difficult to properly cross-match sources at different wavelengths, since central positions from blind catalogues are not well constrained. 
To deal with source multiplicity we choose to perform extraction of 350 $\mu$m and 500 $\mu$m fluxes at exact $\mathit{a-priori}$ position of 250 $\mu$m band detections, allowing much precise identification work. The potential limitation of such a method is that we might be eventually missing some "$500\:\mu$m-risers" that are not included in the prior list after the first iteration of source extraction. We thus run our method iteratively and add new sources that may appear in the residuals at each iteration.

\begin{figure}
	\centering
	\hspace{-0.5cm}
	\includegraphics [width=8.79cm]{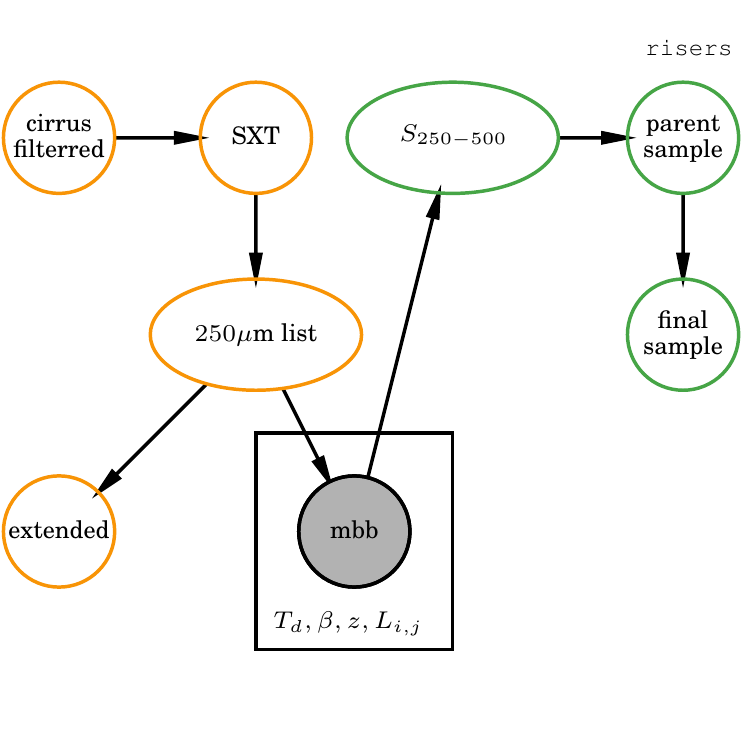}
	\caption{Schematic representation of selection of "$500\:\mu$m-risers" in the HeViCS field. Coloured in orange are segments of source extraction prior to use \texttt{MBB-fitter}. They are (from upper left, following the arrows): 1.) Subtraction of strong cirrus emission 2.) \texttt{SUSSEXtractor} list of initial (threshold=3) 250 $\mu$m detections;  3.) Assumed $\mathit{a-priori}$ list cleaned from extended sources. 
  Enveloped by smaller black square are parameters considered for the fitting procedure with \texttt{MBB-fitter} and it corresponds to all pixels in the map where we have 250 $\mu$m detections; Coloured in green are segments of our selection after performing the photometry, subsequently: 1.) MBB photometry ($S_{250-500}$) at SPIRE wavelengths using 250 $\mu$m priors ; 2.) Parent list of "$500\:\mu$m-risers" (140 sources in total), not cleaned from strong synchrotron contaminants. 3.) Final list (133 in total) of "$500\:\mu$m-risers" after excluding local radio-sources. We apply following selection criteria for the final sample: $S_{500}>S_{350}>S_{250}$, $S_{250}>13.2$ mJy and $S_{500}>$30 mJy.}
\label{fig:Fig.1}
\end{figure}

We use source finding algorithm optimized for isolated point-sources, \texttt{SUSSEXtractor} (\citealt{sussex07}), to create the catalogue of galaxies detected in the $250\:\mu$m map as a prior to extract the flux densities at longer wavelengths. 
We then implement our novel technique (see \hyperref[sec:2.6]{$\rm Section\:2.6$}) where source deblending and single temperature modified blackbody (MBB) fitting are combined in the same procedure.
In following we explain our source extraction pipeline (see \hyperref[fig:Fig.1]{Fig.1}  for graphical description):
\begin{enumerate}
	
	\item We run \texttt{SUSSEXtractor} with use of fully-overlapped HeViCS 250-micron map. The \texttt{SUSSEXtractor} works on the flux-calibrated, Level 2 $\it{Herschel}$ SPIRE maps. We create a point response function (PRF) filtered image, smoothed with the PRF. In \texttt{SUSSEXtractor} PRF is assumed to be Gaussian by default, with full-width-half-maximum (FWHM) provided by the FWHM parameter. We apply values of 17.6", 23.9" and 35.2" at 250, 350 and 500 $\mu$m respectively. Subsequently, pixel sizes at these bands are 6", 10" and 14". 
The algorithm searches for a local maximum which is the highest pixel value within a distance defined by pixel region. The position assigned to the possible source is then refined by fitting a quadratic function to certain pixels in the PSF-filtered image. 
 
	\item To search for even fainter sources that are closer to the confusion limit and usually masked/hidden in highly confused fields like HeViCS, we perform additional step looking for detections in our residual maps. 
	
Applying such additional step, we increase total number of sources by around $4\%$. Errors in the position estimated to be a source are determined as 0.6$\times$(FWHM/signal-to-noise), up to a maximum of 1.0 pixels, as suggested in the literature (\citealt{ivison07}). 

\item We build an initial list of 250 $\mu$m sources selecting all point sources above the threshold=3 (Bayesian criteria in \texttt{SUSSEXtractor}). We also impose the flux density cut choosing the values equal or higher than 13.2 mJy, which corresponds to $S_{250}\gtrsim3\sigma\rm_{conf}$. 
As a result of our source extraction pipeline at 250-micron maps, we listed 64309 sources, similarly distributed among the four (V1-V4) fields.
\item List of 250-detections is then divided in "no-neighbour" list of sources (250 $\mu$m sources without another detection inside the 500 $\mu$m beam) and "close-neighbour" list where we add all sources having another 250 $\mu$m detection within the 500 $\mu$m beam. 
Prior to assign initial 250 $\mu$m list as an input to our modified blackbody fitting procedure, we clean the catalogue from potentially extended sources.

 \end{enumerate}

\subsection{Extended sources}
\label{sec:2.5}
The Virgo cluster is one of the richest local clusters and we expect to have a significant number of extended sources. Objects that are extended on the SPIRE beam scale (see \citealt{wang15} or \citealt{rigby11}) are not expected to be accurately identified with point-source extracting codes, and large galaxies may be misidentified as multiple point sources. To prevail the problem, we implement the same method used in \cite{ciro15}.
They define a mask using the recipe of \texttt{SEXtractor}, which detects a source when a fixed number of contiguous pixels is above a $\sigma$-threshold estimated from the background map. We keep the same value which has been tested in \cite{ciro15} - 70 contiguous pixels above the $1.2\:\sigma$. 
 In this case, most of the sources larger than 0.7 arcmin$^{2}$ are rejected from the sample. Additionally, we cross-match all remaining HeViCS 250 $\mu$m detections with their nearest (within 36") counterpart in the $\mathit{2MASS}$ Extended Source Catalog (\citealt{2mass}). We remove any detection supposed to be a counterpart with a Kron elliptical aperture semi-major larger than 9". In total we suppressed 812 sources from the analysis, thus decrease the number of 250-detections in our parent list from 64309 to 63497.
   
  List of sources detected at 250 $\mu$m down to 13.2 mJy, cleaned from galactic cirrus contaminants and extended sources, is further used as $\mathit{a-priori}$  list for our simultaneous modified blackbody fitting. 


\subsection{Modified blackbody fitter}
\label{sec:2.6}
\begin{figure*}
	\hspace {-2.0cm}
	\includegraphics[width=21.69cm] {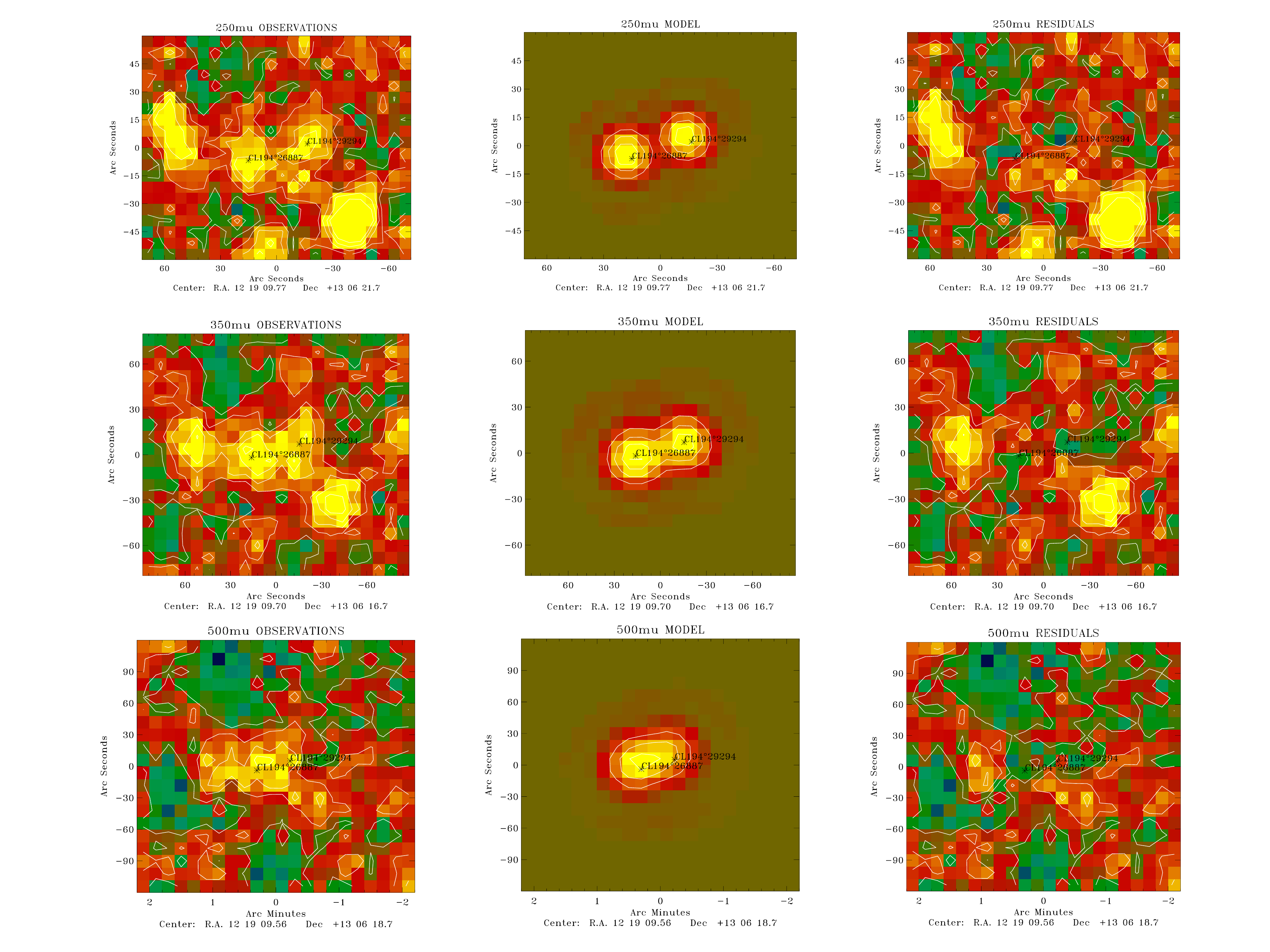}  
	\caption{Example of simultaneous prior-source deblending with \texttt{MBB-fitter}. Sources are taken from our "neighbour" list of 250-detections. Columns from left to right: 250 $\mu$m, 350 $\mu$m an 500 $\mu$m maps of sources (first column), subtracted models (second column) and residuals (third column). Note that in this example two central sources are considered for the fitting.}
	\label{fig:Fig.2}
\end{figure*}
As described in \hyperref[sec:2.1]{$\rm Section\:2.1$}, our maps are limited by confusion noise caused by the high density of sources relative to the resolution of the $\mathit{Herschel}$ instrument. In cases where SPIRE images have multiple sources per beam, measured fluxes might be biased if we treat several blended
sources as one. There are different approaches for source "de-blending" which has been invoked in the literature. We can separate them in three groups: Firstly, they are "de-blending" methods that use only positional priors as an input information (e.g. \citealt{elbaz10}, \citealt{swinbank14}, \citealt{fastphot}, \citealt{roseboom10}). The second group of de-blenders consists of methods which combine positional information with statistical techniques (\citealt{xid+17}, \citealt{safarzadeh15}). For example, \cite{safarzadeh15} and \cite{xid+17} developed the codes based on the Monte Carlo Markov Chain (MCMC) sampling for prior source detections. While \cite{safarzadeh15} used Hubble Space Telescope (HST) H-band sources as a positional argument,  \cite{xid+17} developed a similar MCMC-based prior-extraction for Spitzer 24 $\mu$m detections. Such an algorithm is efficient in obtaining Bayesian PDFs for all priors. However, they might still have some limitations, and the most important one is potential inability of unveiling the true high-$z$ sources in the case they are not included in the initial list. Finally, third group of de-blenders consists from those using SED modelling as an addition to positional arguments (\citealt{mackenzie14}, \citealt{shu16}, \citealt{tphot17}, \citealt{liu17}). 

Due to the lack of $\mathit{Spitzer}$ 24 $\mu$m data for the HeViCS field, and unsufficiently deep existing optical images, we have no opportunity to use positional priors from shorter wavelength surveys, and we based our analysis on SPIRE data instead. We further use models to test the whole procedure of selecting "$500\:\mu$m-risers" (see \hyperref[sec:5]{$\rm Section\:5$}). 

\texttt{MBB-fitter} is a code developed to extract sources from multiwavelength bolometric observations (Boone et al, in prep.). It combines positional priors and spectral information of sources, such that SEDs of fitted galaxies should follow modified blackbody (MBB) shape as defined by \cite{blain03}. 
Our MBB deblending approach is thus very similar to the method applied by \cite{mackenzie14}. 
Although such a model can encapsulate just a segment of the general complexity of astrophysical processes in a galaxy, it can account for the SED data for a wide variety of dusty galaxies. 
The \texttt{MBB-fitter} method is described in detail and is tested on simulations in Boone et al. (in prep.). We summarize here the main points.
In \texttt{MBB-fitter} the map ($M$) of a sky region is described as a regular grid corresponding to the sky flux density distribution convolved by the point spread function (PSF) of an instrument and sampled in pixels. Thus $M_{i}=M(\alpha_{i}, \delta_{i})$ refers to the value of the map at the $i^{\rm th}$ pixel with coordinates $\alpha_{i}, \delta_{i}$. 
We further assume that the sources have a morphology independent of the frequency. The flux density distributions in the 3D space $s_{k}(\alpha, \delta, \nu)$, where $\nu$ is frequency, can therefore be decomposed into a spatial distribution term (morphology) and a SED term:
   \begin{equation}
   	M_{ij}=  M(\alpha_i, \delta_i, \nu_j)=[\sum_{k=1}^{N_{\rm sources}}f_{k}(\nu_{j})\tilde{s}_{k}(\alpha-\alpha_{k}, \delta-\delta_{k})\otimes {PSF_{j}}]_{i},
   	\label{eq.1}
   \end{equation}
where $PSF_{j}$ is the PSF of the map at given frequency ($\nu_{j}$), $N_{\rm sources}$ is a number of sources considered for the fitting, with its reference coordinates $(\alpha_{\kappa}, \delta_{\kappa})$. Here we assume that thermal continuum SED of galaxies in FIR domain can be modelled as a modified blackbody of the form (\citealt{blain03}): 
	\begin{equation}
	f(\nu, T_{\rm d}, \beta, \gamma) \propto
	 \begin{cases}
	  \nu^{3+\beta}/[\rm exp(h\nu/\mathit{kT_{\rm d}})-1],\:if\:\nu<\nu_{\rm w} \\
	\nu^{-\gamma}, {\rm \:if\:\nu\geq\nu_{\rm w}}
	\label{eq.2}
	\end{cases}
	\end{equation}
	where $\nu_{\rm w}$ is the lower boundary of Wien’s regime, $h$ is the Planck constant and $k$ is the Stefan-Boltzmann constant. Thus, \hyperref[eq.1]{Eq.1} describes a general model for a set of maps at different wavelengths, and each source SED is defined by a set of parameters arranged into a vector of following form $P_{\kappa} =  [L_{{\rm IR}_\kappa}, T_{{\rm d}_\kappa}, z_\kappa, \beta_\kappa, \gamma_{\kappa}]$, where $z_\kappa$ is the redshift of given source, $\beta_\kappa$ its emissivity index, $\gamma_{\kappa}$ is the index of the power law to substitute the Wien’s regime and $L_{\rm IR}$ is the IR luminosity of a source. This model is parametrized by the source SED parameters, $P_{\kappa}$, and coordinates $\alpha_{\kappa},\delta_{\kappa}$. The total number of parameters is therefore $N_{\rm par}$=$N_{\rm s}\times (N_{\rm SED_{\rm par}}+2)$, where $N_{\rm SED_{\rm par}}$  is the number of SED parameters and $N_{\rm s}$ is the number of sources.

Introduced set of parameters reflects global properties of the source. There are potential degeneracies, meaning that different values of parameters may give very similar SEDs. This is the most prolific challenge when using the FIR peak as a redshift indicator -  dust temperature and the redshift are completely degenerate (see e.g. \citealt{pope10}). 

Defined model can be compared to a set of observed maps and its fidelity can be assessed with a figure of merit such as the chi-square  exemplified as the sum of the pixel deviations:
	\begin{equation}
\chi(P)^2=\sum_{i}^{N_{\rm pix}}\sum_{j}^{N_{\rm freq}}\frac{(M_{ij}(P)-\tilde{M}_{ij})^2}{\sigma_{ij}^{2}},
	\end{equation}
where $\tilde{M}_{ij}$ is the multiwavelength set of maps, $\sigma_{ij}$ is the Gaussian noise level of the $i$-th pixel of the map at the $j$-th frequency, $N_{\rm pix}$ is the number of accounted pixels, while $N_{\rm freq}$ represents the number of maps, which is equal to the number of different frequencies used. 
	
We use the Levenberg-Marquardt algorithm to find the minimum of the chi-square function in the space of parameters. The algorithm also returns an estimation of the errors on the parameters based on the covariance matrix. We note that all the pixels at maps need not to be included in the chi-square computation, it can be restricted to a subset of pixels (contiguous or not) and sky regions without prior sources can be excluded. 

We impose our list of 250-detected sources as a prior list, and use \texttt{MBB-fitter} to perform two-step photometry: 
(1) simultaneously fitting the sources affected by surrounding source-blend (12135 sources in total), and (2) performing the photometry of sources that are more isolated without a surrounding companion inside the beam (51362 sources in total). We set emission spectral slope and dust temperature fixed at $\beta=1.8$, and $T_{\rm d}=38\pm7$ K. 
These values are chosen to provide a very good description of the data and they are based on our current knowledge of dust temperatures in SPIRE detected sources (e.g. \citealt{yuan15}, \citealt{swinbank14}, \citealt{simeon13}, \citealt{lapi11}, see also \citealt{schreiber17}). An example of simultaneous MBB "deblending" is illustrated in \hyperref[fig:Fig.2]{Fig.2}.

\subsection{Final data sample of "$500\:\mu$m-risers"}
\label{sec:2.7}
	
We select the final list of 133 "$500\:\mu$m-risers" over the area of 55 deg$^{2}$. Selected sources fulfil criteria accepted for the final cut: $S_{500}>S_{350}>S_{250}$, $S_{250}>13.2$ mJy and $S_{500}>$ 30 mJy. The full catalogue is presented in Table \hyperref[Table 8.]{$8$}.
	\begin{figure}[ht!]
		\centering
		\hspace{-0.5cm}
		\includegraphics [width=9.49cm]{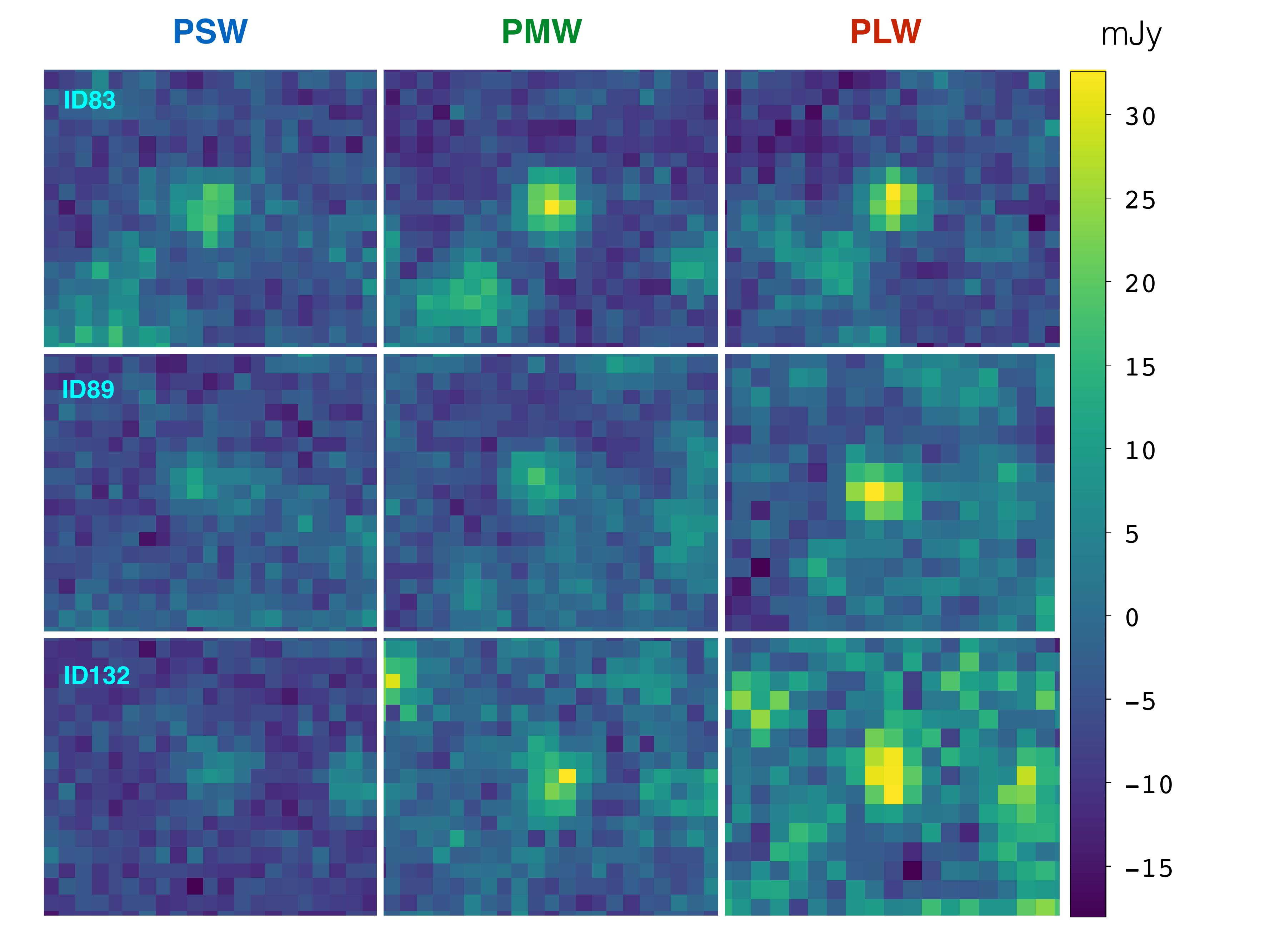}
		\caption{2D image cutouts as an illustration of "$500\:\mu$m-risers" that fulfil our final selection criteria. Presented examples are drawn from our "no-neighbour list", meaning to be isolated, without a clear 250 $\mu$m detections in the radius of at least 36" around their centres. Colorbar shows fluxes measured in mJy.}
		\label{fig:Fig.3}
	\end{figure}
	
	We performed several tests and chose to set flux density cut in 500 $\mu$m band at $S_{500}\simeq30$ mJy, which is related to $4\sigma$ above total noise measured in HeViCS maps. Using this value, we reach the completeness level of $\sim80\%$ at 500 $\mu$m (see \hyperref[fig:Fig.5]{Fig.5}) avoding larger uncertainties in colours due to lower signal-to-noise ratios (see  Table \hyperref[tab:5]{$5$} and \hyperref[sec:5.2]{Section 5.2}). A 250 $\mu$m flux cut ($S_{250}>13.2$ mJy) corresponds to $S_{250}>3\sigma_{\rm conf}$ after applying an iterative $3\sigma$ clipping to remove bright sources (see  \hyperref[sec:2.1]{Section 2.1}). Amongst the final "$500\:\mu$m-risers" (133 in total) we have 11 red candidates with one or two additional 250 $\mu$m detections inside the 500 $\mu$m beam. 
	Example 2D cutouts of several "$500\:\mu$m-risers" from our final sample are presented in \hyperref[fig:Fig.3]{Fig.3}  (see also \hyperref[sec:appC]{Appendix C}). Our final list of "$500\:\mu$m-risers" in the HeViCS field is cleaned from strong radio sources that have flat spectra and very prominent FIR emission (7 objects in total). They may have colours similar to those of dusty, star-forming systems we are interesting to select. Contamination due to radio-bright galaxies is eliminated by cross-matching existing radio catalogues: HMQ (\citealt{hmq15}), NVSS (\citealt{nvss}), FIRST (\citealt{first}) and ALFA ALFA (\citealt{alfaalfa}). Identified radio-bright sources are classified as quasars. 
	\footnote{For the detailed description of this category of dusty, radio sources, see \hyperref[sec:appA]{Appendix A}.} 
	
	\begin{figure}[ht!]
		\vspace{-0.2cm}
		\centering
		\includegraphics [width=9.07cm]{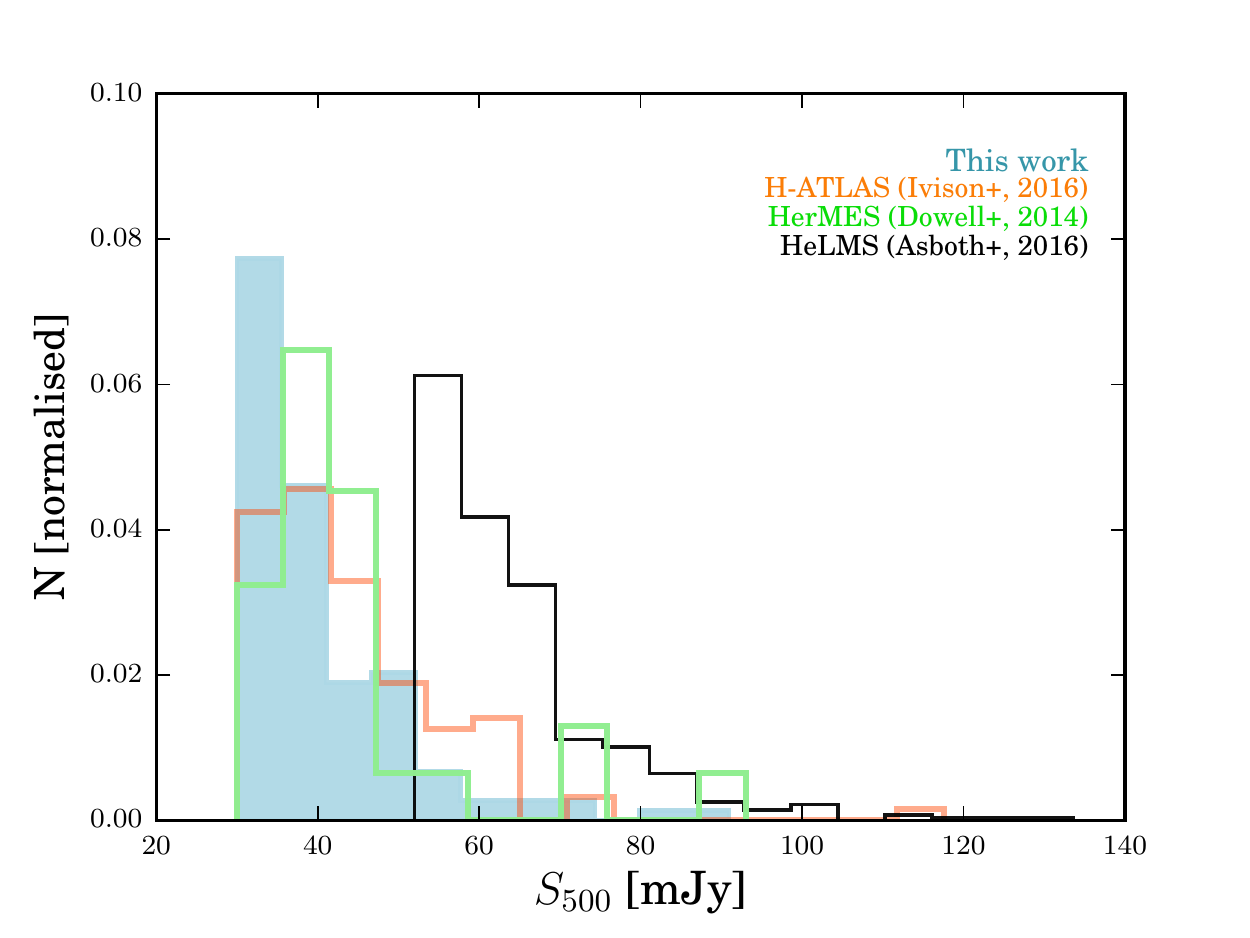}
		\caption{Normalised distribution of 500 $\mu$m fluxes of "$500\:\mu$m-risers". Our sample is represented with filled blue area, while samples of \cite{ivison17}, \cite{dowell14} and \cite{asboth16} are represented by orange, green and black line respectively.}
		\label{fig:Fig.3b}
	\end{figure}


In \hyperref[fig:Fig.3b]{Fig.4} we show actual flux distribution of "500 $\mu$m-risers" selected in this work, along with red sources from other studies. We see that our catalogue contains in average more fainter objects than other existing samples. The median 500 $\mu$m flux of our sample is $38\pm4$ mJy, while median fluxes found in \cite{dowell14}, \cite{asboth16} and \cite{ivison17} are 45 mJy, 65 mJy and 47 mJy respectively. Therefore, our selection is arguably the faintest sample of "500 $\mu$m-risers" available. We note that underlying flux distribution of selected galaxies might be still affected by strong gravitational lensing, which we discuss in detail in \hyperref[sec:6.2] {Section 6.2}. Along with galaxy-galaxy lensing, cluster-lens amplification might affect measured fluxes. The effect is the strongest when the deflector is located half-way between the observer and the lensed object (e.g. \citealt{kneib11}). Strong lensing events are thus more numerous for clusters at $0.1<z<0.5$ (\citealt{johnson14}). Virgo cluster is very close along the line of sight ($z=0.003$) and for this reason we expect negligible impact on SPIRE fluxes of "500 $\mu$m-risers". Colours of sources selected in this work, together with colours of "$500\:\mu$m-risers" with spectroscopic redshifts from other studies are plotted in \hyperref[fig:Fig.27]{Fig.6}. Median observed colour of our sample is $S_{500}/S_{350}=1.11\pm0.10$, whereas median colours from \cite{dowell14}, \cite{asboth16} and \cite{ivison17} are $S_{500}/S_{350}=1.08$, $S_{500}/S_{350}=1.12$ and $S_{500}/S_{350}=1.23$ respectively. In \hyperref[sec:5.1.5] {Section 5.1.5} and \hyperref[sec:5.2] {Section 5.2} we illustrate and discuss how important is the impact of the noise on observed counts and colours. 
\subsection{Completeness and flux accuracy}

We check the quality of our data analysis performing the tests of completeness and flux accuracy. To determine these values, we use real "in-out" simulations.  We calculate completeness by considering the number of injected sources
with certain flux density $S$ which are recovered in the simulated maps as real detections. Since our selection is based on $\mathit{a-priori}$ 250-micron positions, in this test we mostly inspect that band. 
\begin{table}[h]
	\caption{The completeness fraction at SPIRE wavebands measured by injecting artificial sources into real SPIRE maps. Input fluxes are given in first column, while percentages of detected sources (average values per bin) are given in other columns.} 
	\label{tab:2}   
	\centering   
	\begin{tabular}{c c c c c}
		\hline   
		\toprule 
		
		\\
		Flux [mJy] &  V1  &  V2  &  V3  &  V4\\
		& [250 $\mu$m] & [250 $\mu$m] & [250 $\mu$m] & [250 $\mu$m]\\
		\\
		\hline 
		\\
		\centering
		5 &  0.18 & 0.14 & 0.17 & 0.19\\
		10 &  0.37 & 0.40 & 0.39 & 0.38\\
		15 &  0.59 & 0.64 & 0.63 & 0.61\\
		25 &  0.84 & 0.91 & 0.93 & 0.93\\
		35 &  0.94 & 0.97 & 0.96 & 0.97\\
		45 &  0.97 & 0.98 & 0.98 & 0.98\\
		55 &  0.98 & 0.99 & 0.99 & 0.99\\
		65 &  0.98 & 1.0 & 1.0 & 0.99\\
		75 &  0.99 & 1.0 & 1.0 & 1.0\\
		85 &  1.0 & 1.0 & 1.0 & 1.0\\
		95 &  1.0 & 1.0 & 1.0 & 1.0\\
		\\
		& [350 $\mu$m] & [350 $\mu$m] & [350 $\mu$m] & [350 $\mu$m]\\
		\\
		5 &  0.14 & 0.19 & 0.12 & 0.11\\
		10 &  0.30 & 0.43 & 0.29 & 0.28\\
		15 &  0.57 & 0.64 & 0.53 & 0.55\\
		25 &  0.79 & 0.83 & 0.78 & 0.76\\
		35 &  0.92 & 0.94 & 0.93 & 0.93\\
		45 &  0.97 & 0.98 & 0.98 & 0.97\\
		55 &  0.98 & 0.99 & 0.99 & 0.98\\
		65 &  0.99 & 1.0 & 0.99 & 0.99\\
		75 &  0.99 & 1.0 & 1.0 & 1.0\\
		85 &  1.0 & 1.0 & 1.0 & 1.0\\
		95 &  1.0 & 1.0 & 1.0 & 1.0\\
		\\
		& [500 $\mu$m] & [500 $\mu$m] & [500 $\mu$m] & [500 $\mu$m]\\
		\\
		5 &  0.10 & 0.09 & 0.11 & 0.11\\
		10 &  0.21 & 0.20 & 0.18 & 0.18\\
		15 &  0.39 & 0.44 & 0.43 & 0.42\\
		25 &  0.69 & 0.68 & 0.68 & 0.73\\
		35 &  0.91 & 0.88 & 0.89 & 0.94\\
		45 &  0.98 & 0.99 & 0.98 & 0.99\\
		55 &  0.99 & 0.99 & 0.99 & 0.99\\
		65 &  0.99 & 1.0 & 1.0 & 0.99\\
		75 &  0.99 & 1.0 & 1.0 & 1.0\\
		85 &  1.0 & 1.0 & 1.0 & 1.0\\
		95 &  1.0 & 1.0 & 1.0 & 1.0\\
		\\
		\hline     
		\bottomrule          
	\end{tabular} \\
	
\end{table}

\begin{figure}[!htb]
	\vspace{-0.3cm}
	\centering
	\minipage{0.5\textwidth}
	\includegraphics[width=\linewidth]{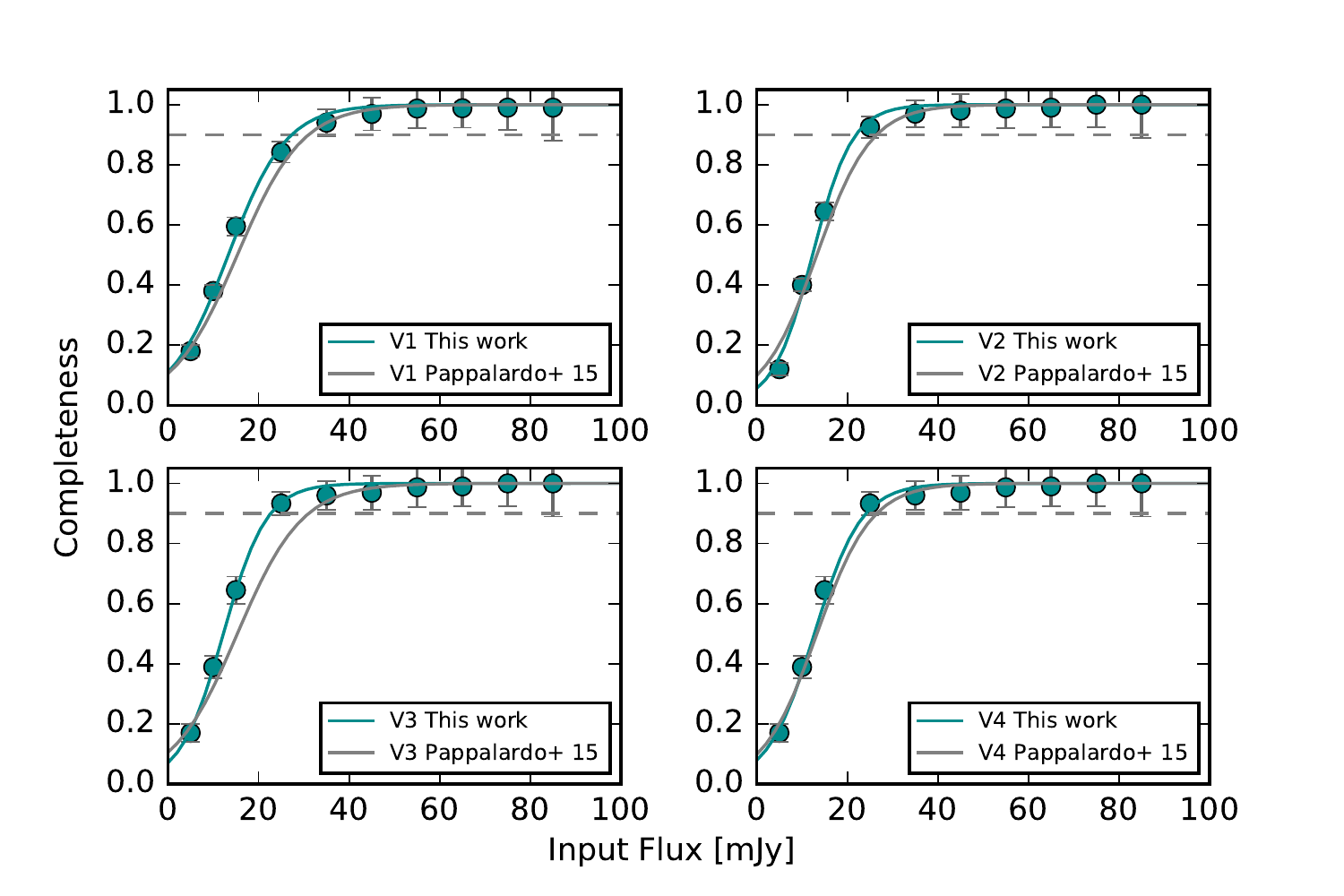}
	\endminipage\hfill
	\minipage{0.41\textwidth}
	\includegraphics[width=\linewidth]{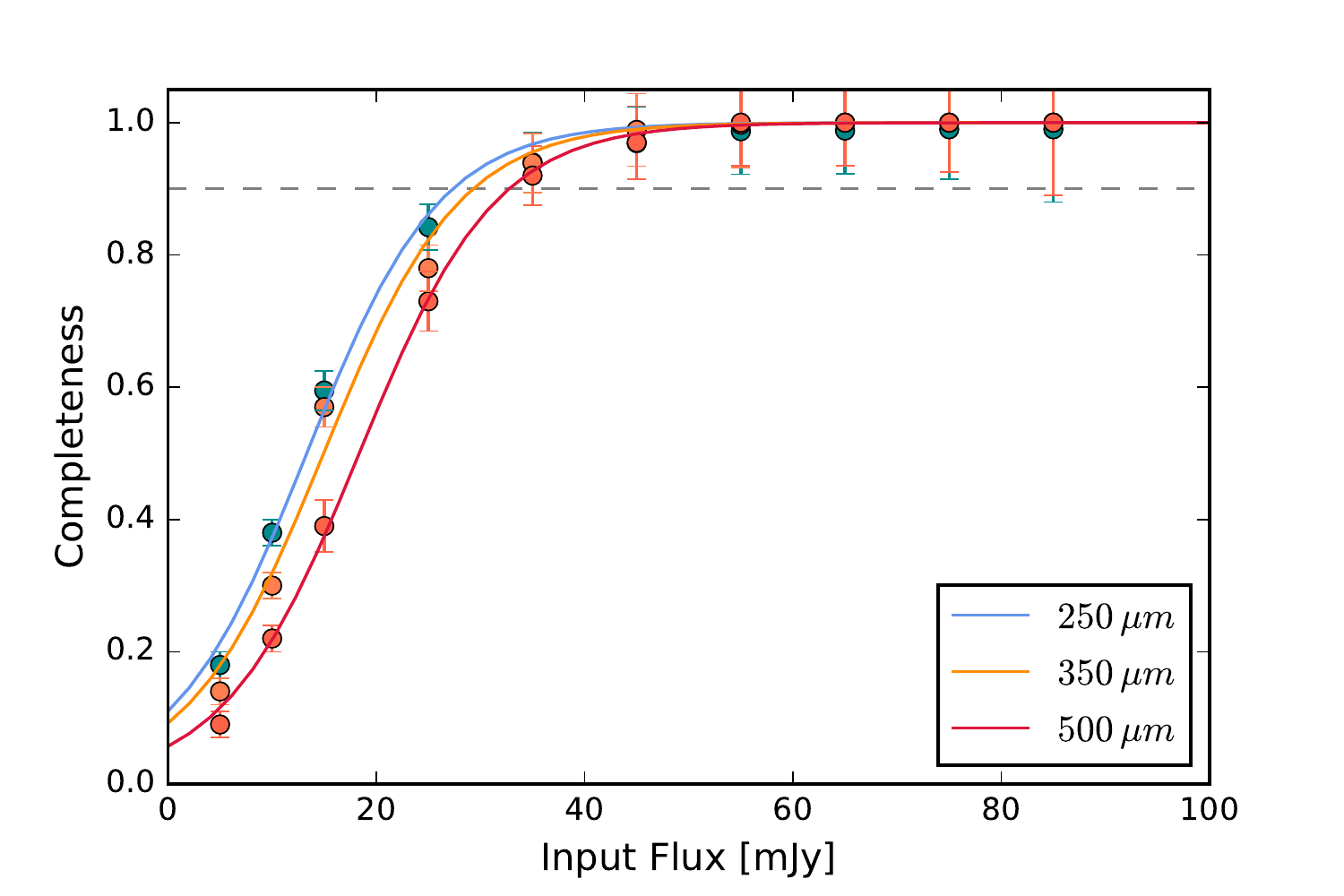}
	\endminipage\hfill
	\minipage{0.41\textwidth}
	\includegraphics[width=\linewidth]{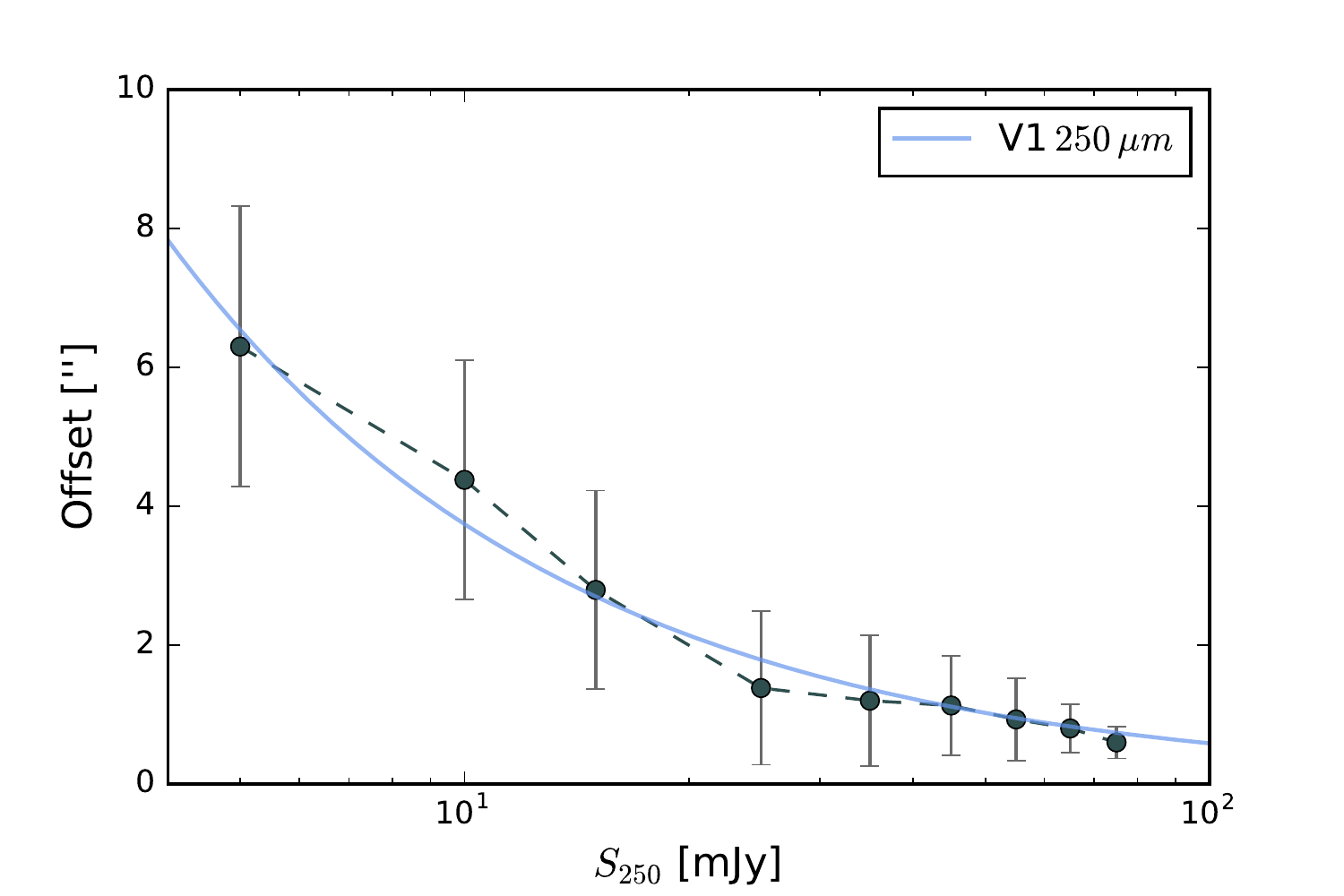}
	\endminipage\hfill
	\minipage{0.41\textwidth}%
	\includegraphics[width=\linewidth]{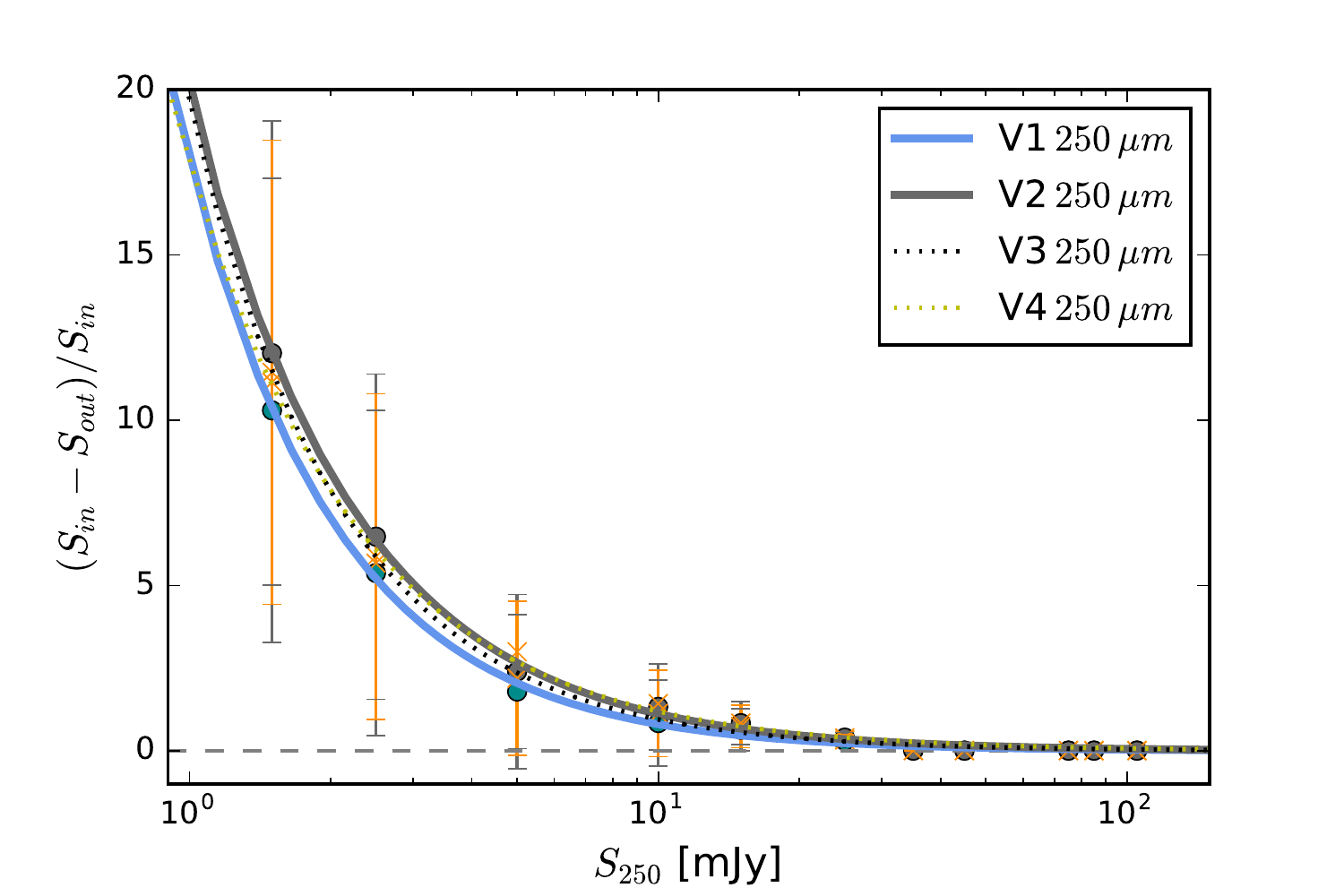}
	\caption{For the HeViCS fields: Completeness at 250 $\mu$m band compared to \citealt{ciro15}, SPIRE completeness for the V1 field, positional accuracy (average radial offset) and flux boosting (the flux difference as a function of input flux) are shown from up to bottom. Simulated (input) flux $S_{\rm in}$ is plotted on $x$-axis for all subplots, while $S_{\rm out}$ refers to observed flux.}
	\label{fig:Fig.4}
	\endminipage
\end{figure}

We use Monte-Carlo simulation, artificially adding Gaussian sources and projecting them at randomized positions to our real 250 $\mu$m data, then convolving with the beam. Simulated sources in 350 $\mu$m and 500 $\mu$m maps are then placed at same positions. FWHM values we used to convolve are 17.6", 23.9" and 35.2", values we consider for our \texttt{SUSSEXtractor} detections. In each HeViCS field we add sources spanning the wide range of flux densities separated in different flux bins (first bin starts from 1-5 mJy, then 5-10 mJy, 10-20 mJy, 20-30 mJy and so, up to the 90-100 mJy). We perform the source extraction pipeline extracting the list of recovered point sources using \texttt{SUSSEXtractor} and searching for matched detections.

The HeViCS field is significantly crowded and we can perform missmatch with an unrelated source close to the input $(x,y)$ coordinate. We match input to output catalogue considering source's distances smaller than half of the 250 $\mu$m beam size (9"), measured from the centre of 250 $\mu$m detection. Quality assessment results are summarized in Table \hyperref[tab:2]{$2$} and \hyperref[fig:Fig.4]{Fig.5}. The plotted completeness curves are the best-fitting logistic functions, describing the completeness as a function of input flux. We can see that positions of recovered sources do not show any significant offset to assigned inputs. More than 70 $\%$ of sources with $S_{250}>13.2$ mJy  have offset lower than 6", which is the value smaller than a 250 $\mu$m pixel size. 
Completeness and flux accuracy are consistent over all four HeViCS fields. 
The top panel in \hyperref[fig:Fig.4]{Fig.5} shows our completeness result compared to one obtained by \cite{ciro15} confirming that our results do not change notably due to different methods of flux estimation. 

With measured numbers in hand, we can adjust the detection cut to see variations in completeness, which is connected to function of input flux density. Our completeness at 250-micron maps shows the trend of fast decrease below the $\sim25$ mJy, and reaches $\sim50\%$ at $\sim13.2$ mJy, the value we choose as the lowest cut for our prior $S_{250}$ list. The fitting logistic function reaches the saturation at the level of 250 $\mu$m fluxes larger than 40 mJy. For the 500 $\mu$m band, our completeness is above $80\%$ at $S_{500}>30$ mJy. This 500 $\mu$m flux we imply as the threshold for the final selection of  "$500\:\mu$m-risers". The bottom panel in \hyperref[fig:Fig.4]{Fig.5} shows that fluxes of sources below $S_{250}=7-10$ mJy are systematically overestimated because of "flux-boosting".

Reliability test is performed to quantify eventual spurious detections. This is important value for measuring the total number counts rather than for selection of "$500\:\mu$m-risers", since we expect that eventually spurious source would be present in one waveband but not in all three SPIRE bands. However, since our prior list is based on as much as close to the confusion limit 250 $\mu$m detections, we make an analysis of possible false detections  injecting fake sources into noise maps. 
\begin{figure*}[ht]
	\vspace{-0.2cm}
	\centering
	\includegraphics [width=17.19cm]{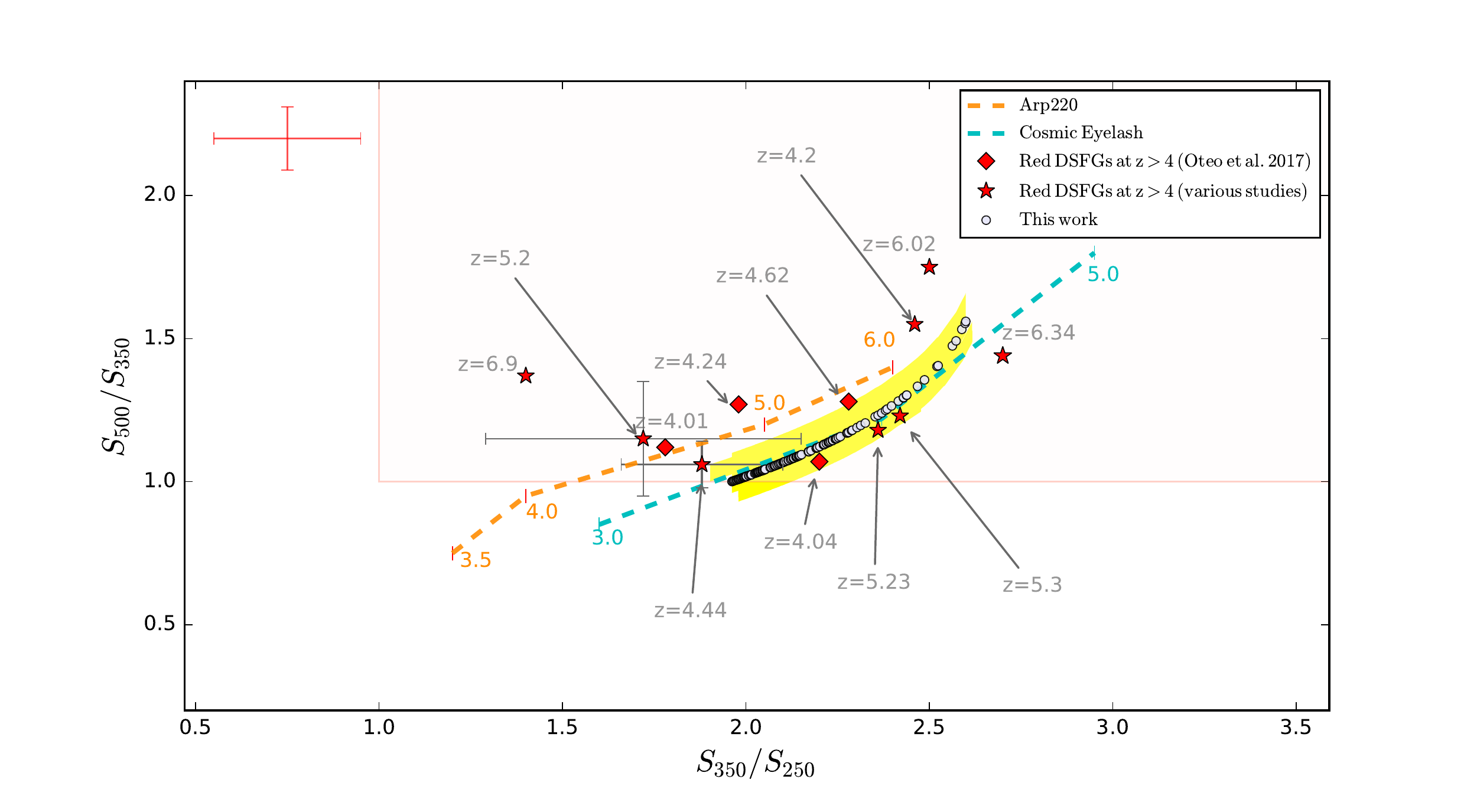}
	\caption{SPIRE colour-colour diagram of "$500\:\mu$m-risers", overlaid with redshift tracks of Arp 220 (\citealt{rangwala11}) and Cosmic Eyelash (\citealt{swinbank10}). Galaxies selected in this work are represented with circles, while yellow shaded region describes uncertainties related to chosen emissivity ($1.5<\beta<2.1$). For comparison we show "$500\:\mu$m-risers" selected in different studies and with known spectroscopic redshifts. Sources marked with red stars are: $z=6.34$ (HFLS3 , \citealt{riechers13}), $z=6.02$ (G09-8988, \citealt{zavala17}), $z=4.44$ (FLS 5, \citealt{dowell14}), $z=5.3$ (FLS 1, \citealt{dowell14}), $z=5.2$ (HELMS-RED4, \citealt{asboth16}), $z=4.04$ (GN20, \citealt{daddi09}), $z=4.2$ (SPT0113-46, \citealt{weiss13}). Sources marked with red diamonds are from \cite{oteo17b}. For some of known sources we plot their colour uncertainties. Representative colour uncertainty for our sample is plotted in the upper left.}
	\label{fig:Fig.27}
\end{figure*}
We also quantify the range of statistical outliers present in our measurements. Outlier contamination is measured as a function of output flux density. Sources are considered as contaminants if their output fluxes are more than $3\sigma$ above the value inferred for injected sources. Considering the lowest detection threshold in 250 $\mu$m band (13.2 mJy) we found the low sample contamination of $2.7\%$. 

\section{Expected $L_{\rm IR}$ and $z$ distribution of "$500\:\mu$m-risers"}
\label{sec:3}
 
Without known interferometric positions and confirmed spectroscopic redshifts of our sources, we are limited to the properties of FIR SEDs of selected "$500\:\mu$m-risers". 
Nevertheless,  our data may be very useful in providing approximate redshift/luminosity distributions of large samples and candidates for follow-ups. 
To fit an MBB model to our photometric data, we should consider a degeneracy between $\beta-T_{\rm d}$, and $T_{\rm d}-z$ (\citealt{blain03}, \citealt{pope10}). The peak of the SED is determined by the $\nu/T_{\rm d}$ term (see \hyperref[eq.2]{Eq.2}), giving that a measurement of the colours alone constrains only  $(1+z)/T_{\rm d}$ ratio. However, assuming reasonable priors 
(in our case $\beta$ and $T_{\rm d}$) it is possible to estimate a qualitative redshift distribution of "$500\:\mu$m-risers" (see $T_{\rm d}-z$ degeneracy illustrated in \hyperref[fig:multimbb]{Fig.B.1}). 

We fit a single temperature MBB model to the data. We fix the power-law slope $\beta=1.8$ and dust temperature $T_{\rm d}=38$ K, and we determine the median redshift of $4.22\pm0.49$. The choice of $\beta$ is arbitrary, likely $1.5<\beta<2.0$ (\citealt{casey14}, \citealt{mackenzie14}, \citealt{roseboom12}), and here we chose the value which is in the middle of that range. Because of aforementioned degeneracy in $(1+z)/T_{\rm d}$ space, decreasing the dust temperature, e.g. from 38 K to 30 K for the fixed $\beta$, peak of the redshift distribution decreases to $z=3.57$ (with 33$\%$ of sources at $z>4$). Modelled redshift tracks for different MBB parameters are provided in Appendix B (see \hyperref[fig:multimbb]{Fig.B.1}). 

To further investigate possible redshift/luminosity range we consider collection of SEDs of extensively studied IR-bright galaxies both from the local and higher-$z$ Universe, namely: the compact local starburst M82, typical interacting, local dusty system Arp220, and the "Cosmic Eyelash", strongly cluster-lensed  dusty galaxy at $\mathit{z}=2.3$ (\citealt{swinbank10}). 

\begin{figure}[ht]
	\centering
	\includegraphics[scale=0.61]{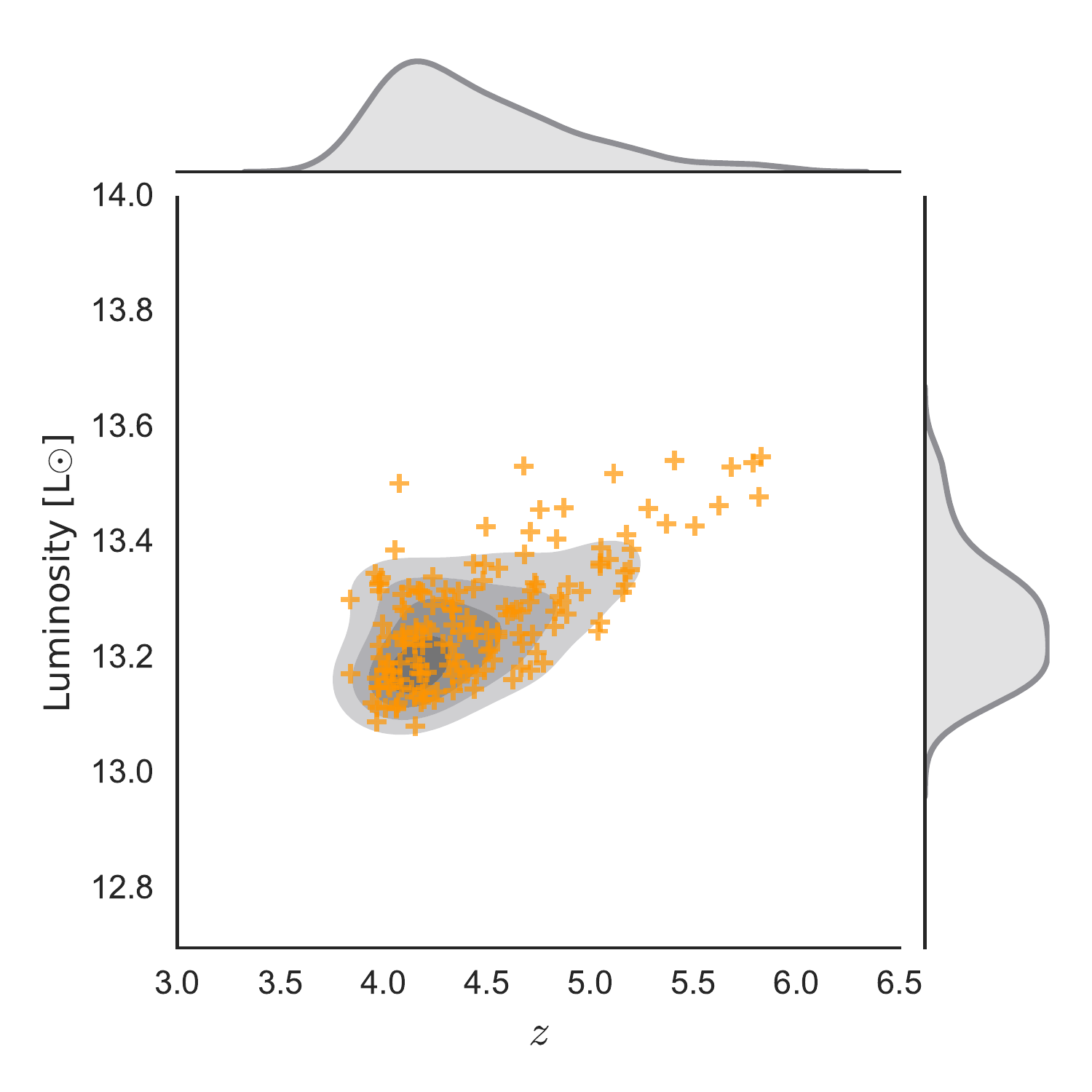}
	\caption{Infrared luminosity of our "$500\:\mu$m-risers" (y-axis) as a function of redshift. Orange crosses are values estimated applying the empirical template made from $\it{H}$-ATLAS galaxies covering the redshift range from $z=0.5$ to $z=4.5$ (\citealt{pearson13}). We imposed the same fitting parameters as used for $\it{H}$-ATLAS galaxies: $T_{\rm cold} = 23.9$ K, $T_{\rm hot} = 46.9$ K, and the cold-to-hot dust mass ratio equal to 30.1.} 
	\label{fig:Fig.6}
\end{figure}

Sample of dusty galaxies at $z>4$ are expected to be a combination of intrinsically luminous starbursts and strongly lensed systems (\citealt{b12}, \citealt{casey14}). 
It has been shown that lack of observational constraints caused some SEDs to express too low fluxes on the long $\lambda$-side of the FIR peak (\citealt{lutz14}, \citealt{elbaz10}). Another important point considered here is that Ultra-luminous Infrared Galaxies (ULIRGs) at higher redshifts express a wider variance in dust temperatures comparing to their local analogues (e.g. \citealt{smith14}, \citealt{simeon13}). To overcome these differences we use empirical template representative for $\it{H}$-ATLAS galaxies (\citealt{pearson13}).
The template is built for subset of 40 $\it{H}$-ATLAS survey sources with accurately measured redshifts covering the range from 0.5 to 4.5.
It adopts two dust components with different temperatures and it is already used in the literature as a statistical framework for characterizing redshifts of sources selected from SPIRE observations (see \citealt{ivison17}). 
With different templates in hand we can better characterize the systematics and uncertainties of measured redshifts. Shifting these templates to fit with our FIR fluxes, we forecast redshift ranges of "$500\:\mu$m-risers" in the HeViCS field. Probability distribution function (PDF) is built based on fitted $\chi^{2}$ values and it allows us to derive an estimate of the median redshift. Median redshift of "$500\:\mu$m-risers" is 3.87 if we apply "Cosmic Eyelash" template, 4.14 for Arp220 and 4.68 for M82. 
Next, we apply $H$-ATLAS empirical template accepting the following best-fit parameters: $T_{\rm cold} = 23.9$ K, $T_{\rm hot} = 46.9$ K, and the ratio between the masses of cold and warm dust of 30.1, as same as in \cite{pearson13}. The result is shown in \hyperref[fig:Fig.6]{Fig.7}. We determine a median redshift $\hat{z}=4.28$, and an interquartile range of 4.08-4.75. This estimate matches 1-sigma uncertainty range of photometric redshifts calculated by \cite{dowell14} and \cite{asboth16}. To determine an average photometric redshift for red sources selected from several HerMES fields, \cite{dowell14} used the affine invariant method (\citealt{emcee}) to fit the MBB model with fixed emissivity and rest-frame wavelength peak for each source. They found $\langle{z}\rangle=4.7\pm0.9$. The median photometric redshift estimated in \cite{ivison17} is somewhat lower ($\hat{z}=3.7$), but note that not all of sources in their catalogue of high-$z$ candidate DSFGs are "$500\:\mu$m-risers". In \hyperref[sec:5.2] {Section 5.2 }we analyse how the noise and resolution effects might impact the observed colours of DSFGs and thus the redshift distribution of selected "$500\:\mu$m-risers". 

To compute the IR luminosity, which is defined as the integral over the rest frame spectrum between 8 $\mu$m to 1000 $\mu$m, we rely on the same approach. Our results favour the presence of very luminous systems with no regard of chosen template - in $90\%$ of cases selected "$500\:\mu$m-risers" have $L_{\rm IR}\geq10^{13}L_{\odot}$, with minimum and maximum values $L_{\rm IR}=8.1\times10^{12}L_{\odot}$ and $L_{\rm IR}=5.1\times10^{13}L_{\odot}$ respectively. Concerning the MBB model, we obtain $\hat{L_{\rm IR}}=2.14\times10^{13}L_{\odot}$, but it is worth to note that the model accounts for mid-IR excess using a very simplified method (see \hyperref[eq.2]{Eq.2}). 
Furthermore, we apply the $H$-ATLAS representative template to determine corresponding median rest-frame IR luminosity $\hat{L_{\rm IR}}=1.94\times10^{13}L_{\odot}$. The result is again in consistency with \cite{dowell14} and \cite{ivison17}. These studies performed SED modeling of the sources with known photometric or spectroscopic redshifts, referring that "$500\:\mu$m-risers" have in average $L_{\rm IR}\geq10^{13}L_{\odot}$. For example, \cite{ivison17} report the median value $L_{\rm IR}=1.3^{+0.7}_{-0.5}\times10^{13}L_{\odot}$. The corresponding luminosities in their sample range from $L_{\rm IR}=6.0\times10^{12}L_{\odot}$ to $L_{\rm IR}=5.8\times10^{13}L_{\odot}$. 

In \hyperref[fig:Fig.6]{Fig.7} we illustrate the $L_{\rm IR}$-redshift trend of our "$500\:\mu$m-risers". We see that expected redshift distribution is not fully uniform, expressing the heavy-tail towards the higher redshifts. It indicates that very red submm galaxies are present at redshfits above the interquartile range, but it could also be due to strong gravitational lensing or unresolved blending. We should account with the possibility to have a much wider range of $L_{\rm IR}-z$ values under larger uncertainties in dust temperatures. The IR luminosity emitted by a MBB source at a given $T_{\rm d}$ depends on the emissivity and size of radiation area. Therefore, eventual gravitationally lensed candidates would lead to an apparent boost in dust luminosity at a given $T_{\rm d}$  due to a larger magnification (\citealt{greve14}). We consider and quantify effects of strong lensing in our selection (see \hyperref[sec:5.1.5] {Section 5.1.5} and \hyperref[sec:6.2] {Section 6.2}). We should emphasize here that the proper "training" of selected templates requires addition of higher-resolution data, and just in that case we can estimate systematic uncertainties and reject unreliable templates.
\subsection{Comparison to red sources from the literature }

In \hyperref[fig:Fig.27]{Fig.6} we show how the sources from this work relate to some of "$500\:\mu$m-risers" with known redshifts selected from the literature. Redshift tracks of Arp 220 and Cosmic Eyelash are added for comparison. We see that colours of our sources agree very well with a large number of red DSFGs at $z>4$ found in different studies. There are some exceptions, and one of them is the most distant "$500\:\mu$m-riser", SPT0311-58 at $z=6.9$ (\citealt{strandet17}). The possible reason for this might be additional contribution from AGN, since it has been shown
that SPT0311-58 experiences additional mechanical energy input, probably via AGN driven outflows. There are also examples of sources with redshifts somewhat lower than what would be expected from their very red colours. These sources usually have much lower $T_{\rm d}$ compared to median dust temperature of "$500\:\mu$m-risers" at $z>4$ . Such an example is "$500\:\mu$m-riser" at $z=4.2$ (\citealt{weiss13}). It is strongly lensed red source with colours very similar to those of known "$500\:\mu$m-risers" at $z>6$. However it has estimated $T_{\rm d}=30$ K. For comparison, all known "$500\:\mu$m-risers" at $z>6$ have dust temperatures higher than 50 K (\citealt{zavala17}, \citealt{strandet17}, \citealt{riechers13}).

We draw attention to some interesting "$500\:\mu$m-risers" selected in this work. According to MBB model introduced in  \hyperref[sec:2.6] {Section 2.6} , it is plausible to expect a source at $z\sim6$ if it has $T_{\rm d}\geq45$ K and satisfies colour criteria $S_{500}/S_{350}>1.3$ and $S_{350}/S_{250}>2.4$ (see \hyperref[fig:multimbb]{Fig.B.1} in Appendix B). In the literature there are 2 out of 3 "$500\:\mu$m-risers" at $z>6$ that fulfil these, so-called "ultra-red" requirements (\citealt{zavala17}, \citealt{riechers13}). We have 7 such sources in our sample, and we highlight them as suitable very high-redshift candidates. They are catalogued as HVS 5, HVS 6, HVS 21, HVS 35,  HVS 47, HVS 75, HVS 85 and HVS 94. We also stress 5 of our "$500\:\mu$m-risers" (HVS27, HVS60, HVS 64, HVS115 and HVS 130) that reside in potential overdense regions unveiled by $\mathit{Planck}$ (\citealt{ade16}).  If these overdensities are not just part of random fluctuations in the galaxy number density within the $\mathit{Planck}$ beam, it makes them candidate protoclusters of high-$z$ DSFGs (\citealt{greenslade}, \citealt{smolcic17}, \citealt{oteopc}, \citealt{clements16}).

\vspace{0.3cm}
\section{Differential number counts}
\label{sec:4}
\begin{figure*}[ht]
	\centering
	\includegraphics [width=15.3cm]{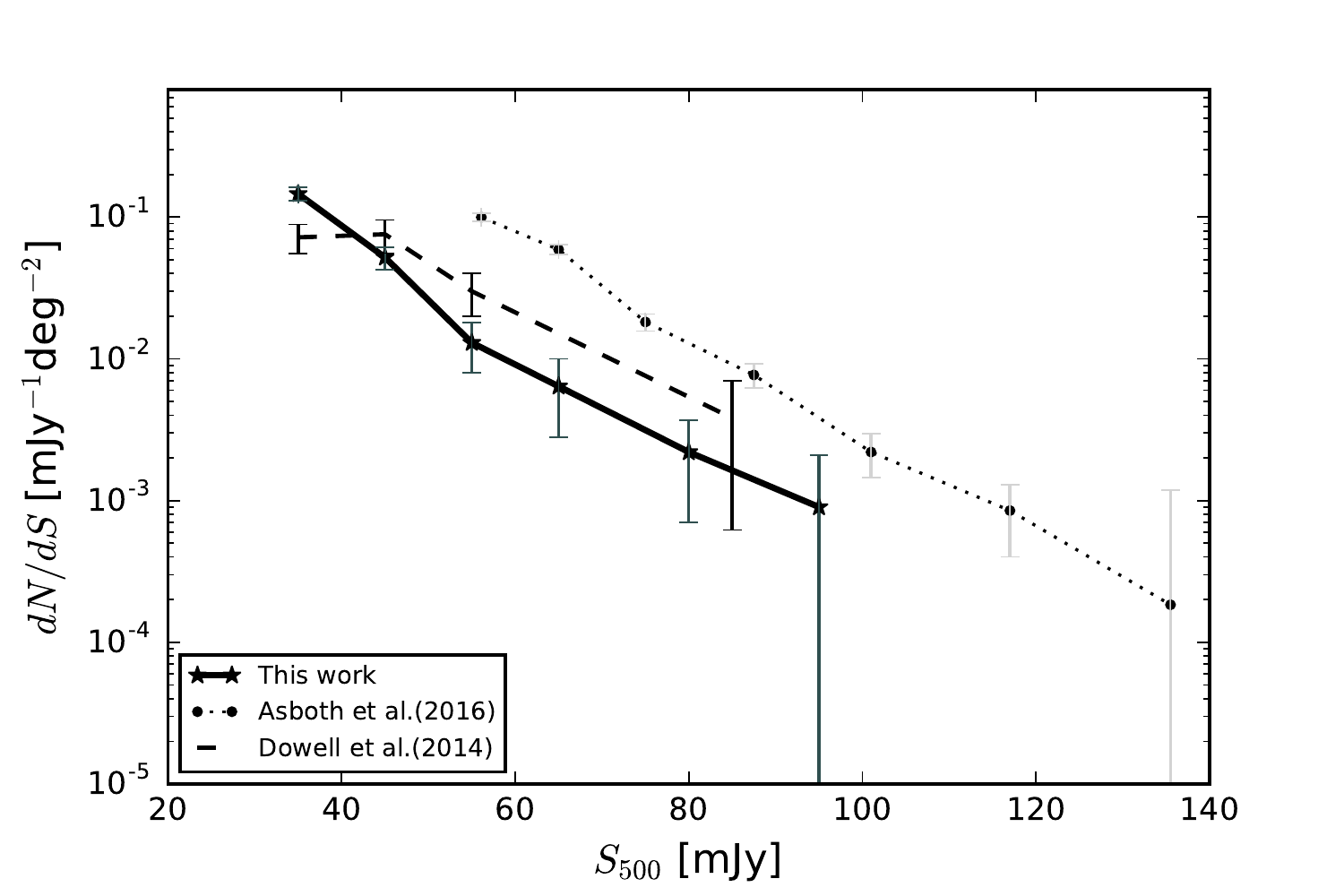}
	\caption{Raw differential number counts of red sources from several studies. HeViCS number counts are presented with black stars, connected with a full, black line. Counts from \cite{asboth16} are presented with black circles and connected with dotted line, whilst observational findings of \cite{dowell14} are presented with dashed black line. Filled points have Poisson $1\sigma$ error bars added, and for the brightest flux bins with low statistics of sources, we plot the upper $95\%$ confidence levels.}
	\label{fig:Fig.7}
\end{figure*}

We measure the raw 500 $\mu$m differential number counts and test our selection in respect to previous studies (this section), but also models of galaxy evolution (\hyperref[sec:5] {Section 5}). We determine our counts placing 133 "$500\:\mu$m-risers" in seven logarithmic flux bins between 30 mJy and 100 mJy. The resulting plot is shown in \hyperref[fig:Fig.7]{Fig.8}. The total raw number density of "$500\:\mu$m-risers" in 55 deg$^{2}$ of HeViCS is $\mathit{N}=2.41$ sources per deg$^{2}$, with the corresponding $1\sigma$ uncertainty of 0.34. 
Note that this value is uncorrected for non-red contaminants and completeness. In \hyperref[sec:5] {Section 5.2} we fully inspect and quantify biases that affect observed distribution, correcting for these effects. Furthermore, the presented differential number counts ignore eventual source multiplicity. Without interferometric data at longer wavelengths our counts should be interpreted as the density of sources in respect to the beam resolution. Potential impact of SPIRE source multiplicity on our "FIR-riser" selection is discussed in \hyperref[sec:6]{Section 6}.  

In Table \hyperref[tab:counts]{$3$} we presented raw differential and integral number counts. In \hyperref[fig:Fig.7]{Fig.8} we plot our differential number counts along with values from studies of \cite{asboth16} and \cite{dowell14}. They applied so-called difference map method  ($\mathit{DMAP}$) to select red sources. The method homogenizes beams to the size of 500 $\mu$m, and creates the $\mathit{DMAP}$ where SPIRE images are all smoothed to an identical angular resolution. Prior to source identification, weighted combination of the SPIRE maps is formed applying the relation $\mathit{D}=0.920\times{M_{500}}-0.392\times{M_{250}}$, where $M_{500}$ and $M_{250}$ are the smoothing values after matching SPIRE maps to the 500 $\mu$m resolution. To select "$500\:\mu$m-risers", they performed red colour map-based search, starting from lowest resolution 500 $\mu$m maps. 
\begin{table}[ht]
	\caption{Differential and integral number counts}
	\label{tab:counts} 
	\centering   
	\vspace{0.03cm}   
	\begin{tabular}{c c c | c} 
		\hline
		\toprule 
	    Flux (500 $\mu$m) & & Diff. counts &  Int.counts $^{(1)}$ \\[-7pt]\\
		$S_{\rm min}$ &  $S_{\rm max}$ & $dN/dS$ & $N(>S_{\rm min})$\\[-7pt]\\
		~[mJy] & [mJy] & [$\times10^{-4}$mJy$^{-1}$deg$^{-2}$]   & [deg$^{-2}$]\\[-3pt]\\
		\hline  
		\\
		\centering
		30 & 40  & 1527$\pm$ 164 & 2.41$\pm$ 0.22\\
		\centering
		40 & 50 & 581$\pm$ 90 & 0.91$\pm$0.14\\
		\centering
		50 & 60 & 145$\pm$51 & 0.29$\pm$0.09\\
		60 & 70 & 72$\pm$36 & 0.14$\pm$0.05\\
		70 & 90  & 27$\pm$15 & 0.07$\pm$0.03\\
		90 & 110 & 9$\pm$5 & 0.02$\pm$0.01\\
		\hline    
		\bottomrule               
	\end{tabular} \\
	\vspace{0.5cm} 
	\caption*{(1) In order to calculate the cumulative (integral) number counts we have sum over all the "$500\:\mu$m-risers" above the specified flux density ($S_{\rm min}$). Here we presented raw differential and integrated number counts.} 
\end{table}	
From \hyperref[fig:Fig.7]{Fig.8} it can be seen that differential number counts notably differ, especially in the lowest two flux bins where our measurements are in consistency with \cite{dowell14}. The slope obtained by \cite{dowell14} in the first two flux bins is almost flat, which is different from HeViCS curve, where a steeper decrease is present. While our number counts show fairly good agreement with those measured by \cite{dowell14} especially in lower flux bins, they are well below the values measured by \cite{asboth16}. Depending on a chosen flux bin discrepancy factor ranges from 3 to 10. We note that discrepancy between our number counts and those from \cite{asboth16} are due to several effects. \cite{asboth16} included possible blends in their final catalogue under assumption that at least one component of a blend is a red DSFG. This introduces a non-negligible number of contaminants, which is the effect we analyse in \hyperref[sec:5.2]{Section 5.2}. The effect of blending on our number counts is reduced since we apply a new technique of source extraction based on positional and SED priors. Additionally, \cite{bethermin17} simulated the same process of source extraction and selection as in \cite{asboth16} and found number of observed "$500\:\mu$m-risers" higher by factor of 8 compared to the number of genuine (modelled) "$500\:\mu$m-risers". They concluded that combination of noise, resolution effects, and the steepness of flux density distributions produce numerous red artefacts that match "$500\:\mu$m-risers" criteria. In \hyperref[sec:5.1.5]{Section 5.1.5} and \hyperref[sec:5.2]{Section 5.2} we analyse in details the effect of noise and explain how it affects our selection.
	
Apart from samples mentioned above, it is possible to find other "$500\:\mu$m-risers" in the literature, but they are not selected in uniform manner (e.g. \citealt{casey12}, \citealt{miettinen16}, \citealt{negrello17}). "$500\:\mu$m-risers" are serendipitously detected in relatively wide and shallow surveys that use wavelengths longer than 500 $\mu$m for initial detection (e.g. South Pole Telescope, see \citealt{vieira13}). These samples contain some of brightest "$500\:\mu$m-risers", all of them significantly amplified by gravitational lensing ($S_{500}>100$ mJy). 
\begin{table*}[ht]
	\caption{Comparison of models used in our analysis.}
	\label{tab:3} 
	\centering   
	\begin{tabular}{c c c c} 
		\hline
		\toprule 
		\\[-12pt]\\
		&&Models\\[-13pt]\\\\
		& B12 & B17 & S16\\[-7pt]\\
		&\cite{b12} &\cite{bethermin17} & \cite{s16}\\
		\\[-7pt]\\
		\hline 
		\\
		\centering
		Formalism$^{(1)}$ &  2SFM & 2SFM  & 2SFM  \\[-3pt]\\
		\centering
		sSFR$^{(2)}$ & evolves up to $z=2.5$ & evolves up to $z=4$ & evolves continuously \\[-3pt]\\
		\centering
		Dispersion ($\sigma_{\rm MS}$)$^{(3)}$ & 0.15 dex & 0.3 dex & 0.3 dex \\[-3pt]\\
		Strong lensing & Yes & Yes & No \\[-3pt]\\
		Passive galaxies & Yes & Yes & Yes \\[-3pt]\\
		Evolution of $T_{\rm d} $ & up to $z=2$ & up to $z=4$ & continuous \\[-3pt]\\
		AGN contribution & Yes & Yes & No \\[-3pt]\\
		\hline    
		\bottomrule               
	\end{tabular} \\
	\vspace{0.5cm} 
	\caption*{(1) All the models are based on two SF mode formalism. Stellar mass function (SMF) is described by a double Schechter function: $\phi(\rm M_{\ast})d(\rm M_{\ast})=exp(-\frac{\rm M_{\ast}}{M^{\ast}})[\varPhi^{\ast}_{1}(\frac{\rm M_{\ast}}{M^{\ast}})+\varPhi^{\ast}_{2}(\frac{\rm M_{\ast}}{M^{\ast}})]$, where $M^{\ast}$ is the characteristic mass of the Schechter break. Later value is redshift invariant in \cite{b12} model, and evolves with redshift in other two presented models. For redshifts $z>4$, model of \cite{bethermin17} assumes single Schechter function fixing the $\varPhi^{\ast}_{1}$=0, while model of \cite{s16} adopts double Schechter fitting to results of \cite{grazian15} for $4.5<z<7$. (2) The specific star-formation rate, defined as sSFR=SFR/$M_{\ast}$. In \cite{b12} sSFR increases with redshift up to $z=2.5$ and than flattens. This trend is independent of chosen range of stellar masses. In later two models the evolution is different, see Eq. 6 in \cite{bethermin17}; (3) Modelled width of the main-sequence (as log-normal scatter). }
	
\end{table*}
  \section{Comparison to models of galaxy evolution}
  \label{sec:5}
	In this section we compare our results with expectations from the most recent models of galaxy evolution. To evaluate our selection method, we perform simulations by generating realistic SPIRE maps from catalogues of simulated galaxies. 

	\subsection{Models}
	\label{sec:5.1}

In the literature there is a number of methods aiming to predict the evolution of number counts at different IR wavelengths. In this work we consider galaxy evolution models based on multiband surveys. In general such models rely on the combination of observed SED templates of galaxies and luminosity functions. It has been shown (\citealt{dowell14}) that some galaxy evolution models, mostly from "pre-Herschel era", underpredict either the total number of high-$z$, red sources (\cite{LeBorgne09}, or the number of red sources which lie at $z>4$ (\citealt{bethermin11}). There are some exceptions (e.g. \citealt{franceschini10}), that anticipate larger number of red, high-$z$ galaxies, but predicts at the same time large amount of very low-$z$ ($z<2$) red objects which seems unphysical. Additionally, we expect that "almost red" sources (described as sources having a peak in their FIR SEDs at wavelengths between 350 and 500 $\mu$m) contaminate the sample of "500 $\mu$m-risers", making observational artefacts caused by the noise. It is then very important to have models that predict significant number of such sources, to check the systematics in the detection rate of these contaminants. 
In this work we consider phenomenological models of \cite{b12}, \cite{s16} and \cite{bethermin17}, hereafter denoted as B12, S16 and B17 respectively. These models were built on $\mathit{Herschel}$ data and accurately match the total IR number counts. Summary of models and their main ingredients is given in Table \hyperref[tab:3]{$3$}. Common for all models is their usage of stellar mass function (SMF) as a starting point from which properties of galaxies are generated. They share the same general description of star-forming galaxies, with star formation rate (SFR) assigned based on the dichotomy model of \cite{sargent12}. It decomposes bolometric FIR-luminosity function with main-sequence (MS) and starburst (SB) galaxies. The MS galaxies are described as secularly evolving galaxies tightly relating stellar mass and SFR, while SB galaxies show an offset from the main sequence, expressing very high specific star formation rates (sSFR = SFR/$\rm M_{\star}$). 
Shape of the SEDs is controlled by the galaxy type (MS or SB) and mean intensity of the radiation field $\langle{U}\rangle$, which couples with the temperature of dust. Differences between the models are described briefly in following subsections - the most important are scatter from the main sequence, parameters chosen to fit stellar masses, and redshift evolution of a dust temperature. Models presented here have also slightly different description of mergers. 

\subsubsection{\cite{b12} model}
	\label{sec:5.1.1}

The B12 model describes the stellar mass function (SMF) with a Schechter function with redshift invariant characteristic mass parameter (see eq. 4 in \citealt{b12}, but also \citealt{peng10}). IR SED template are based on \cite{DL07} models, with the mean radiation field $\langle{U}\rangle$ evolving with redshift up to $z=2.0$ (\citealt{magdis12}). The dispersion of the MS log-normal distribution is $\sigma_{\rm MS}$=0.15 dex, following \cite{sargent12} and \cite{salmi12}. The model takes into account strong lensing effects reckoning the magnification rate PDF from \cite{hezaveh11}. These lensed sources contribute $\sim20\%$ to the bright-end submm counts ($\rm PLW\gtrsim100$ mJy). 
The AGN contribution is statistically associated based on results from \cite{aird12} and \cite{agn11}. 

	\subsubsection{\cite{bethermin17} model}
	\label{sec:5.1.2}
The B17 model is based on similar prescriptions as B12, but implies several improvements in order to match recent interferometric results that disclosed notable overestimation of submm number counts due to the source blending (\citealt{karim13}, \citealt{simpson15}). The model accounts for detailed consideration of clustering and resolution effects. To describe the physical clustering an abundance matching procedure is used to populate the dark-matter halos of a light cone constructed from the Bolshoi-Planck simulation (\citealt{darkmatter16}).
The MS-scatter is updated ($\sigma_{\rm MS}$=0.3 dex) in order to match the measurements of \cite{ilbert15} and \cite{schreiber15}. The model uses a new parametric form to fit the redshift evolution of the radiation field $\langle{U}\rangle$. Evolution of dust temperature of a main sequence SF galaxies stops at $z=4$ (instead at $z=2.0$ as in B12) and remains constant at higher values. Contribution of AGNs and strong lensing effects are modelled as in B12. The weak lensing regime is modelled with magnifications that are randomly drawn from a Gaussian distribution. Their width and mean values are derived based on a cosmological simulation of \cite{hilbert07}.

	\subsubsection{\cite{s16} model}
	\label{sec:5.1.3}

Similarly to B12 and B17, the S16 model is based on the MS-SB dichotomy. MS galaxies are modelled based on a stellar mass and redshift from \cite{schreiber15} and \cite{pannella15}. The scatter of the main sequence $\sigma_{\rm MS}$=0.3 dex. Randomly-selected galaxies ($5\%$) are placed in the SB mode by enhancing their SFR by a factor of $\sim5$. The distribution of stellar masses is described by double power-law Schechter fit, with parameters evolving with redshift. These parameters are chosen accordingly to observational data from \cite{schreiber15} (see also \citealt{grazian15}). A new set of template SEDs is used to model the dust emission of star-forming galaxies. These SEDs are based on the physically-motivated dust model of \cite{galliano11}. Redshift evolution of a dust temperature is modelled as
\begin{equation}
    T_{\rm d}[\rm K] =
    \begin{cases}
     4.65\times(z-2)+31, & \text{for}\:\rm MS \\
     T_{\rm d}^{\rm MS}+6.6\times\log_{10}(\rm R_{SB}), & \text{for} \:\rm SB 
    \end{cases}
\end{equation}

The "starburstiness" term $\rm R_{SB}$ is used to quantify the SFR offset between MS and SB galaxies ($\rm R_{SB} = SFR/SFR_{MS}$). Dependence of sSFR of the stellar mass shows that sSFR is constant at lower masses, and drops at the highest masses, $M>10^{10.5}M_{\ast}$ (e.g. \citealt{schreiber15}, \citealt{whitaker15}, \citealt{magnelli14}). This trend is similar to the one used in B17, but different from the fit used in B12. To summarize, in S16 both sSFR and $T_{\rm d}$ evolve continuously with redshift, which is not the case for other two models (see Table \hyperref[tab:3]{$3$}). Effects of strong lensing and AGNs are not included in the model. 
   
   \begin{figure*}[ht]
   	\vspace{-0.2cm}
   	\hspace{-1.89cm}
   	\includegraphics [width=22.39cm]{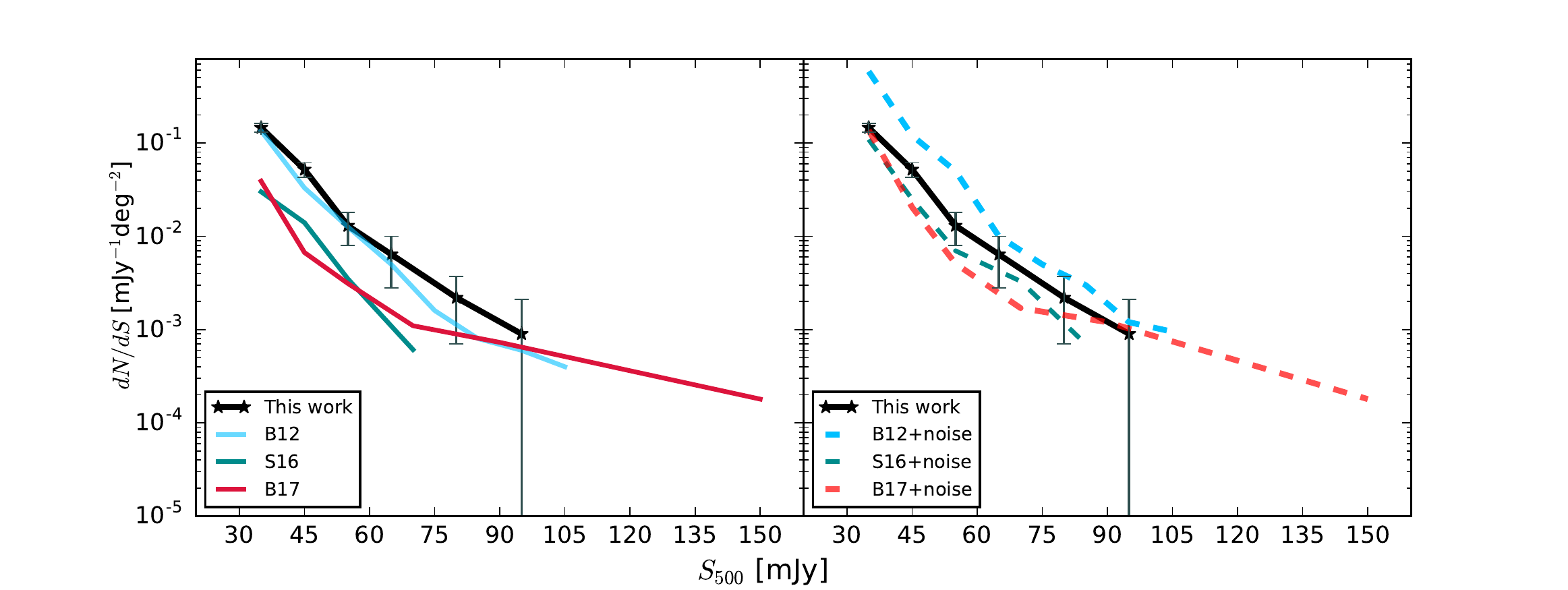}
   	\caption{Observed differential number counts in the HeViCS field (black line) confronted to models. Expected values from models are overplotted as coloured lines. 
   	$Left$: Comparison between observed and modelled counts. Models are represented by full, coloured lines: cyan for B12 (\citealt{b12}), red for B17 (\citealt{bethermin17} and green for S16 (\citealt{s16}). Effect of simulated noise is ignored. $Right$:  Comparison between observed and modelled counts if the effect of noise is simulated. The effect of confusion and instrumental noise is simulated by adding a random Gaussian noise to the modelled fluxes. Differential counts are then represented by dashed, coloured lines.} 
   	\label{fig:Fig.8}
   \end{figure*}

	\subsubsection{Mock catalogues}
	\label{sec:5.1.4}

We set mock catalogues based on models described in previous section. Our goal is to compare the number counts of observed "500 $\mu$m-risers" to the ones predicted by models.\footnote{ In the following, we make the distinction between "intrinsic" and "observed" quantities. The former ones are the true, modelled properties of galaxy, free of measurement errors and systematics. The latter are measured values, affected by biases.} 
The catalogues based on B12, B17 and S16 cover sufficiently large areas to offer accurate inspection of our observational criteria in the HeViCS field. 
Namely, for the B12 model we used the catalogue which is a result of 500 deg$^{2}$ simulations (\citealt{b12}). We also used the catalogue created by the B17 model covering the area of 274 deg$^{2}$ (Simulated Infrared Dusty Extragalactic Sky, SIDES, \citealt{bethermin17}). The size of simulated area is equal to the size of HeLMS field, thus perfectly suited for a comparison of our selection technique to one performed by \cite{asboth16}. 
	With S16 model, we generated a mock catalogue covering the size of the HeViCS field (55 deg$^{2}$).
	
We further apply our "$500\:\mu$m-risers" selection criteria ($S_{500}>S_{350}>S_{250}$, $S_{250}>13.2$  mJy and $S_{500}>$30 mJy) on modelled sources. Left panel in \hyperref[fig:Fig.8]{Fig.9} shows observed counts of HeViCS-selected "$500\:\mu$m-risers" plotted against the model predictions. 
We see that observed counts are in a fairly good agreement with models, while corresponding power-law slope show no significant difference from the slopes anticipated by models. Observed values are steep at the fainter-end and flatter at the bright-end ($S_{500}>80-90$ mJy) which is due to flux magnification by gravitational lensing. Our HeViCS differential number counts have the perfect agreement with B12, but we have to note there are recent evidences (see \citealt{bethermin17} for more details) that the simulated catalogue based on the B12 model overpredicts the number counts at 500 $\mu$m fluxes below 50 mJy by a factor of 2-3. The number of sources satisfying our selection in B17 and S16 are 0.45 deg$^{-2}$ and 0.40 deg$^{-2}$ respectively. Predictions of B17 and S16 are consistent with the $1\sigma$ error bars of observed counts in the higher flux bins (60 mJy$<S_{500}<100$ mJy). In lower flux bins (30 mJy$<S_{500}<60$ mJy) discrepancy factors are between 1 and 6 depending on a chosen bin. In \hyperref[sec:5.1.5]{Section 5.1.5} we discuss in more details potential reasons of a present difference. However, it is clear that we have a much better agreement with empirical models than previous studies (\citealt{asboth16}) who claim observational discrepancy of an order of magnitude.  

\subsubsection{Effects of noise on measured counts}
	\label{sec:5.1.5}

The results described in \hyperref[sec:5.1.4]{Section 5.1.4} neglect the effect of noise on the number counts of "$500\:\mu$m-risers". Therefore, we simulate the effect of both confusion and instrumental noise by adding a random Gaussian noise drawn from the values measured in the HeViCS field (\hyperref[sec:2.1]{Section 2.1}). 
The comparison between observations and models with simulated Gaussian noise are illustrated in the right panel in \hyperref[fig:Fig.8]{Fig.9}. The significant increase of counts in the lowest two flux bins is clearly seen, while their slopes  do not express significant change compared to intrinsic ones (left panel in \hyperref[fig:Fig.8]{Fig.9}). 

Considering the addition of noise, the number of simulated sources that appear as "$500\:\mu$m-risers" in B17 jumps to 473, resulting to an increase of number counts from 0.45 per deg$^2$ to 1.73 per deg$^2$. We obtain a very similar result with S16 - a number density increases from 0.4 per deg$^2$ to 1.54 per deg$^2$. 
As a consequence, observed HeViCS values match predictions of the B17 and S16 model even in the faintest 500 $\mu$m flux regime. \footnote{In B17 star formation rate (SFR) is limited to a maximum 1000 M$_{\odot}\rm yr^{-1}$.} 
\begin{table}[h]
	\caption{Comparison of modelled number counts before and after adding the Gaussian noise. Counts are calculated by imposing our criteria to select "$500\:\mu$m-risers".}
	\label{tab:4} 
	\centering   
	\vspace{0.03cm}   
	\begin{tabular}{c c c} 
		\hline
		\toprule 
		& Predicted density of & Faint\\[-7pt]\\
		&  $500\:\mu$m-risers $^{(1)}$  & $500\:\mu$m-risers $^{(2)}$\\[-3pt]\\
		&& ($S_{500}<30$ mJy) \\[-7pt]\\
		\hline 
		\\
		\centering
		B17 &  0.45 deg$^{-2}$ (51$\%$ lensed) & -\\[-3pt]\\
		\centering
		B17 + noise & 1.73 deg$^{-2}$  (24$\%$ lensed) & 52 $\%$ \\[-3pt]\\
		\centering
		S16 & 0.4 deg$^{-2}$  & - \\[-3pt]\\
		S16 +noise & 1.54 deg$^{-2}$  & 67 $\%$  \\[-3pt]\\
		\hline    
		\bottomrule               
	\end{tabular} \\
	\vspace{0.5cm} 
	\caption*{(1) Modelled density of $500\:\mu$m-risers before and after the addition of random Gaussian noise. Criteria used to select "$500\:\mu$m-risers" are: $S_{500}>S_{350}>S_{250}$, $S_{250}>13.2$ mJy and $S_{500}>30$ mJy. Numbers reported in brackets refer to the percentage of strongly lensed galaxies; (2) Contribution of intrinsically red but faint sources ($S_{500}<30$mJy) to the population of modelled sources that match our "$500\:\mu$m-risers" selection criteria  after the addition of modelled Gaussian noise.} 
\end{table}
We find that noise often tends to increase the number of genuinely red sources that are slightly below our 500 $\mu$m flux cut. These galaxies have modelled flux densities $S_{500}<30$ mJy, but they can pass our final "500 $\mu$m-risers" selection after the addition of a Gaussian noise. As shown in Table \hyperref[tab:4]{$5$} contribution of these sources to the modelled "500 $\mu$m-risers" vary between 52-67$\%$. In \hyperref[sec:4]{Section 4}, we mentioned that for the wide and shallow HeLMS field, \cite{bethermin17}  found an increase of observed number counts by factor of 8 compared to their modelled values. This strong increase is caused by the combination of noise and clustering, and it is two times higher than noise-driven increase of density of detected "500 $\mu$m-risers" in the HeViCS field (see Table \hyperref[tab:5]{$5$}). We thus conclude that the effect of noise is more prominent for shallower fields, while for deeper fields (e.g. HeViCS and deep HerMES fields analysed in \citealt{dowell14}) the noise has just a mild impact on measured number counts.

The effect of noise also changes the relative contribution of modelled lensed and unlensed "$500\:\mu$m-risers". 
The percentage of weakly lensed and non-lensed "risers" in B17 increases from 49$\%$ (modelled values) to 76$\%$ (modelled values+Gaussian noise). A closer inspection suggests that weak lensing ($1<\mu<2$) is a 
contributor to the flux budget for a non-negligible number of sources (166 out of 473, or 35$\%$). Fainter, weakly lensed red sources can pass our final "500 $\mu$m-risers" criteria more often if they are on a positive fluctuations of the magnification. This has an important role in producing observed "500 $\mu$m-risers" without a support of strong lensing. 
\begin{figure*}[ht]
	\centering
	\includegraphics [width=17.7cm]{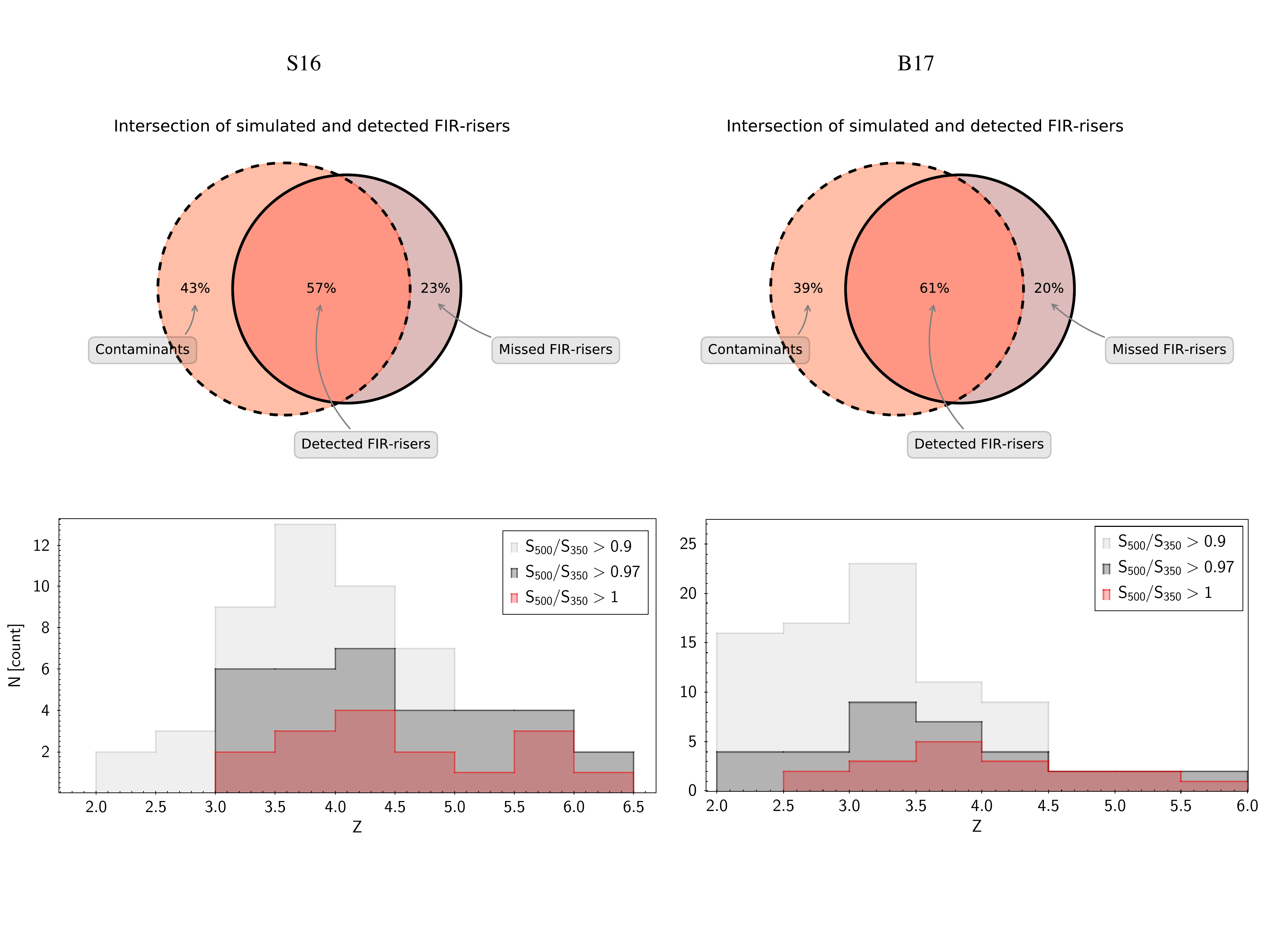}
	\vspace*{-13mm}
	\caption{Left and right panel attribute to S16 and B17 respectively. $\mathit{Upper\: panels}$: Galaxies detected in mock maps. "FIR-riser" criteria are imposed as for real HeViCS maps ($S_{500}>S_{350}>S_{250}$, $S_{250}>13.2 \:\rm mJy$, $S_{500}>30\:\rm mJy$). Intersected area coloured in dark orange depicts recovered red sources. 
	Light orange area represents detected contaminants. 
	Violet area represents missed sources. Those genuinely red galaxies are present in our catalogue, but not as "$500\:\mu$m-risers". 
	$\mathit{Lower\:panels}$: Redshift distribution of modelled galaxies. Different $S_{500}/S_{350}$ colour cuts are imposed. 
	Those cuts are related to statistical properties of the confusion noise, which are the most responsible for colour uncertainties. (see also Table \hyperref[tab:5]{$6$}).}
	\label{fig:Fig.9}
	
\end{figure*}

	\subsection{Simulated maps}
	\label{sec:5.2}
	

Mock mapping simulations are needed to explore the exact nature of selected sources and possible biases of our $\mathit{Herschel}$ SPIRE selection. These are mostly induced by the limited angular resolution combined with the noise effects.  
Having used two different models, we can uncover systematic uncertainty about the purity of detected sources. This is a crucial step if we want to investigate our photometric pipeline of "$500\:\mu$m-risers", since it is necessary to count with the numbers of sources we potentially miss and to determine fraction of contaminants. 

We generate a set of simulated SPIRE maps with use of Empirical Galaxy Generator Code (\texttt{EGG}, \citealt{egg16}). We filled our mock maps with the sources that has been drawn from S16 and B17 mock catalogues. We convert mock catalogues into maps, assigning theoretical $\mathit{Herschel}$ PSF and the same noise properties as measured in our real HeViCS images. 
We chose to simulate clustering. The mock galaxies from the catalogues are then placed on the sky at coordinates with a fixed angular two-point correlation function to implement this effect. Positions of galaxies in S16 are clustered by default with use of \cite{cluster1} algorithm, which produces a two-point correlation function with a power-law shape. To assign clustered coordinates for mock maps created from B17, we follow prescriptions from simulation described in \cite{bethermin17}. 
It adopts sophisticated clustering model with the sub-halo abundance matching procedure used to populate the dark-matter halos of a light cone constructed from the Bolshoi-Planck simulation (\citealt{darkmatter16}).
With this we ensure that simulated maps contain a realistic background of clustered sources that can contribute to the confusion noise. 
 
Due to computational dependencies, we generated maps covering the area of one of the four HeViCS fields (16 deg$^{2}$ each). As shown in \hyperref[sec:2.4]{Section 2.4}, our source extraction lead to an homogeneous distribution of sources over all HeViCS fields, so we can easily extrapolate our results for the whole observed area of 55 deg$^{2}$. We use the methods described above and perform the same source detection pipeline as we did in our raw HeViCS maps. We also check positional and flux accuracies in the same way we did for real HeViCS images. 

Results from our simulations are summarized and presented in \hyperref[fig:Fig.9]{Fig.10}. We detected all modelled "$500\:\mu$m-risers" via our detection procedure, but not all of them are characterised with their intrinsic colours.
We thus disclose three groups of sources described below.

\begin{enumerate}
\item $\mathit{Recovered\:\:"500\:\mu\rm m}$-$risers"$

Dark orange region in Fig. 9 represents the fraction of recovered "$500\:\mu$m-risers".  These are intrinsically red sources selected as "$500\:\mu$m-risers" when we applied the same selection criteria as for the real HeViCS maps ($S_{500}>S_{350}>S_{250}$, $S_{250}>13.2$  mJy and $S_{500}>$30 mJy). The percentage of recovered "$500\:\mu$m-risers" is around 60$\%$ for both models (57$\%$ for S16 and 61$\%$ for B17). 
\\
\\
As expected from the mock catalogue analysis (Section \hyperref[sec:5.1.4]{$5.1.4$}), population of recovered risers is a mix of both lensed and unlensed galaxies. We confirm that our criteria to select "$500\:\mu$m-risers" unveil significant group of dusty and potentially very high-$z$ sources that are not biased toward higher magnifications.  
\\
\\
As introduced in Section \hyperref[sec:5.1.5]{$5.1.5$} noise effects might produce colours somewhat redder than original ones. We quantify this as a "reddening", namely:
\begin{equation}
\Delta=\left(\frac{S_{500}+\sigma_{PLW}}{S_{350}+\sigma_{PMW}}\right)-\frac{S_{500}}{S_{350}},
\end{equation}
where $\sigma_{PLWm}$ and $\sigma_{PMW}$  are 1-sigma total noise ratios measured for appropriate bands in the HeViCS field (see \hyperref[sec:2.1]{Section 2.1}). The differences between intrinsic and observed colours of recovered $500\:\mu$m-risers are presented in Table \hyperref[tab:5]{$6$}. We see that the largest difference 
is for the lowest flux bin, where we also expect largest flux uncertainties due to the lower S/N ratio. Redshift distribution of recovered "$500\:\mu$m-risers" in \hyperref[fig:Fig.9]{Fig.10} reflects differences between models. Following S16, all red galaxies are at $z>3$, and most of them ($70\%$) at $z>4$ . The mean redshift value is $\langle{z}\rangle=4.56\pm0.94$.  The B17 model has a comparable high redshift tail, 
but this model peaks at lower redshift, causing the mean value ($\langle{z}\rangle=3.89\pm0.9$) to be lower than in S16. 
\\

\item $\mathit{Contaminants}$

Light-orange region in \hyperref[fig:Fig.9]{Fig.10}  represents a fraction of contaminants. Sources are detected as "$500\:\mu$m-risers", but having assigned their modelled colours and fluxes, they should not pass our selection criteria. The total percentage of contaminants is $39\%$ for B17 and $43\%$ for S16. They consist of two different populations of sources: the first one are genuinely red, but fainter galaxies, with modelled 500 $\mu$m fluxes not bright enough for the final catalogue inclusion ($S_{500}<30$ mJy). Nonetheless, their observed fluxes and colours are sufficiently high to pass "$500\:\mu$m-risers" criteria. The relative contribution of these contaminants is $60\%$ for B17 and $72\%$ for S16. The second population of contaminants are "non-red" sources with narrowly ranged intrinsic colours between 0.95<$S_{500}/S_{350}$<1.0. They are weakly lensed or unlensed "almost red" sources. We also find that non-red contaminants are accompanied with sources clustered at $1<z<2$. The largest 500 $\mu$m excess is caused if clustered neighbours are at $z=1.3\pm0.1$. These sources are massive galaxies with $M_{\ast}>10^{9.8}M_{\odot}$. 
It confirms the conclusion from Section \hyperref[sec:5.1.4]{$5.1.4$} that combined with weak lensing, confusion which arises from clustering and noise has a tendency to increase the number of observed red sources. 
\\
\\
Having exploited both models, we see that all detected contaminants are high-$z$ dusty galaxies, with just one source at $z<3$. Redshift distribution of contaminants peaks at 3.48 for B17 ($z_{min}=2.2$ and $z_{max}=5.11$), and 4.17 for S16 ($z_{min}=3.28$ and $z_{max}=5.86$). Additionally, according to both models, contaminants are galaxies with warm dust temperatures ($40\:{\rm K}<T_{\rm d}<50\:{\rm K}$). \newline

\item $\mathit{Missed\:\:"500\:\mu\rm m}$-$risers"$
\\
With our extraction procedure we detected all modelled "$500\:\mu$m-risers" from simulated maps. However there is a fraction of sources for which we failed to properly characterize their colours with \texttt{MBB-fitter}. Violet region in \hyperref[fig:Fig.9]{Fig.10}  represents sources having 
$S_{500}/S_{350}<1$ or $S_{500}<30$ mJy. In this way we missed to select around $20\%$ of genuine "$500\:\mu$m-risers". According to models, missed "$500\:\mu$m-risers" have 
colours and fluxes very close to the lowest selection threshold. In our simulated maps, they often immerse in complex blends, having a nearby, unresolved blue source at $0.5<z<1.0$ inside the 250 $\mu$m beam. Therefore, missed sources passes both 250 $\mu$m and 500 $\mu$m flux selection cuts but their observed colours are non-red. We quantify that the difference between observed and modelled colours of missed "risers" is shifted bluewards by a factor of $\sim0.03$, thus giving negative $\Delta$. Interestingly, not just observed, but the modelled $\Delta$ of missed sources exhibits negligible change due to noise effects. It tends to make their colours marginally redder or even bluer compared to the average value we measure for "$500\:\mu$m-risers" ($\langle\Delta\rangle=0.01$ for the whole population, while for missed sources $\langle\Delta\rangle<0.003$ having consulted both models). The final colour-criterion is thus highly sensitive to multiple effects including the strong clustering of blue sources. The redshift distribution of missed "$500\:\mu$m-risers" shows that $50\%$ of them are weakly lensed galaxies at $z>4$. It becomes vital to account for the missed population of red sources while searching for $z>4$ dusty galaxies (see \hyperref[sec:5.2.1]{Section 5.2.1}).
\end{enumerate}

\begin{table}[h]
	\caption{Intrinsic vs. observed colours of simulated "500 $\mu$m-risers". Catalogue and maps are based on B17.}
	\label{tab:5} 
	\centering   
	\vspace{0.03cm}   
	\begin{tabular}{c c c c} 
		\hline
		\toprule 
		Flux bin & $S_{500}/S_{350} ^{(1)}$ &  $S_{500}/S_{350} ^{(2)}$ & $\Delta^{(3)}$\\[-7pt]\\
		&   [B17$_{\rm cat}$]          & [B17$_{\rm map}$] & \\[-3pt]\\
		\hline 
		\\
		\centering
		30-40 mJy &  1.03$\pm$ 0.09 & 1.07$\pm$ 0.1 & 0.04 (0.012)\\[-3pt]\\
		\centering
		40-50 mJy & 1.10$\pm$ 0.05 & 1.12$\pm$0.06 & 0.02 (0.009)\\[-3pt]\\
		\centering
		$>50$ mJy & 1.15$\pm$0.11 & 1.16$\pm$0.18 & 0.01 (0.008)\\[-3pt]\\
		\hline    
		\bottomrule               
	\end{tabular} \\
	\vspace{0.5cm} 
	\caption*{(1) Catalogued colours from B17. Noise is not added to initial flux values; (2) Measured colours in simulated maps.; (3) Colour difference between observed and simulated flux values. It is quantified as "reddening" ($\Delta=\left(\frac{S_{500}\:+\:\sigma_{PLW}}{S_{350}\:+\:\sigma_{PMW}}\right)-\frac{S_{500}}{S_{350}}$). Values in brackets show the difference between catalogued fluxes before and after the additon of modelled Gaussian noise. }
	
\end{table}
\subsubsection{Modifying criteria to select  galaxies at $z>4$ ?}
	\label{sec:5.2.1}

Percentage of simulated red sources recovered by our detection pipeline is not as high as the values presented in \cite{asboth16} and \cite{dowell14}. These studies reported purity of almost 90 per cent. Yet, there is an important and significant difference between our simulations. While we created mock images assuming the colours of sources that have been drawn directly from the simulated catalogues, \cite{asboth16} and \cite{dowell14} omitted modelled red sources from their maps, injecting "risers" with SPIRE colours fixed to the median value measured in their raw maps. Such an approach may lead to a bias, since we found (see \hyperref[sec:5.1]{Section 5.1}) that most of unlensed "500 $\mu$m-risers" have $S_{500}/S_{350}$ very close to one. Additionally, S16 and B17 predict large number of "almost red" FIR sources (e.g. $0.9<S_{500}/S_{350}<1.0$ and $S_{250}>13.2$ mJy). The modelled number density of such defined sources is between 8-10 per deg$^{2}$. They are included in our initial "prior" catalogue, and due to effects of noise, clustering and/or lensing, some of them might be observed "redwards", thus being responsible for lower purity index. 

Considering the colour differences (parameter $\Delta$ in Table \hyperref[tab:5]{$6$}), it appears clear that distinguishing between the population with $S_{500}/S_{350}=0.97$ and $S_{500}/S_{350}=1$ is very troublesome in crowded environments, even for fields with significantly reduced instrumental noise.  As expected, colour uncertainties increase towards the lower fluxes. 
We conclude that some flexibility on a colour threshold is needed to account for noise and environmental effects. 
\begin{table}[h]
	\caption{Relation of different colour cuts to the number of observed/missed red sources and their redshift distributions. 
	Sources observed in simulated SPIRE maps are divided in three columns according to different colour limits ($S_{500/350} >0.9,\: S_{500/350}>0.97, \:S_{500/350}>1.0$). In each of three cells left column ($m$) substitute to percentage of missed, genuine "$500\:\mu$m-risers", while the right column represents the percentage of $z>4$ sources contained in selected populations.} 
	\label{tab:6}   
	\begin{tabular}{l||l|l||l|l||l|l}
		\toprule
		Models&\multicolumn{5}{c}{Colour-cuts}\\[-7pt]\\
		&\multicolumn{2}{c}{$S_{500/350}>0.9$}&\multicolumn{2}{c}{$S_{500/350}>0.97$} &\multicolumn{2}{c}{$S_{500/350}>1.0$}\\
		\cline{2-7}\\[-11pt]\\
		&$\mathit{m}$ & $z>4$ & $\mathit{m}$ & $z>4$ & $\mathit{m}$ &$z>4$\\
		\hline\hline
		\\
		S16 &3$\%$&43$\%$&18$\%$&62$\%$&23$\%$&72$\%$\\[-3pt]\\
		B17 &0$\%$&23$\%$&15$\%$&33$\%$&20$\%$&47$\%$\\[-3pt]\\
		\hline
		\bottomrule
	\end{tabular}
	
\end{table}
Here we test several colour cuts. 
The goal of such a test is twofold:  to find the colour threshold that increases the number of true (modelled) "$500\:\mu$m-risers", but introducing in the same time low contamination of $2<z<4$ sources. We quantify what is the contribution of these lower-$z$ objects for each colour limit. 
The result of the analysis is summarized in Table \hyperref[tab:6]{$7$} and histograms in the lower panel of \hyperref[fig:Fig.9]{Fig.10}. Having imposed $S_{500}/S_{350}>0.9$ we anticipate that almost all "$500\:\mu$m-risers" will be found independently of chosen models. Nevertheless, contribution of $2<z<4$ sources is significant, and vary from $57\%$ (S16) to $77\%$ (B17). Redshift distribution peaks at $z=3.25$ for B17 and $z=3.92$ for S16. 
If we increase the colour cut requirement to $S_{500}/S_{350}>0.97$, percentage of detected simulated intrinsic "$500\:\mu$m-risers" ranges between the 82$\%$ and 85$\%$. Contamination is heavily reduced, and just a few galaxies are at $z<3$ (e.g. in S16 all the galaxies that satisfy this colour cut are at $z>3$). Furthermore, redshift peak shifts to higher values, $z=3.68$ for B17 and $z=4.29$ for S16. 

Following these results, we suggest that a cut at $S_{500}/S_{350}>0.97$ might be imposed to search for $z\gtrsim4$ galaxies in the future. It is a good compromise that accounts for several effects which decrease the percentage of recovered red galaxies. In the same time, it works well against the lower-$z$ contaminants, especially those at $z<3$, limiting their contribution to a maximum 15$\%$. 
 Recent follow-up observations of red sources selected in H-ATLAS field (\citealt{ivison17}) also support our quest for modifying "$500\:\mu$m-risers" colour criteria.  
They report just $30\%$ of sources selected as "risers" are lying at $z>4$, claiming that such a low fraction is perhaps due to significant number of spurious-red contaminants in their catalogue. To account for a high level of noise in the H-ATLAS field, authors agreed that some refined selection technique is needed. In this paper we keep our final selection based on traditional threshold $S_{500}/S_{350}>1$ due to easier comparison to existing studies. 

\section{What are "$500\:\mu$m-risers"?}
\label{sec:6}
\subsection{Problem of multiplicity}
\label{sec:6.1}

\noindent The potential problem in performing refined selection criteria for "$500\:\mu$m-risers" is our limited knowledge of source multiplicity. 
Different rates of multiplicities are claimed by studies found in the literature (e.g. \citealt{cowley15}, \citealt{simpson15}). Their conclusions are tightly related to selection criteria and instrument used to measure the flux (single dish or interferometer). Some observations suggest that multiplicity fraction may be dependent on the observed FIR flux, where is more liable for brighter sources to have multiple components (\citealt{karim13}, \citealt{bussmann15}). This assumption includes strongly lensed sources as well, since they are known to be very bright in SPIRE bands. More recent SPIRE multiplicity study is done by \cite{scudder16}. They used improved version of XID+ code (\citealt{xid+17}) to search for many potential contributing sources per 250 $\mu$m detection. 
They have considered dusty galaxies covering a wide range of photometric redshifts and $250\:\mu$m flux densities. For 250 $\mu$m fluxes between 30 mJy and 45 mJy they found that combination of the two brightest components accounts for approximately 90$\%$ of the flux in the 250 $\mu$m source. Additionally, the brightest component contributes with more than $60\%$ to the total flux.

\begin{figure}[h]
		\centering
		\includegraphics [width=9.3cm]{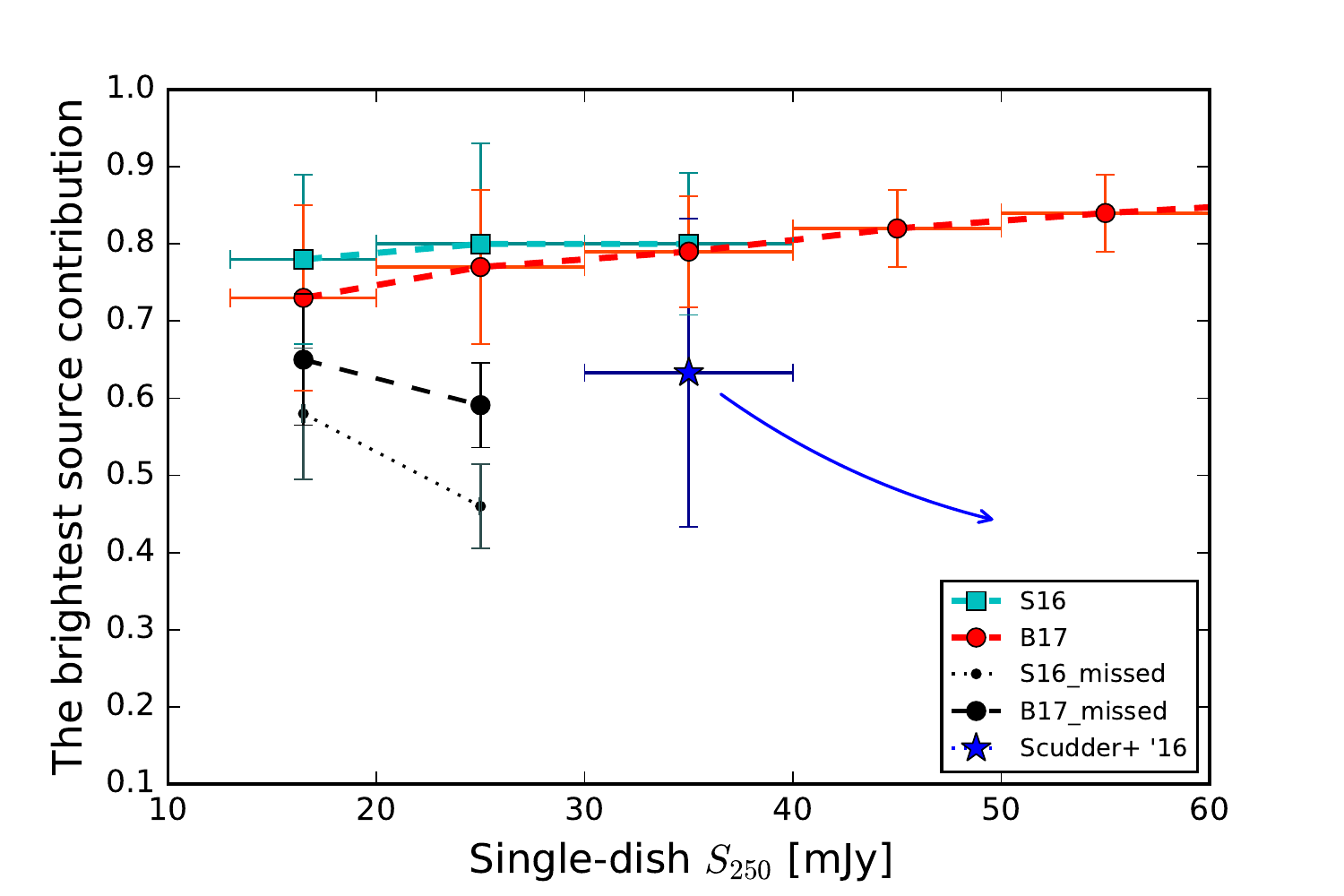}
		\caption{Multiplicity of red sources detected in mock maps. Average fraction of the 250 $\mu$m flux density emitted by the brightest galaxy in the $\mathit{Herschel}$ beam is plotted as a function of total, single-dish 250 $\mu$m flux measured in our simulated maps. Horizontal error bars indicate the width of the chosen flux bin, while vertical error bars show the standard deviation of the distribution. 
			Blue star refers to the study of \cite{scudder16}, and blue arrow indicates observed trend they found towards brighter fluxes.
			}
		\label{fig:Fig.10}
		
	\end{figure}

Since we work with SPIRE data only, our multiplicity analysis is based on simulations. We perform photometry on simulated SPIRE maps and compare the flux emitted from the brightest galaxy to the total single-dish flux ratio. For each source detected as "$500\:\mu$m-riser" in our mock maps, we assign the brightest component in the radius no larger than the half of 250 $\mu$m beam (9"). We then measure the ratio between the flux of this brightest galaxy from mock catalogues and the measured single-dish flux in our simulated maps. We place computed values in several flux bins. 
Our results are presented in \hyperref[fig:Fig.10]{Fig.11}. For galaxies detected as "$500\:\mu$m-risers", we found that 72$\%$ to 85$\%$ of observed 250 $\mu$m flux is emitted by the brightest galaxy (78$\%$ in average). 
	
Overall, we see almost insignificant change for S16 and B17, with the trend expressing slight increase towards higher 250 $\mu$m fluxes. Repeating the same analysis for 500 $\mu$m beam we find the similar trend with somewhat lower brightest galaxy fraction ($64\:\% $ in average) due to stronger resolution effects. 
Even from such a simple multiplicity analysis, it seems plausible to expect that selection of red sources from prior 250-micron detections does not experience strong resolution effects. Situation is somewhat different for missed red sources - the contribution of brightest components is lower (47-65$\%$) with the tendency of decreasing towards larger 250 $\mu$m fluxes. This is due to a presence of "blue" blends close to position of "$500\:\mu$m-risers" ($r<9''$).  

Fraction of the brightest 250 $\mu$m component to the total flux measured in simulated maps is higher than one found in \cite{scudder16}. They use deep multiwavelength data and apply different definition of the flux density fraction. Additionally, their analysis consider sources regardless of SPIRE colours making it difficult for a direct comparison here. Our multiplicity predictions appear to be in a good agreement with the measurements of Greenslade et al. (in prep.). They performed Submillimeter Array (SMA) follow up program for 36 of "$500\:\mu$m-risers" selected from various fields and found the multiplicity rate of $33\%$ with brightest components contributing $50-75\%$ to SMA flux. To shed a new light on this issue, the contamination of sources at intermediate redshifts ($1<z<2$) seen in deep NGVS and PACS images will be discussing in our following paper. 
\subsection{Lensing and clustering}
\label{sec:6.2}
\begin{figure*}[ht]
	\vspace{-0.2cm}
	\centering
	\includegraphics [width=9.9cm]{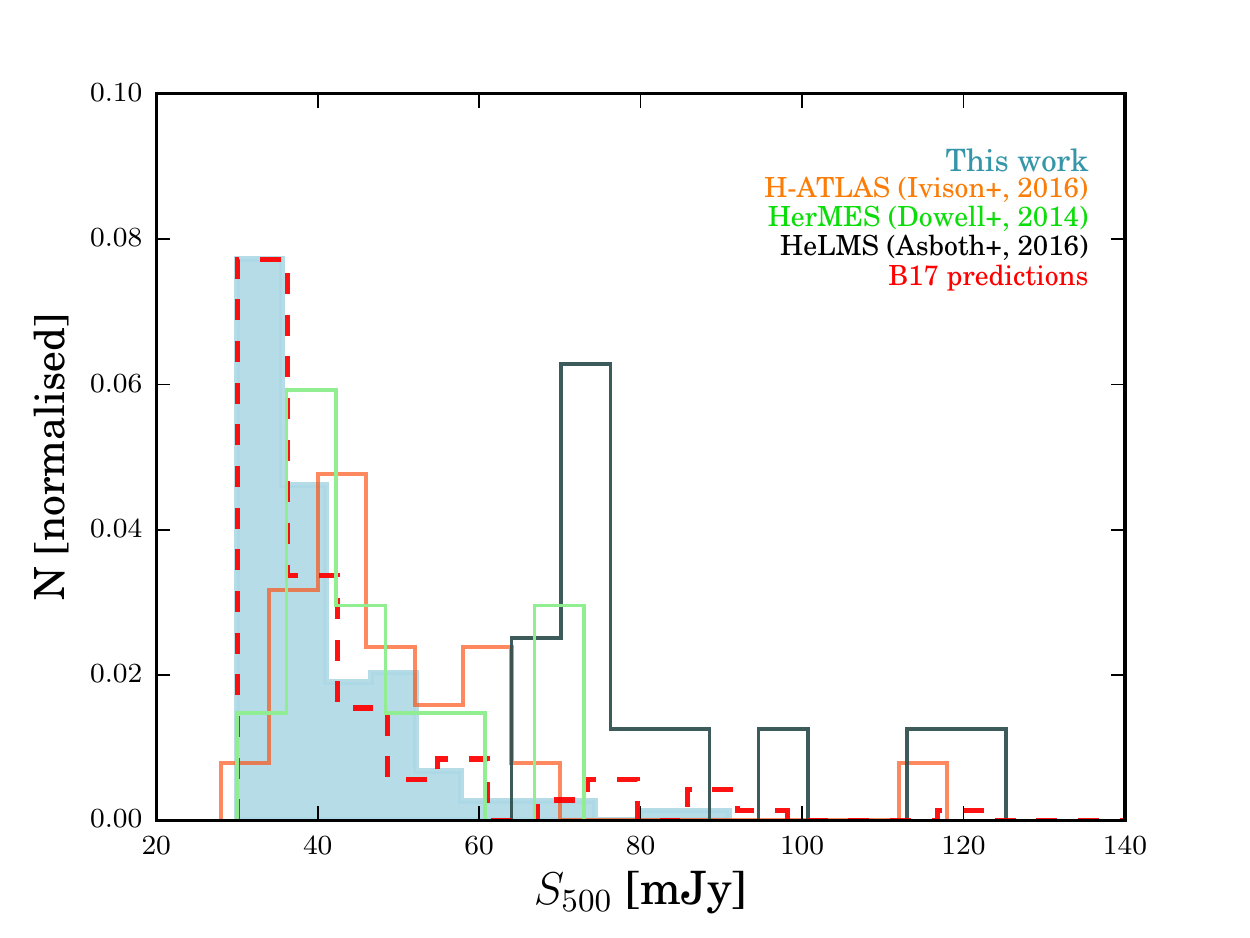}
	\caption{Flux distribution of "$500\:\mu$m-risers" from this work (coloured blue area), compared to "$500\:\mu$m-risers" considered for a submm, interferometric follow-up in different studies. Orange, black and green stepfilled lines represent sub-samples selected from \cite{ivison17}, \cite{asboth16} and \cite{dowell14} respectively. Flux distribution of red sources from the B17, that are modelled applying the same selection criteria presented in this work, is represented by dashed, red line.}
	\label{fig:Fig.17}
\end{figure*}
\begin{table*}[ht]
	\caption{A relative contribution of strongly lensed sources to "$500\:\mu$m-risers" selected in different studies (1):} 
	\label{tab:7} 
	\centering   
	\vspace{0.03cm} 
	\begin{tabular}{c c c c} 
		\hline
		\toprule 
		& Median & Lensed &  Lensed \\[-7pt]\\
		& $S_{500}$ [mJy] & [observed] & [modelled  (B17)] \\[-3pt]\\
		\hline 
		\\
		\centering
		$\it{H}$-ATLAS & $45\pm14$ & $23_{-11}^{+13}\:\%$ (5/21) &  $28_{-9}^{+8}\:\%$\\[-3pt]\\
		\centering
		HeLMS & $78\pm19$ & $69_{-25}^{+21}\:\%$ (9/13) &  $75_{-13}^{\:+10}\:\%$ \\[-3pt]\\
		\centering
		HerMES & $47\pm9$ & $40_{-8}^{+10}\:\%$ (4/10) & $36_{-7}^{+9}\:\%$ \\[-3pt]\\
		\hline
		\\
		This work & $38\pm4$  & - & $24_{-5}^{+6}\%$ \\[-3pt]\\
		\hline    
		\bottomrule               
	\end{tabular} \\
	\vspace{0.5cm} 
	\caption*{(1) Flux distribution of "$500\:\mu$m-risers" from this work compared to 44 red DSFGs presented in \cite{oteo17b}. These are  $\mathit{H}$-ATLAS, HerMES and HeLMS sub-samples observed with ALMA and NOEMA at 870 $\mu$m, 1 mm and 3 mm (\citealt{oteo17b}, \citealt{fudamoto17}). The second column shows median 500 $\mu$m fluxes along with the median 500 $\mu$m flux from this work. Relative contribution of observed, strongly lensed sources in various selections is shown in the third column. The fourth column shows relative contribution of modelled, strongly lensed sources from B17. A relative contribution of strongly lensed sources is modelled applying the same selection cuts of "$500\:\mu$m-risers" for tabulated extragalactic fields.} 
\end{table*}

Considering the increase with redshift of the optical depth to lensing and the magnification bias, we account for the possibility that some of our "$500\:\mu$m-risers" are strongly lensed (\citealt{bethermin17}, \citealt{negrello17}, \citealt{cai13}). Without interferometric data which would allow us to identify the true lensing fraction of our "$500\:\mu$m-risers", we cannot precisely correct for the lensed population in our final sample. We thus investigate their relative contribution making the comparison with recent ALMA/NOEMA follow-up results of "$500\:\mu$m-risers" selected in other fields ($H$-ATLAS, HerMES and HeLMS, \citealt{oteo17b}, see also \citealt{fudamoto17}), and using predictions from B17. Our findings are summarized in \hyperref[fig:Fig.17]{Fig.12} and Table \hyperref[tab:7]{$8$}. 

Sample used for high-resolution ALMA follow-up in \cite{oteo17b} contain 44 red DSFGs. They were chosen from complete samples of red DSFGs presented in \cite{ivison17}, \cite{asboth16} and \cite{dowell14}. The sources span the wide range of 500 $\mu$m fluxes, from 30 mJy to 162 mJy. We note here that the sample is highly incomplete, accounting just for the reddest sources whose colours are consistent with high photometric redshifts, estimated to be $z\sim4-6$. Observed total fraction of strongly lensed galaxies in \cite{oteo17b} is 40 $\%$ (18 out of 44). However, most of the lensed sources are in the bright flux regime of red DSFGs ($S_{500}>52$ mJy) \footnote{Here we distinguish between fainter ($S_{500}<52$ mJy), and brighter  "500 $\mu$m-risers" ($S_{500}>52$ mJy). The chosen separation is based on model predictions (B17), where highly-magnified sources are contributing more than 50$\%$ to the population of red, DSFGs at fluxes higher than $S_{500}>52$ mJy.} To make the sample suitable for comparison with our sources and model predictions, we split it into three subsamples according to methods used for their selections: 21 sources from \cite{ivison17}, 13 sources from \cite{asboth16} and 10 sources from \cite{dowell14}. In \hyperref[fig:Fig.17]{Fig.12} we show 500 $\mu$m flux distribution of our sample alongside the sub-samples drawn from \cite{oteo17b}. We see that "500 $\mu$m-risers" selected in this work have average fluxes fainter than other presented galaxies. We show observed number of strongly lensed objects in $H$-ATLAS, HerMES and HeLMS  in the second column of Table \hyperref[tab:7]{$8$}. The sub-sample of "500 $\mu$m-risers" in HeLMS contains the largest fraction of lensed sources (75$\%$), while the lowest observed fraction is for $H$-ATLAS (23$\%$).

We further use B17 and consider the same selection criteria from initial studies (\citealt{ivison17}, \citealt{asboth16}, \citealt{dowell14}). For each sub-sample we examine the range of fluxes identical to what is presented in \cite{oteo17b} and simulate the effect noise as explained in Section \hyperref[sec:5.1.5]{$5.1.5$}.  From Table \hyperref[tab:7]{$8$} we see there is a very good match between model predictions and observations. We estimate the predicted contribution of strongly lensed sources to our sample of "500 $\mu$m-risers". Applying our selection criteria we find that B17 predicts $24^{+6}_{-5}\%$ of strongly lensed "500 $\mu$m-risers" in our sample. More precisely, 17$\%$ of strongly magnified galaxies are expected to lie at fainter flux regime ($S_{500}<52$ mJy) where most of our sources reside (119 out of 133, or 89$\%$). The median redshift of modelled, strongly lensed sources is $\hat{z}=4.2\pm0.4$, whilst median magnification differs from $\hat{\mu}=5.3\pm3$ and $\hat{\mu}=8.1\pm5$ for the fainter and brighter flux regime respectively. 
It implies that strong lensing can affect the observed luminosity function of our "500 $\mu$m-risers" and lowers the fraction of intrinsically bright sources. After correcting for lensing, we expect to have 24$\%$ of galaxies with luminosities falling in the range $1.3\times10^{12}{L_{\odot}}<L_{\rm IR}<10^{13}{L_{\odot}}$ (classified as ULIRGs), and 76$\%$ of sources with $L_{\rm IR}>10^{13}{L_{\odot}}$, classified as hyperluminous IR galaxies (HyLIRGs). 

Apart from strong lensing, clustering might be responsible in producing observed excess in number counts of "$500\:\mu$m-risers". 
Several studies of clustering of submm-bright galaxies at $1<z<3$ (e.g. $M_{\ast}>10^{11}M_{\odot}$, and $L_{\rm IR}>10^{12}L_{\odot}$, see \citealt{farrah06}, \citealt{wilkinson17}) have been shown they are hosted in moderately strongly clustered massive halos. $\mathit{Herschel}$ detected star-forming galaxies are in average more strongly clustered at higher-redshifts ($z\sim2$) than nearby objects (\citealt{bethermin14}, \citealt{ono14}). Since the beam size in $\mathit{Herschel}$ is a direct function of the wavelength, we expect largest SPIRE beam (500 $\mu$m) to be more affected than other two beams. \cite{bethermin17} have studied in details influence of clustering on FIR continuum observations. They found that the confusion which arises from clustering increases the number of
observed "500 $\mu$m-risers". The number of such contaminants causes the peak of intrinsic redshift distribution of "500 $\mu$m-risers" is somewhat lower than what would be expected from observed SPIRE colours. \footnote{\cite{ivison17} and \cite{duivenvoorden} found that more than a half of DSFGs from their samples have $z_{\rm phot}<4$ . Photometric redshifts are estimated from SPIRE and SCUBA-2/LABOCA data.} Therefore, future spectroscopic confirmations of full samples of selected "500 $\mu$m-risers" are needed to quantify and compare exact fraction of sources at $z>4$. 

Simulations presented in Section \hyperref[sec:5.2]{$5.2$} have some limitations meaning that they are affected by cosmic variance beyond simple Poissonian, thus containing under- or overdensities at some redshifts. On top of that, pure probabilistic treatment is used to model the lensing, lacking the physical connection to the overdensity of lower-$z$ systems. To improve our understanding of how lensing and clustering might shape "500 $\:\mu$m-risers" selection, different solutions can be considered. Along with deeper, interferometric observations, more complex cosmological simulations are needed. They should include larger halo volumes, high-resolution to unveil fainter galaxies responsible for confusion, environmental effects and refined treatment of a lens modelling. 

\subsection{Star-formation rate density}

\begin{figure}[!ht]
	\centering
	\includegraphics [width=9.89cm]{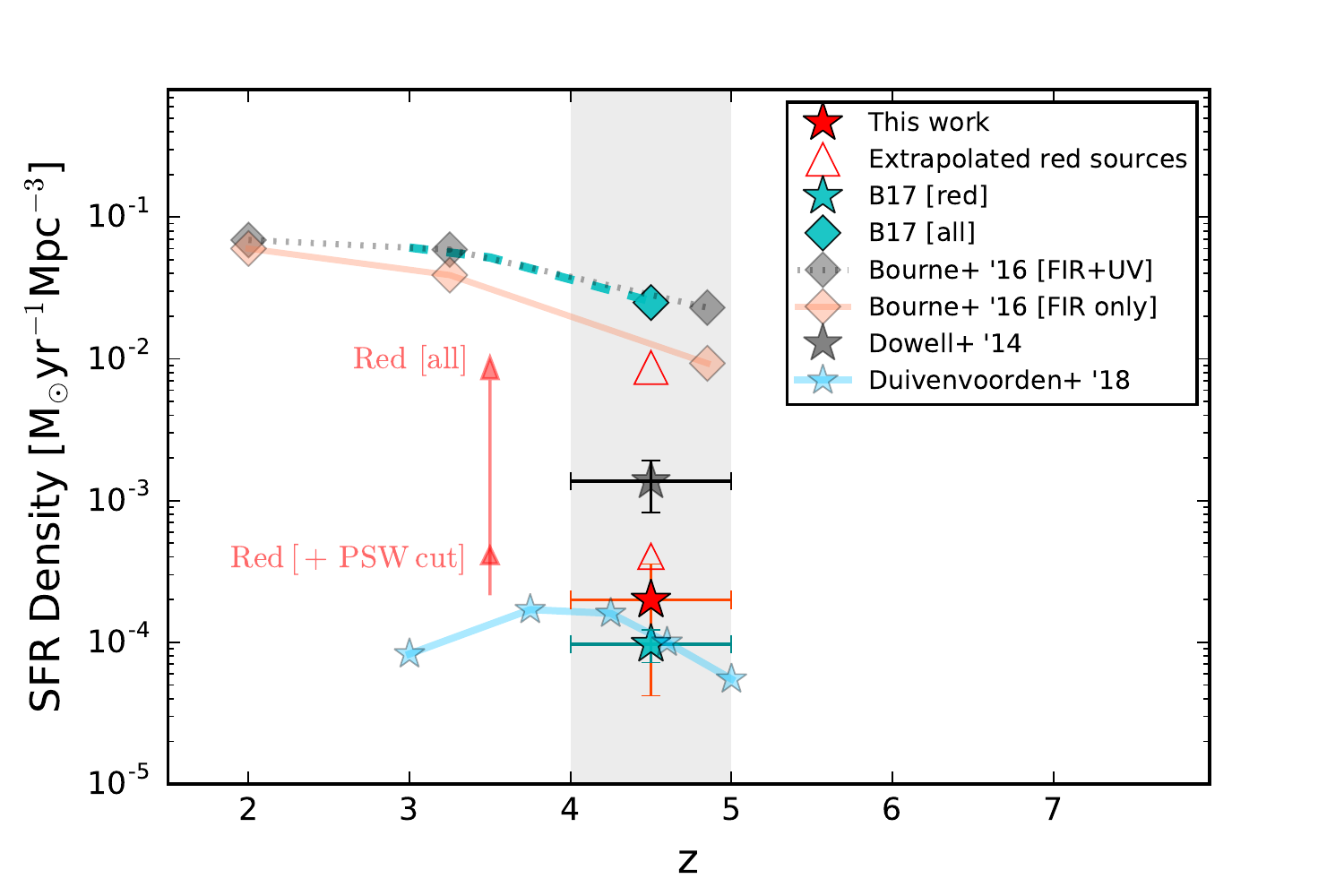}
	\caption{SFRD as a function of redshift. Filled red star represents our measurement. The estimated SFRDs of extrapolated "500 $\mu$m-risers" are shown with empty red-edged triangles: smaller when 250 $\mu$m cut is imposed ($S_{250}>13.2$ mJy), and larger triangle if all flux cuts are removed. Flux corrections are annotated by arrows. We show for comparison results from other 
			"500 $\mu$m-risers" selections: B17 (cyan), \cite{dowell14} (black) and \cite{duivenvoorden} (blue). 
			Horizontal bars reflect the bin size. 
			FIR and total FIR+UV observed SFRD (\citealt{bourne17}), as well as total simulated SFRD (B17), are marked with diamonds.}
	\label{fig:Fig.11}
	
\end{figure}

The bright individually-detected "$500\:\mu$m-risers" play an important role in understanding the evolution of massive systems. Here we test if such a selection is suitable to measure the total star formation rate density (SFRD). We make a rough estimate of the SFRD assuming the statistical properties of selected "$500\:\mu$m-risers". 
We determine SFRD at $4<z<5$ applying the equation (\citealt{sfrd}, \citealt{hogg99}): 
\begin{equation}
\rho_{\ast}{(z)}=\frac{\sum{\rm SFR_{\rm IR}}}{\frac{4\pi}{3}\int_{z=4}^{z=5}\frac{c/H_{0}}{\sqrt{\Omega_{\rm M}(1+z)^{3}+\Omega_{\Lambda}}}dz},
\label{eq:Eq.6}
\end{equation}
where $\rho_{\ast}{(z)}$ is SFRD, defined as a total sum of SFRs per co-moving volume, while $\rm SFR_{IR}$ is determined from IR luminosity. Bottom number (denominator) in \hyperref[eq:Eq.6]{$\rm Eq.6$} is the co-moving volume contained within $4<z<5$, and IR luminosity is converted to SFR applying the standard conversion formula from \cite{kennicutt98}: 
\begin{equation}
{\rm SFR_{IR}(M_{\odot}\:yr^{-1}})=1.71\times10^{10}\:L_{\rm IR}\:(L_{\odot})
\end{equation}
We apply our selection criteria to models (B17). Since B17 provides magnification factor for each source, we further use the modelled, lensing-corrected luminosities and redshift distribution of "$500\:\mu$m-risers". 
We then correct for the number of missed and contaminant sources applying the result from our simulations (\hyperref[fig:Fig.10]{Fig.10}, \hyperref[sec:5.2]{Section 5.2}).  
We place IR luminosities in a wide bin by the co-moving volume per deg$^{2}$. Scaling with the observed area and applying cosmological parameters $\Omega_{\rm M}=0.307, \:\Omega_{\Lambda}=0.693$ and $H_{0}=67.8$ km/s/Mpc (\citealt{planck16}), we find that $\rm SFRD=1.99\times10^{-4}$ M$_{\rm \odot}\:yr^{-1}$ Mpc$^{-3}$, for "$500\:\mu$m-risers" at $4<z<5$. This result is shown with a filled, red star in \hyperref[fig:Fig.11]{Fig.13}. Large uncertainties ($\pm1.4\times10^{-4}$) determined by Monte Carlo bootstrapping are due to the Poisson noise and the fact that we make constrains with SPIRE data only. We additionally cross-check this result and make the same estimation of SFRD but adopting median IR luminosity ($1.94\times10^{13}\:L_{\odot}$) from observed $L_{\rm IR}-z$ distribution (see \hyperref[sec:3]{Section 3.}). 
 Based on expected fraction of strongly amplified objects found in \hyperref[sec:6.2]{Section $6.2$} (24$\%$), we randomly chose sources from our catalogue to correct for lensing, namely $17\%$ of sources at $S_{500}<52$ mJy, and 83$\%$ at $S_{500}>52$ mJy. These percentages correspond to predictions from B17. We then adopt a median magnifications from B17 that correspond to chosen flux-regime ($\mu=5.3$ for fainter, and $\mu=8.1$ for brighter "$500\:\mu$m-risers"), and produce the observed, lensing-corrected luminosity distribution. We find $\rm SFRD = 2.87\times10^{-4}$ M$_{\rm \odot}\:yr^{-1}$ Mpc$^{-3}$. Our model-based SFRD estimation is consistent with recent observational findings from \cite{duivenvoorden}. Note also that offset between the study of \cite{dowell14} and our own may be explained by the fact that estimation by \cite{dowell14} assumed substantially higher IR luminosities and purity ratios.

Furthermore, we can use the shape of luminosity function to correct for the fainter red sources. We extrapolate the contribution of "$500\:\mu$m-risers" to SFRD - firstly we remove 500 $\mu$m flux limit (thus accounting for a red sources fainter than $S_{500}<30$ mJy), and then we remove lower 250 $\mu$m flux cut, thus accounting for all "$500\:\mu$m-risers" below the $\mathit{Herschel}$ sensitivity limit. Extrapolated contributions are depicted by empty red triangles in \hyperref[fig:Fig.11]{Fig.13}). 

Here we assume the luminosity distribution from B17, but we should mention that very similar result (higher by a factor of 1.2) we obtain applying the luminosity function from S16. As expected, the largest portion of sources responsible for a stellar budget at $z>4$ are missed due to sensitivity limitations of $\mathit{Herschel}$ instruments and not because of colour selection. 

The contribution of the emission of dusty star-forming galaxies (DSFGs) to the total infrared (IR) luminosity functions up to $z=5$ is partially investigated (e.g. \citealt{koprowski17}, \citealt{bourne16}, \citealt{mrr16}, \citealt{madau14}, \citealt{gruppioni14}, \citealt{burgarella13}). However, there is no consensus about whether or not ultraviolet (UV) estimates have notably underestimated the contribution of dust-enshrouded star formation in DSFGs because our knowledge at $z>3$ is severely incomplete, mostly due to optical obscuration. %
Attempting to extend the cosmic SFRD up to $z=6$ \cite{mrr16} exploited existing 500 $\mu$m $\mathit{Herschel}$ selections of "$500\:\mu$m-risers". They report no decline in dust-obscured SFRD between $z=3-6$, and very high SFRD value at $4<z<5$ order of magnitude higher than our estimation. However, our result is in a much better agreement with studies of \cite{bourne16} (marked with diamonds in \hyperref[fig:Fig.11]{Fig.13}), who used very deep millimetre imaging to reduce the effects of confusion on cosmic SFRD measurements. 

\section{Conclusions}

We have performed a systematic search of dusty, background galaxies onto 55 deg$^2$ area. We introduce an innovative method to select "500 $\mu$m-risers" ($S_{500}>S_{350}>S_{250}$). 
In order to estimate the 250-500 $\mu$m flux of blended sources, we use the multiwavelength \texttt{MBB-fitter} code. We assign positions of galaxies detected in the 250 $\mu$m map as a prior to extract the flux densities at longer wavelengths, iteratively fitting their SEDs as modified blackbodies.  

\begin{itemize}
	\item We select 133 "500 $\mu$m-risers" which fulfil following criteria: $S_{500}>S_{350}>S_{250}$, $S_{250}>13.2$  mJy and $S_{500}>$30 mJy. We reject all strong synchrotron sources by cross-matching available radio catalogues.
	\item We use statistical properties of selected galaxies and estimate a median redshift value $\hat{z}=4.28$ and the corresponding median rest-frame IR luminosity $\hat{L_{\rm IR}}=1.94\times10^{13}L_{\odot}$.
	\item The total raw number density of selected sources is $2.41\pm0.34$ deg$^{-2}$. To evaluate our results, we retrieve mock catalogues based on models of \cite{b12}, \cite{bethermin17} and \cite{s16}. Differential number counts are fairly consistent with model predictions. Namely, towards the brighter end ($S_{500}>70$ mJy), HeViCS counts have both slope and values identical to ones anticipated by models. At the lowest flux end, discrepancy is the effect of noise and lensing. We find that noise tends to increase number counts of "500 $\mu$m-risers". It may be high enough to fully match with observations within 1-sigma uncertainty. 
We confirm that selecting "500 $\:\mu$m-risers" applying our selection criteria is a direct way to detect significant population of unlensed, dusty, high-$z$ galaxies. 
	\item We build simulated maps. 
	Clustering and lensing are modelled in order to fully resolve effects that produce colour uncertainties. 
	We propose an modified colour criterion ($S_{500}/S_{350}>0.97$) that should be probed in the future selections of potentially $z>4$ dusty sources. 
	Motivation to apply this criterion is twofold: (1) it would account for colour uncertainties that arise from effects of noise, clustering and lensing; (2) it increases the sample of candidate $z>4$ galaxies. Anticipated contribution of $2<z<3$ contaminants satisfying this colour cut is small, and reach the maximum of 13$\%$. 
	\item We inspect the effect of multiplicity working with simulated SPIRE data. The brightest galaxy inside the beam contributes in average 
	75$\%$ and 64$\%$ to the total single-dish flux measured at 250 $\mu$m and 500 $\mu$m respectively. 
	\item We use the model of \cite{bethermin17} and find that $24^{+6}_{-5}\%$ of "500 $\mu$m-risers" selected with our criteria would be strongly lensed. After correcting measured luminosities for lensing, we expect 24$\%$ of sources with luminosities $10^{12}{L_{\odot}}<L_{\rm IR}<10^{13}{L_{\odot}}$ (classified as ULIRGs) and $76\%$ of sources with $L_{\rm IR}>10^{13}L_{\odot}$ (classified as HyLIRGs). 
	\item We use statistical properties of our galaxies to determine the role of "500 $\mu$m-risers" to the SFRD for $4<z<5$. We show that after correcting for fainter sources, projected contribution of "500 $\mu$m-risers" is 
	factor of 2-3 below the total, dust-corrected SFRD. 
\end{itemize}

\begin{acknowledgements}
We are thankful to the anonymous referee for very constructive comments and points which significantly improved this paper. We would like to acknowledge David Corre, Denis Burgarella, Alessandro Boselli, Tom Bakx and Eric Jullo for useful discussions. The project leading to this publication has received funding from Excellence Initiative of Aix-Marseille University - A*MIDEX, a French "Investissements d'Avenir" programme (ANR-11-IDEX-0001-02) and from the OCEVU Labex (ANR-11-LABX-0060). 
d’Avenir” French government program managed by the ANR. %
C.P. acknowledges support from the Science and Technology Foundation (FCT, Portugal) through the Postdoctoral Fellowship SFRH/BPD/90559/2012, PEst-OE/FIS/UI2751/2014, PTDC/FIS-AST/2194/2012,  and  through  the  support to the IA activity via the UID/FIS/04434/2013 fund.
\\
	HeViCS is a $\mathit{Herschel}$  open time key program (PI Davies, \citealt{davies10}). All the data are publicly available through Herschel Science Archive \url{http://archives.esac.esa.int/hsa/whsa/#home}.
\\
	$\mathit{Herschel}$ is an ESA space observatory with science instruments provided by European-led Principal Investigator consortia and with important participation from NASA. SPIRE  has  been  developed  by  a  consortium  of  institutes led by Cardiff University (UK) and including Univ. Lethbridge (Canada);  NAOC  (China);  CEA,  LAM  (France);  IFSI,  Univ. Padua (Italy); IAC (Spain); Stockholm Observatory (Sweden); Imperial  College  London,  RAL,  UCL-MSSL,  UKATC,  Univ.Sussex (UK); and Caltech, JPL, NHSC, Univ. Colorado (USA). This development has been supported by national funding agencies:  CSA  (Canada);  NAOC  (China);  CEA,  CNES,  CNRS (France); ASI (Italy); MCINN (Spain); SNSB (Sweden); STFC
(UK); and NASA (USA). This publication  makes  use  of  public data  products  from  the Two Micron All Sky Survey (2MASS),  which is a joint project of the  University  of  Massachusetts  and the Infrared Processing and Analysis Center California Institute of Technology, funded by the National Aeronautics and Space Administration and the National Science Foundation. We express are gratitude to \texttt{Daft} \url{https://github.com/dfm/daft}, a probabilistic graphical model available under the MIT License.
\end{acknowledgements}

\bibliographystyle{aa}
\bibliography{ddrisers}



\clearpage
\newpage
	\onecolumn
	\begin{longtable}
		{rrrrrrrrrr}
			\caption{List of selected sources} 
			\label{Table 8.}\\
				\hline\hline
		\\
		ID &Source name &RA [J2000 ]&Dec [J2000] & $S_{250}$&$S^{\rm err}_{250}$&$S_{350}$&$S^{\rm err}_{350}$&$S_{500}$&$S^{\rm err}_{500}$ \\
		
		(HVS)& &[h m s] & [$^{\circ}$ $'$ $''$] & [mJy] & [mJy]& [mJy]& [mJy]& [mJy]& [mJy] \\

		\hline
	\\
	\endfirsthead
    \caption{continued.}\\
	\hline\hline
	\\
	ID &Source name &RA [J2000 ]&Dec [J2000] &$S_{250}$&$S^{\rm err}_{250}$&$S_{350}$&$S^{\rm err}_{350}$&$S_{500}$&$S^{\rm err}_{500}$ \\
	
	(HVS)& & [h m s] & [$^{\circ}$ $'$ $''$]  & [mJy] & [mJy]& [mJy]& [mJy]& [mJy]& [mJy] \\
	\hline
	\\
	\endhead
	\hline
		\endfoot

    1 & HVS J120715+1453.9 & 12:07:15.07 & +14:53:17.9 & 17.81 & 1.75 & 35.07 & 3.46 & 35.94 & 3.54\\
    2 & HVS J120724+1430.0 & 12:07:24.76 & +14:30:14.0 & 20.96 & 1.87 & 40.52 & 3.61 & 40.9 & 3.64\\
    3 & HVS J120909+1433.7 & 12:09:09.18 & +14:33:26.7 & 14.03 & 1.67 & 32.46 & 3.87 & 38.84 & 4.63\\
    4 & HVS J120936+1518.0 & 12:09:36.01 & +15:18:36.0 & 17.95 & 1.69 & 37.63 & 3.55 & 40.07 & 3.78\\
    5 & HVS J121140+1416.8 & 12:11:40.83 & +14:16:10.8 & 9.86 & 2.23 & 25.62 & 5.86 & 39.8 & 5.68\\
    6 & HVS J121540+1307.9 & 12:15:40.18 & +13:07:36.9 & 11.82 & 1.61 & 28.72 & 3.91 & 37.17 & 5.06\\
    7 & HVS J121847+1537.3 & 12:18:47.43 & +15:37:39.3 & 18.22 & 1.73 & 36.82 & 3.49 & 38.61 & 3.67\\
    8 & HVS J122017+1535.9 & 12:20:17.55 & +15:35:14.9 & 18.14 & 1.86 & 34.69 & 3.56 & 34.69 & 3.56\\
    9 & HVS J122217+1534.7 & 12:22:17.80 & +15:34:11.7 & 14.62 & 1.74 & 29.46 & 3.52 & 30.18 & 3.61\\
    10 & HVS J123037+1207.1 & 12:30:37.15 & +12:07:15.1 & 16.34 & 1.75 & 33.94 & 3.64 & 35.82 & 3.84\\
    11 & HVS J122654+1214.6 & 12:26:54.33 & +12:14:22.6 & 15.88 & 1.79 & 31.50 & 3.55 & 31.82 & 3.59\\
    12 & HVS J122445+0930.7 & 12:24:45.12 & +09:30:14.7 & 13.34 & 1.65 & 29.88 & 3.70 & 34.92 & 4.32\\
    13 & HVS J123304+1309.0 & 12:33:04.61 & +13:09:16.0 & 15.45 & 1.77 & 31.86 & 3.65 & 34.05 & 3.90\\
    14 & HVS J123027+1237.7 & 12:30:27.61 & +12:37:39.7 & 13.59 & 2.04 & 28.34 & 3.63 & 30.01 & 3.85\\
    15 & HVS J122438+0953.5 & 12:24:38.28 & +09:53:41.6 & 17.27 & 1.73 & 35.37 & 3.54 & 37.56 & 3.76\\
    16 & HVS J122646+1152.1 & 12:26:46.27 & +11:52:35.0 & 17.76 & 2.09 & 35.35 & 3.49 & 36.57 & 3.61\\
    17 & HVS J122416+1208.1 & 12:24:16.90 & +12:08:16.1 & 23.64 & 1.79 & 46.08 & 3.49 & 46.82 & 3.56\\
    18 & HVS J122938+1355.7 & 12:29:38.92 & +13:55:49.7 & 21.81 & 1.81 & 42.91 & 3.56 & 43.95 & 3.65\\
    19 & HVS J121814+1529.1 & 12:18:14.42 & +15:29:36.1 & 15.73 & 1.57 & 36.9 & 3.68 & 45.05 & 4.49\\
    20 & HVS J122358+0725.1 & 12:23:58.59 & +07:25:34.1 & 13.43 & 1.67 & 32.19 & 3.92 & 41.36 & 5.04\\
    21 & HVS J121740+1613.0 & 12:17:40.07 & +16:13:26.0 & 9.3 & 1.72 & 24.17 & 4.48 & 37.71 & 6.99\\
    22 & HVS J121723+0418.0 & 12:17:23.84 & +04:18:58.0 & 14.86 & 1.75 & 30.38 & 3.55 & 31.27 & 3.65\\
    23 & HVS J122816+0824.3 & 12:28:16.15 & +08:24:02.3 & 13.41 & 1.58 & 31.65 & 3.74 & 39.0 & 4.61\\
    24 & HVS J123138+1020.0 & 12:31:38.77 & +10:20:59.0 & 11.99 & 1.58 & 28.29 & 3.73 & 35.44 & 4.67\\
    25 & HVS J123027+1237.3 & 12:30:27.76 & +12:37:37.3 & 14.32 & 2.14 & 31.91 & 3.56 & 36.42 & 3.70\\
    26 & HVS J123232+0701.9 & 12:32:32.99 & +07:01:32.9 & 18.56 & 2.05 & 37.31 & 3.51 & 38.07 & 3.58\\
    27 & HVS J123435+0839.9 & 12:34:35.03 & +08:39:47.9 & 15.65 & 1.73 & 31.67 & 3.49 & 33.25 & 3.68\\
    28 & HVS J121613+0253.1 & 12:16:13.80 & +02:53:11.1 & 36.75 & 1.78 & 70.84 & 3.44 & 71.31 & 3.46\\
    29 & HVS J121744+0721.5 & 12:17:44.72 & +07:21:27.5 & 21.99 & 1.85 & 46.42 & 3.46 & 49.75 & 3.39\\
    30 & HVS J121409+0516.7 & 12:14:09.70 & +05:16:37.7 & 15.53 & 1.73 & 31.68 & 3.52 & 32.85 & 3.65\\
    31 & HVS J121801+0348.4 & 12:18:01.13 & +03:48:10.4 & 14.03 & 2.11 & 31.22 & 3.73 & 35.51 & 4.25\\
    32 & HVS J121016+0527.1 & 12:10:16.25 & +05:27:22.1 & 15.28 & 2.16 & 31.37 & 3.62 & 32.72 & 3.77\\
    33 & HVS J122526+0414.7 & 12:25:26.88 & +04:14:52.7 & 13.68 & 2.15 & 29.71 & 3.62 & 32.83 & 3.98\\
    34 & HVS J122521+0407.8 & 12:25:23.23 & +04:07:18.8 & 12.45 & 2.38 & 28.51 & 3.84 & 33.64 & 4.53\\
    35 & HVS J122821+0407.4 & 12:28:21.13 & +04:07:32.4 & 12.81 & 1.60 & 32.32 & 4.04 & 45.43 & 5.67\\
    36 & HVS J123917+1112.4 & 12:39:17.85 & +11:12:42.4 & 13.24 & 2.21 & 28.33 & 3.58 & 30.82 & 3.91\\
    37 & HVS J122854+1328.6 & 12:28:54.43 & +13:28:05.6 & 12.72 & 1.65 & 28.21 & 3.67 & 32.57 & 4.23\\
    38 & HVS J122649+0502.2 & 12:26:49.90 & +05:02:01.2 & 13.34 & 1.65 & 29.03 & 3.62 & 32.78 & 4.11\\
    39 & HVS J121755+0721.2 & 12:17:55.74 & +07:21:00.2 & 12.69 & 1.63 & 29.85 & 3.87 & 36.63 & 4.73\\
    40 & HVS J122505+0735.6 & 12:25:05.61 & +07:35:44.6 & 24.77 & 1.77 & 50.85 & 3.25 & 53.04 & 3.07\\
    41 & HVS J121654+1604.8 & 12:16:54.97 & +16:04:14.8 & 14.05 & 1.73 & 29.65 & 3.65 & 31.78 & 3.91\\
    42 & HVS J122446+0540.0 & 12:24:46.52 & +05:40:18.0 & 14.22 & 1.79 & 29.09 & 3.67 & 30.25 & 3.81\\
    43 & HVS J122031+1551.5 & 12:20:31.49 & +15:51:26.5 & 16.36 & 1.65 & 37.25 & 3.72 & 43.61 & 4.35\\
    44 & HVS J123419+1054.2 & 12:34:19.79 & +10:54:00.2 & 17.46 & 1.57 & 40.73 & 3.66 & 49.37 & 4.48\\
    45 & HVS J122810+0512.6 & 12:28:10.78 & +05:12:02.6 & 15.66 & 1.73 & 31.94 & 3.58 & 33.11 & 3.66\\
    46 & HVS J123200+0821.9 & 12:32:00.58 & +08:21:37.9 & 18.42 & 1.86 & 41.18 & 3.65 & 47.19 & 3.67\\
    47 & HVS J121443+0344.0 & 12:14:43.75 & +03:44:29.0 & 10.08 & 1.63 & 25.98 & 4.19 & 38.98 & 6.3\\
    48 & HVS J123220+1214.0 & 12:32:20.88 & +12:14:37.0 & 13.46 & 1.67 & 28.0 & 3.53 & 30.18 & 3.80\\
    49 & HVS J121851+0301.8 & 12:18:51.09 & +03:01:53.8 & 16.38 & 1.69 & 34.8 & 3.59 & 37.58 & 3.87\\
    50 & HVS J122000+0422.0 & 12:20:00.48 & +04:22:01.0 & 18.28 & 1.61 & 40.2 & 3.54 & 45.09 & 3.97\\
    51 & HVS J122744+0948.1 & 12:27:44.28 & +09:48:01.1 & 12.12 & 1.59 & 27.86 & 3.66 & 33.04 & 4.34\\
    52 & HVS J122623+0621.1 & 12:26:23.79 & +06:21:26.1 & 16.81 & 1.55 & 40.81 & 3.78 & 52.72 & 4.88\\
    53 & HVS J122939+1355.5 & 12:29:39.00 & +13:55:49.5 & 18.62 & 1.73 & 39.84 & 3.53 & 44.21 & 3.64\\
    54 & HVS J122000+0524.9 & 12:20:00.60 & +05:24:12.9 & 17.27 & 1.77 & 34.4 & 3.54 & 34.89 & 3.59\\
    55 & HVS J122124+0911.4 & 12:21:24.48 & +09:11:46.4 & 18.33 & 1.65 & 37.64 & 3.40 & 39.28 & 3.55\\
    56 & HVS J120934+1526.0 & 12:09:34.04 & +15:26:47.0 & 29.55 & 1.69 & 60.62 & 3.48 & 63.18 & 3.62\\
    57 & HVS J122728+0939.0 & 12:27:28.12 & +09:39:00.0 & 16.81 & 1.80 & 35.48 & 3.51 & 38.03 & 3.48\\
    58 & HVS J123348+0703.5 & 12:33:48.69 & +07:03:55.5 & 17.86 & 1.69 & 37.32 & 3.54 & 39.62 & 3.76\\
    59 & HVS J122248+1635.6 & 12:22:48.00 & +16:35:19.6 & 18.42 & 1.68 & 41.2 & 3.64 & 47.15 & 4.17\\
    60 & HVS J121346+0438.2 & 12:13:46.22 & +04:38:36.2 & 19.86 & 1.79 & 41.28 & 3.67 & 43.59 & 3.88\\
    61 & HVS J122101+1449.9 & 12:21:01.77 & +14:49:40.9 & 22.0 & 1.64 & 47.46 & 3.46 & 52.08 & 3.79\\
    62 & HVS J122026+0731.4 & 12:20:26.24 & +07:31:28.4 & 21.89 & 2.18 & 43.96 & 3.37 & 44.9 & 2.89\\
    63 & HVS J122157+0526.6 & 12:21:57.97 & +05:26:35.6 & 18.95 & 1.73 & 38.51 & 3.52 & 39.79 & 3.64\\
    64 & HVS J121349+0438.1 & 12:13:49.72 & +04:38:28.1 & 31.41 & 1.58 & 73.99 & 3.69 & 90.97 & 4.54\\
    65 & HVS J123044+0950.7 & 12:30:44.45 & +09:50:49.7 & 14.14 & 2.09 & 28.58 & 3.51 & 29.99 & 3.69\\
    66 & HVS J122547+1529.7 & 12:25:47.42 & +15:29:32.7 & 21.98 & 2.13 & 45.11 & 3.53 & 47.04 & 3.68\\
    67 & HVS J122142+1611.4 & 12:21:42.97 & +16:11:50.4 & 21.81 & 2.10 & 42.79 & 3.53 & 42.81 & 3.53\\
    68 & HVS J123312+0713.5 & 12:33:12.72 & +07:13:27.5 & 15.85 & 2.27 & 30.32 & 3.45 & 30.34 & 3.46\\
    69 & HVS J122634+1400.0 & 12:26:34.07 & +14:00:10.0 & 18.18 & 2.11 & 38.23 & 3.6 & 40.85 & 3.84\\
    70 & HVS J121742+0351.1 & 12:17:42.83 & +03:51:28.1 & 11.77 & 2.19 & 26.34 & 3.69 & 30.18 & 4.23\\
    71 & HVS J122148+1443.8 & 12:21:48.69 & +14:43:52.8 & 17.55 & 2.15 & 34.65 & 3.27 & 34.86 & 3.29\\
    72 & HVS J122415+0958.8 & 12:24:15.19 & +09:58:41.8 & 22.87 & 2.23 & 46.32 & 3.50 & 47.68 & 3.60\\
    73 & HVS J121815+0307.5 & 12:18:15.98 & +03:07:43.5 & 14.73 & 1.75 & 31.08 & 3.39 & 33.31 & 3.35\\
    74 & HVS J122854+1328.3 & 12:28:54.46 & +13:28:05.3 & 15.08 & 1.54 & 33.27 & 3.41 & 37.45 & 3.84\\
    75 & HVS J121900+1550.6 & 12:19:00.96 & +15:50:19.6 & 14.52 & 1.56 & 36.32 & 3.90 & 49.9 & 5.36\\
    76 & HVS J121949+1610.5 & 12:19:49.92 & +16:10:09.5 & 17.93 & 1.81 & 37.84 & 3.42 & 40.56 & 3.41\\
    77 & HVS J121827+1544.7 & 12:18:27.36 & +15:44:15.7 & 16.02 & 1.87 & 32.89 & 3.49 & 34.31 & 3.43\\
    78 & HVS J121911+1618.4 & 12:19:11.52 & +16:18:09.4 & 13.02 & 1.67 & 28.65 & 3.67 & 32.79 & 4.21\\
    79 & HVS J123032+1014.2 & 12:30:32.38 & +10:14:07.2 & 23.69 & 2.06 & 47.57 & 3.38 & 48.59 & 3.02\\
    80 & HVS J123000+0415.5 & 12:30:00.59 & +04:15:21.5 & 19.6 & 1.61 & 40.25 & 3.31 & 42.0 & 3.47\\
    81 & HVS J122024+1434.2 & 12:20:24.48 & +14:34:23.2 & 12.5 & 1.64 & 28.23 & 3.72 & 33.34 & 4.39\\
    82 & HVS J122610+1027.9 & 12:26:10.56 & +10:27:02.9 & 20.16 & 1.78 & 42.07 & 3.62 & 45.49 & 3.92\\
    83 & HVS J123346+1007.0 & 12:33:46.32 & +10:07:03.0 & 13.78 & 2.17 & 28.29 & 3.60 & 30.1 & 3.84\\
    84 & HVS J121604+1246.7 & 12:16:04.08 & +12:46:21.7 & 13.34 & 1.75 & 28.85 & 3.68 & 32.37 & 4.14\\
    85 & HVS J123126+1249.7 & 12:31:26.16 & +12:49:43.7 & 8.93 & 1.76 & 23.24 & 4.61 & 36.93 & 7.34\\
    86 & HVS J123120+1412.2 & 12:31:20.88 & +14:12:38.2 & 13.72 & 1.68 & 28.88 & 3.55 & 30.87 & 3.81\\
    87 & HVS J122627+0336.6 & 12:26:27.53 & +03:36:31.6 & 16.27 & 1.79 & 32.58 & 3.58 & 33.9 & 3.72\\
    88 & HVS J123155+1420.9 & 12:31:55.20 & +14:20:17.9 & 13.41 & 1.71 & 28.45 & 3.63 & 30.66 & 3.91\\
    89 & HVS J121037+0517.2 & 12:10:37.19 & +05:17:22.2 & 15.93 & 1.86 & 32.7 & 3.40 & 34.11 & 3.28\\
    90 & HVS J122212+1141.7 & 12:22:12.24 & +11:41:33.7 & 15.94 & 1.86 & 32.72 & 3.41 & 34.13 & 3.35\\
    91 & HVS J122527+1153.0 & 12:25:27.12 & +11:53:28.0 & 13.64 & 1.72 & 28.83 & 3.63 & 31.58 & 3.98\\
    92 & HVS J122449+0951.2 & 12:24:49.90 & +09:51:12.2 & 15.92 & 1.71 & 32.89 & 3.53 & 34.51 & 3.74\\
    93 & HVS J122727+0525.6 & 12:27:27.06 & +05:25:29.6 & 23.19 & 2.13 & 46.57 & 3.40 & 47.57 & 2.97\\
    94 & HVS J121601+1231.4 & 12:16:01.18 & +12:31:11.4 & 14.78 & 2.13 & 36.54 & 3.84 & 48.96 & 5.18\\
    95 & HVS J122845+0929.8 & 12:28:45.36 & +09:29:34.8 & 14.58 & 1.73 & 29.13 & 3.46 & 30.23 & 3.59\\
    96 & HVS J123007+1338.1 & 12:30:07.92 & +13:38:45.1 & 21.29 & 2.05 & 42.75 & 3.30 & 43.67 & 2.90\\
    97 & HVS J122508+0756.7 & 12:25:08.64 & +07:56:02.7 & 11.9 & 1.63 & 26.49 & 3.63 & 30.72 & 4.21\\
    98 & HVS J121843+0328.8 & 12:18:43.68 & +03:28:10.8 & 18.27 & 1.81 & 35.72 & 3.54 & 36.38 & 3.61\\
    99 & HVS J121713+1202.9 & 12:17:13.96 & +12:02:42.9 & 18.93 & 2.09 & 38.0 & 3.42 & 38.82 & 3.57\\
    100 & HVS J122519+0518.1 & 12:25:19.92 & +05:18:40.1 & 16.7 & 1.78 & 31.94 & 3.42 & 31.96 & 3.62\\
    101 & HVS J122709+0856.1 & 12:27:09.42 & +08:56:26.1 & 17.52 & 1.68 & 38.42 & 3.69 & 42.96 & 4.13\\
    102 & HVS J122156+1527.0 & 12:21:56.03 & +15:27:57.0 & 18.44 & 2.02 & 37.86 & 3.41 & 39.49 & 3.77\\
    103 & HVS J123003+1318.3 & 12:30:03.80 & +13:18:04.3 & 22.1 & 1.77 & 46.63 & 3.47 & 49.98 & 3.69\\
    104 & HVS J122619+0443.7 & 12:26:19.44 & +04:43:37.7 & 14.24 & 1.82 & 30.05 & 3.47 & 32.21 & 3.77\\
    105 & HVS J121735+0413.8 & 12:17:35.04 & +04:13:42.8 & 15.58 & 1.98 & 31.98 & 3.52 & 33.36 & 4.37\\
    106 & HVS J121451+0412.5 & 12:14:51.58 & +04:12:44.5 & 11.74 & 1.67 & 26.36 & 3.75 & 30.31 & 4.32\\
    107 & HVS J121617+0408.0 & 12:16:17.76 & +04:08:21.0 & 13.08 & 1.62 & 29.6 & 3.66 & 35.0 & 4.33\\
    108 & HVS J123427+0714.0 & 12:34:27.98 & +07:14:54.0 & 25.39 & 2.03 & 50.99 & 3.31 & 52.08 & 3.93\\
    109 & HVS J121543+0619.5 & 12:15:43.92 & +06:19:12.5 & 16.35 & 1.75 & 33.65 & 3.61 & 35.9 & 3.85\\
    110 & HVS J123007+1332.1 & 12:30:07.93 & +13:32:07.1 & 18.17 & 1.92 & 37.3 & 3.42 & 38.91 & 3.54\\
    111 & HVS J122414+1014.6 & 12:24:14.05 & +10:14:36.6 & 21.98 & 2.02 & 45.12 & 3.38 & 47.06 & 3.78\\
    112 & HVS J121304+0519.0 & 12:13:04.30 & +05:19:28.0 & 11.33 & 1.66 & 25.84 & 3.74 & 30.3 & 4.42\\
    113 & HVS J122517+1121.3 & 12:25:17.75 & +11:21:56.3 & 13.98 & 2.53 & 30.82 & 3.46 & 34.68 & 3.87\\
    114 & HVS J122100+1306.9 & 12:21:00.68 & +13:06:51.9 & 30.94 & 2.13 & 62.14 & 3.38 & 63.47 & 3.95\\
    115 & HVS J122709+0540.2 & 12:27:09.60 & +05:40:00.2 & 20.09 & 2.86 & 38.74 & 3.52 & 39.02 & 3.57\\
    116 & HVS J123213+0740.8 & 12:32:13.44 & +07:40:30.8 & 12.36 & 2.68 & 27.12 & 3.70 & 30.92 & 4.21\\
    117 & HVS J122109+0946.9 & 12:21:09.49 & +09:46:33.9 & 30.53 & 2.12 & 61.3 & 3.30 & 62.62 & 3.83\\
    118 & HVS J121904+1338.9 & 12:19:04.56 & +13:38:53.9 & 25.03 & 1.96 & 51.39 & 3.44 & 53.6 & 3.21\\
    119 & HVS J123619+0909.5 & 12:36:19.04 & +09:09:44.5 & 26.06 & 2.13 & 52.33 & 3.31 & 53.45 & 3.84\\
    120 & HVS J121818+0506.4 & 12:18:18.40 & +05:06:48.4 & 35.85 & 2.05 & 71.98 & 3.38 & 73.53 & 3.04\\
    121 & HVS J122611+1518.8 & 12:26:11.82 & +15:18:33.8 & 24.69 & 2.02 & 50.7 & 3.41 & 52.88 & 3.23\\
    122 & HVS J123032+0641.8 & 12:30:32.44 & +06:41:54.8 & 31.33 & 2.04 & 62.91 & 3.29 & 64.26 & 3.93\\
    123 & HVS J121110+1220.3 & 12:11:10.88 & +12:20:35.3 & 24.74 & 2.06 & 49.68 & 3.30 & 50.75 & 3.90\\
    124 & HVS J123113+1220.1 & 12:31:13.14 & +12:20:43.1 & 16.38 & 2.14 & 33.62 & 3.44 & 35.07 & 3.29\\
    125 & HVS J122646+1049.3 & 12:26:46.02 & +10:49:06.3 & 16.22 & 1.66 & 34.18 & 3.50 & 36.57 & 3.75\\
    126 & HVS J121900+1516.5 & 12:19:00.34 & +15:16:49.5 & 21.75 & 1.92 & 44.65 & 3.41 & 46.57 & 3.23\\
    127 & HVS J122806+0928.7 & 12:28:06.04 & +09:28:11.7 & 39.83 & 2.03 & 79.97 & 3.32 & 81.69 & 3.97\\
    128 & HVS J123000+0803.6 & 12:30:00.33 & +08:03:26.6 & 21.04 & 1.99 & 43.19 & 3.41 & 45.05 & 3.15\\
    129 & HVS J121609+1601.2 & 12:16:09.60 & +16:01:59.2 & 14.11 & 1.64 & 31.25 & 3.64 & 36.04 & 4.19\\
    130 & HVS J122241+0708.4 & 12:22:41.71 & +07:08:26.4 & 21.1 & 1.71 & 44.93 & 3.65 & 49.44 & 4.03\\
    131 & HVS J122335+1108.0 & 12:23:35.52 & +11:08:56.0 & 12.35 & 1.63 & 26.86 & 3.61 & 30.33 & 4.08\\
    132 & HVS J123304+1326.6 & 12:33:04.80 & +13:26:27.6 & 14.75 & 1.68 & 32.51 & 3.70 & 37.26 & 4.25\\
    133 & HVS J123317+1235.2 & 12:33:17.76 & +12:35:31.2 & 14.29 & 1.77 & 28.74 & 3.58 & 30.03 & 3.74\\
  

	\end{longtable}
	
Column descriptions: 
	 
Column 1: Catalogue numbers (HVS);

Column 2: Full source name (source names are based on their positions for the J2000).

Columns 3-4: Coordinates of sources (RA, Dec);

Column 4-9: SPIRE fluxes ($S_{250}$, $S_{350}$, $S_{500}$) and accompanied errors in the flux estimation ($S^{\rm err}$). Flux errors are estimated as explained in \hyperref[sec:2.6] {Section 2.6}. Note, however, that errors are strongly correlated, since we performed "forced" MBB photometry at SPIRE wavelengths.

	\twocolumn
\begin{appendix} 
\section{Dusty radio sources}
\label{sec:appA}
As mentioned in \hyperref[sec:2.5]{Section 2.5}, our final list of "$500\:\mu$m-risers" is cleaned from strong radio sources and quasars (QSOs, 7 objects in total). We measure significant FIR emission close to position of these sources, and here we briefly investigate its nature. Radio emission from  quasars may be due to star formation in the quasar host galaxy, to a jet launched by the supermassive black hole, or to relativistic particles accelerated in a wide-angle radiatively-driven outflow. Our MBB fitting approach "protect" quiet well star-forming selection against the flat (in terms of SED) and dusty synchrotron sources. When \texttt{MBB-fitter} photometry of such flat sources is performed, most of  $\chi^2$ values lie outside the range we accept as good to our final selection ($\chi^2<3.84$, see \hyperref[sec:appB]{$\rm Appendix\:B$}). Between 7 objects removed from the final list, 3 of them are defined as radio-loud quasars (QSO) or blazars, with catalogued L-band 1.4 GHz radio emission larger than 1 mJy (5.04 mJy, 799.97 mJy and 6.2 mJy respectively), and their redshifts span the range from 0.6 to 0.965. An example is the brightest radio source catalogued as HVS387 in our catalogue or J122222.5+041316 in the FIRST database. It has IR excess and very strong UV and gamma-ray emission. Source has a spectroscopic redshift of $z=0.964$ and according to the classical PSF-fitting photometry performed with e.g \texttt{SUSSEXtractor} or \texttt{DAOphot} it could passes our red source criteria, but it cannot be fitted accurately using the \texttt{MBB-fitter}. Since they may have colours similar to those of high-$z$ DSFGs, heterogeneously selected  "$500\:\mu$m-risers" that exist in the literature thus should be considered with caution, since they may be polluted by synchrotron-dominated sources whose IR energy is not star-formation related. Another possibility is that some of those QSOs act as a foreground lens, amplifying the signal from the higher-redshift, dusty system.

Another 4 QSO-like sources are found by cross-matching with the Half Million Quasar catalogue (HMQ, \citealt{hmq15}). They are optically bright, high-$z$ QSO-like sources close to the position of our SPIRE detection (r<3"). They have high photometric redshifts - in average $z=2.74$, with one sources lying at $z>4$. 
To fully inspect relation of these objects to the population we aim to select in the current study, in our following paper we consider optical and PACS data, together with archival WISE data. Here we can determine expecting radio emission of our final group of 133 "$500\:\mu$m-risers" implementing FIR-radio correlation: 
\begin {equation}
\label{eq:qir}
q_{\rm IR}	=\log\left(\frac{F_{\rm FIR}}{3.75\times10^{12}{\rm Wm^{-2}}}\right)-\log\left(\frac{S_{1.4}}{\rm Wm^{-2}Hz^{-1}}\right)
\end {equation}
where $S_{1.4}$ is the continuum radio emission flux at 1.4 GHz per frequency such that 
$S_{1.4} \propto \nu^{-\alpha}$ and $\alpha$ is the radio spectral index, positive in vast majority of sources. 
$F_{\rm FIR}$ is the rest-frame FIR dust emission flux.
There is considerable evidence that local star-forming galaxies express a tight FIR-radio correlation. For example, \cite{yun01} analyzed the sample of 1800 {\it IRAS} galaxies and measured this value to be  $q_{\rm IR}=2.37\pm0.01 $  with a dispersion of 0.25 dex. From this we estimate that for our $z\sim4$ candidates 1400 MHz flux densities are in the range between 25 $\mu$Jy to 90 $\mu$Jy. Assuming the faintest level of existing radio surveys covering Virgo ($\sim$0.75-1 mJy at 1.4 GHz), we can reveal just local cluster members and radio bright quasars, but not dusty background, star-forming galaxies. 

However, FIR-radio correlation cannot be implemented accurately in order to estimate expected radio flux of optically selected quasars which have radio flux below the FIRST (NVSS) limit. As we add QSO contribution
to both radio and IR, correlation might break down. It could be particularly due to the origin of the FIR/submm emission which includes a small contribution from the AGN torus, being predominantly linked to (kpc-scale) heated dust by the AGN. Recent studies have enlarged sample of $\mathit{Herschel}$ detected optically selected QSOs (e.g. \citealt{qso16}). They, however, claimed statistically large group of objects for which SEDs are due to the cold-dust components within the host galaxies, consistent with being heated by active star formation. Lacking the $L_{IR}$ correlation with the black hole masses or the X-ray QSO luminosities of the quasars, it could support scenario that their thermal SED is not AGN-driven. Thus, our subsample of QSO-like objects might be very interesting case study for probing the connection between AGN activity and host galaxy star-formation.
Existing large sky radio catalogues are not deep enough to distinguish between very dusty radio dominated galaxies at higher redshifts and star-forming systems. This issue will be addressed with deep radio observations performed over the area covering Virgo. Evolutionary Map of the Universe (EMU survey, \cite{EMU} will make a deep (10", $rms$=$10\:\mu$Jy/beam) continuum maps. 

It is clear that in addition to FIR/submm data, radio data are of significant importance to further uncover true nature of selected dusty sources. Since photometric redshifts calculated using the SPIRE data alone are highly uncertain, adding of radio data (via FIR-radio correlation), one might expect photometric redshift accuracy will be improved, for example to $\Delta{z}/(1+z)=0.15$ (\citealt{roseboom12}). From this, it is vital to understand true nature of FIR to radio correlation at redshifts higher than 2, since latest results claimed much larger scatter and even break at redshifts larger than 4 (e.g. \cite{miettinen16}, \cite{delhaize17}, see also \citealt{schober17}). Thus it cannot be used straightforwardly to break temperature-redshift degeneracy as suggested in most of previous papers found in the literature (e.g. \citealt{pope10}). Approaching existing and future interferometers (JVLA, SKA), one we can put constrains on photometric redshifts of our red sources, but also uncover sizes of their star-forming regions and investigate how they compare to other tracers such are FIR or CO.  

\section{Color-redshift diagram related to the MBB model}
\label{sec:appB}
In the FIR regime, SED peak is provoked by thermal emission from dust of different temperatures, and it can be used in extragalactic surveys such as $\it{Herschel}$ to identify objects in a specific redshift range. To test our approach that selecting "$500\:\mu$m-risers" is the way to select $z\gtrsim4$ candidates, we run Monte Carlo simulation to sample all possible colours as a function of redshift. We test which range of redshifts we should expect assuming the MBB model with the fixed various ranges of parameters. We confirm that if sources are not sufficiently cold, they reside at $z\gtrsim4$. Thus, we are particularly interested to unveil MBB-fitted sources having $S_{500}/S_{350}>1$ and $S_{500}/S_{250}>2$. Plotted in Fig. B.1 are such examples, where we applied different combinations of fixed $T_{\rm d}$ and  $\beta$. At redshifts $z>6$ the heating of dust by the cosmic microwave background (CMB) could be more significant and we take that into account in our modelling of observed SED following \cite{daCunha13}. 
  	
  \begin{figure*}[ht]
  	\centering
  	\includegraphics[width=17.39cm] {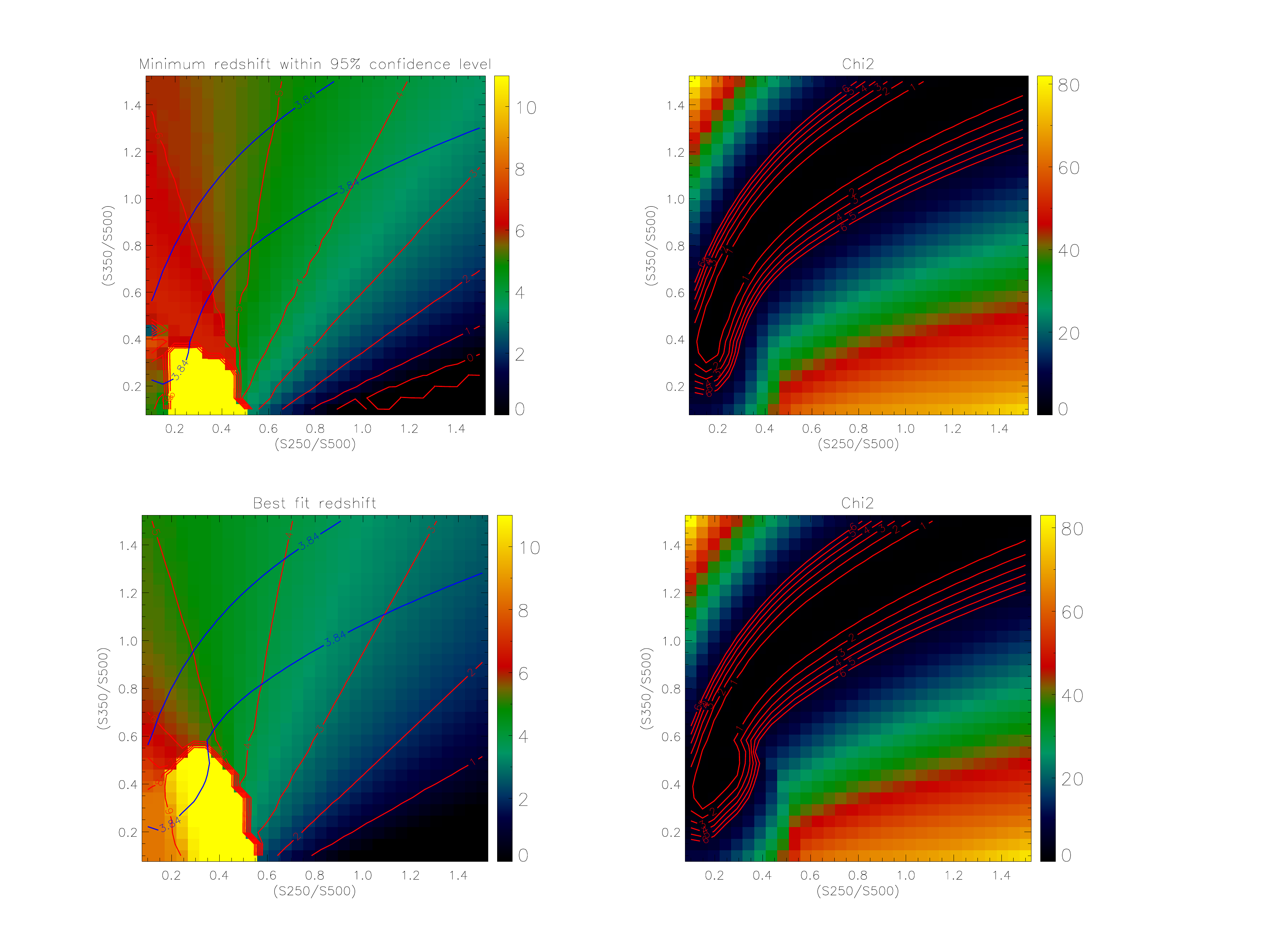} 

  	\caption{Criteria for selecting "$500\:\mu$m-risers". Plotted is SPIRE colour diagram related to the modified blackbody (MBB) model. The plot is a result of a Monte Carlo simulation in which we start with a single-temperature modified blackbody with the flux  $S_{500}$=30 mJy. The mock SEDs are generated to account different fixed $T_{\rm d}$ and $\beta$ illustrating temperature-redshift degeneracy. Plotted against all possible colours are simulated sources with $T_{\rm d}$=45 K and  $\beta=1.8$ (upper panel) and $T_{\rm d}$=38 K and $\beta=1.8$ (lower panel).  $\mathit{Left:}$ Colour-redshfit plot. Red lines represent redshift tracks rising from right to left. The coloured background indicates the average redshift in these colour-colour spaces. Blue lines define the 95$\%$ confidence limit region. 
   Correction factor due to modelled effect of CMB heating on colours is applied. However, our simulations shown that we do not expect major contribution from that effect at sources lying below $z\sim7$; $\mathit{Right:}$ Different chi-square fitting values reflected on SPIRE colour-colour ratios. For our final "$500\:\mu$m-risers" selection we kept just those sources whose chi-square reside inside 95$\%$ confidence limit region ($\chi^{2}<3.84$).}
  	\label{fig:multimbb}
  	\end{figure*}	

\section{An example of faint and deblended "$500\:\mu$m-risers"}
\label{sec:appC}

\begin{figure*}
	\centering
	\includegraphics[width=17.89cm] {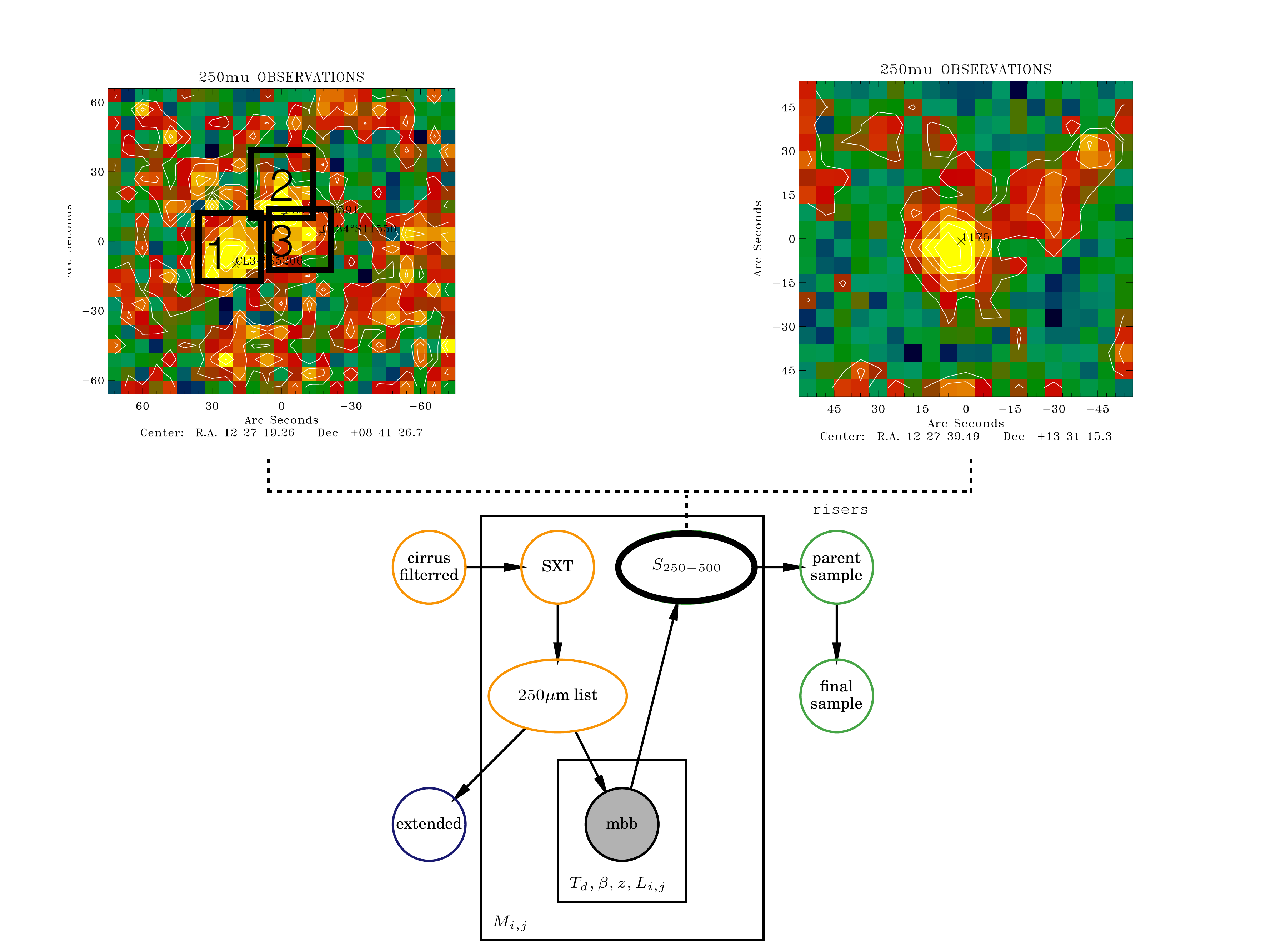}  
	\caption{Example 2D cutouts of confused (left) and isolated point source detections (right) from our 250 $\mu$m catalogue. The first group of sources are fitted simultaneously, while second one is fitted as a single source. }
	\label{fig:cluster}
\end{figure*}
\begin{figure*}[ht]
	\centering
	\includegraphics[width=10.09cm] {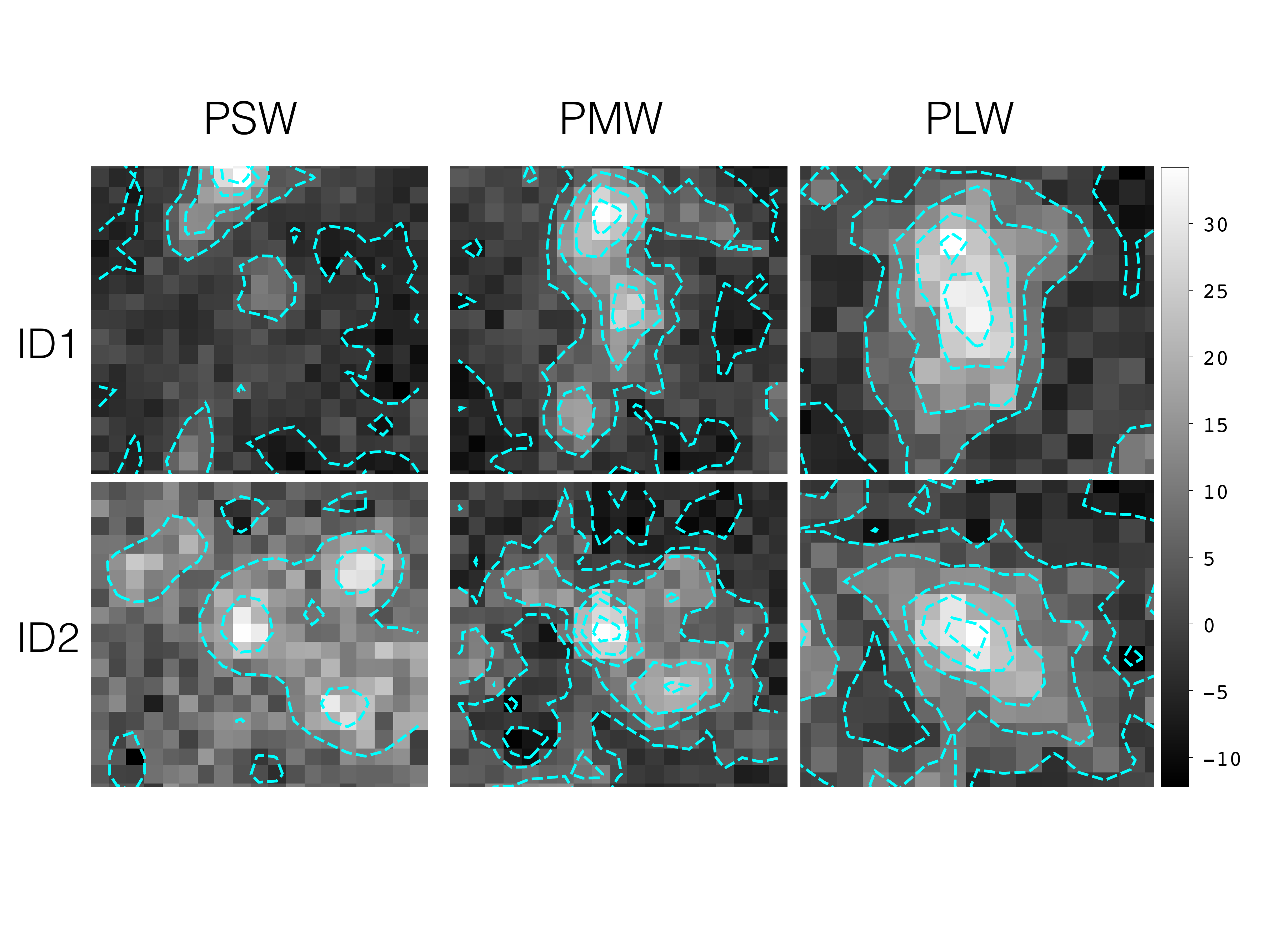}  
	\caption{Example 60"x 60" cutouts of "$500\:\mu$m-risers" from our final sample. Both sources are detected in the second iteration of our extraction procedure, when we add sources that appeared bright in our residual maps. Upper panel: 250 $\mu$m, 350 $\mu$m and 500 $\mu$m cutout of an multisource case where two sources are de-blended for their 350 $\mu$m and 500 $\mu$m emission. Lower panel: An isolated case, without a prominent 350 $\mu$m or 500 $\mu$m blend in respect to the beam.}.
	\label{fig:risers}
\end{figure*}


\end{appendix}
\end{document}